\documentclass[prb,onecolumn,showpacs,superscriptaddress,preprintnumbers,amssymb,nofootinbib]{revtex4-2}
\usepackage{graphicx}
\usepackage[bookmarks=true,colorlinks=true,linkcolor=magenta, urlcolor=blue, citecolor=blue]{hyperref}
\usepackage{bm}
\usepackage[normalem]{ulem}
\usepackage[section]{placeins}
\usepackage{amsmath}
\usepackage{booktabs}
\usepackage{multirow}
\usepackage{xcolor}
\usepackage{mwe}
\usepackage{changes}
\usepackage[export]{adjustbox}
\usepackage{tikzscale}
\makeatletter 
\let\l@subsubsection\@gobbletwo
\makeatother
\definechangesauthor[name={KS}, color=turquoise]{KS}
\newcommand{\h}[1]{\hat{#1}}

\newcommand{\td}{\tilde{t}}
\newcommand{\bn}{\mathbf{n}}
\newcommand{\mm}{\mathbf{m}}
\newcommand{\bd}{\mathbf{d}}
\newcommand{\beq}{\begin{equation}}
\newcommand{\eeq}{\end{equation}}
\newcommand{\beqn}{\begin{eqnarray}}
\newcommand{\eeqn}{\end{eqnarray}}

\newcommand{\vect}[1]{{\bm{#1}}}
\newcommand{\tphi}{\tilde{\phi}}

\newcommand{\ra}{\rightarrow}

\newcommand{\cL}{ {\cal L} }

\newcommand{\cE}{ {\cal E} }
\renewcommand{\vect}[1]{{\bm{#1}}}
\newcommand{\ii}{\mathrm{i}}

\newcommand{\hphi}{\hat{\phi}}
\newcommand{\llangle}{\langle\!\langle}
\newcommand{\rrangle}{\rangle\!\rangle}

\newcommand{\cx}[1]{{\color{black} #1}}

\usepackage[utf8]{inputenc}

\usepackage{amsmath}
\usepackage{amstext}
\usepackage{amssymb}
\usepackage{amsfonts}
\usepackage{amsthm}
\usepackage{mathtools} 

\usepackage{physics}
\usepackage{enumitem}
\usepackage{bbm}

\usepackage{caption}
\usepackage{overpic}

\usepackage{multirow}
\usepackage{array}

\usepackage{tikz}
\usetikzlibrary{positioning}
\usetikzlibrary{calc}
\def\[#1\]{\begin{align}#1\end{align}}

\newcommand{\calL}{\mathcal{L}}

\newcommand{\one}{\mathbbm{1}}

\renewcommand{\ket}[1]{|#1\rangle}
\renewcommand{\braket}[2]{\langle#1|#2\rangle}
\renewcommand{\mel}[3]{\langle#1|#2|#3\rangle}

\RenewDocumentCommand{\v}{s m}{%
  \IfBooleanTF{#1}%
    {\bm{\hat{#2}}}%
    {\bm{#2}}%
}

\NewDocumentCommand{\D}{s m}{%
  \IfBooleanTF{#1}%
    {\mathrm{d}#2}%
    {\mathrm{d}#2\,}%
}

\newcommand{\kett}[1]{| #1 \rangle\rangle}
\newcommand{\braakett}[2]{\langle\langle #1 | #2 \rangle\rangle}

\makeatletter
\newcommand{\phantomlabel}[2]{
    \protected@write\@auxout{}{
        \string\newlabel{#2}{
            {\@currentlabel#1}{\thepage}
            {\@currentlabel#1}{#2}{}
        }
    }
    \hypertarget{#2}{}
}
\makeatother
\usepackage{amsmath}
\DeclareMathOperator*{\argmax}{arg\,max}

\begin{document}
\title{Strong-to-Weak Symmetry Breaking in Open Quantum Systems: From Discrete Particles to Continuum Hydrodynamics}

\author{Jacob Hauser}
\thanks{These authors contributed equally to this work.}
\affiliation{Department of Physics, University of California, Santa Barbara, California 93106, USA}

\author{Kaixiang Su}
\thanks{These authors contributed equally to this work.}
\affiliation{Department of Physics, University of California, Santa Barbara, California 93106, USA}

\author{Hyunsoo Ha}
\thanks{These authors contributed equally to this work.}
\affiliation{Department of Physics, Princeton University, Princeton, New Jersey 08544, USA}

\author{Jerome Lloyd}
\thanks{These authors contributed equally to this work.}
\affiliation{Department of Theoretical Physics, University of Geneva, Geneva, Switzerland}

\author{Thomas G. Kiely}
\affiliation{Kavli Institute for Theoretical Physics, University of California, Santa Barbara, CA 93106, USA}

\author{Romain Vasseur}
\affiliation{Department of Theoretical Physics, University of Geneva, Geneva, Switzerland}

\author{Sarang Gopalakrishnan}
\affiliation{Department of Electrical and Computer Engineering, Princeton University, Princeton NJ 08544, USA}

\author{Cenke Xu}
\affiliation{Department of Physics, University of California, Santa Barbara, California 93106, USA}

\author{Matthew P. A. Fisher}
\affiliation{Department of Physics, University of California, Santa Barbara, California 93106, USA}

\begin{abstract}

We explore the onset of spontaneous strong-to-weak symmetry breaking (SW-SSB) under $U(1)$-symmetric (i.e., charge-conserving) open-system dynamics. We define this phenomenon for quantum states and classical probability distributions, and explore it in three complementary models, one of which exhibits nontrivial quantum coherence at short times. Our main conclusions are as follows. In one dimension, the strong symmetry is not spontaneously broken at any finite time; however, correlators probing strong-to-weak symmetry breaking develop order on length scales that grow linearly in time, parametrically faster than charge diffusion. We provide numerical evidence for this scaling in multiple distinct probes of SW-SSB, and derive it from a field-theory analysis. Moreover, we relate this scaling to the problem of inferring the charge inside a subregion by measuring its surroundings, and construct explicit decoding protocols that illustrate its origin. In two dimensions, field theory and numerical simulations support a finite-time Berezinskii--Kosterlitz--Thouless-like SW-SSB transition. Within continuum hydrodynamics, by contrast, SW-SSB happens at infinitesimal time in two or more dimensions. The SW-SSB transition time can thus be interpreted as marking the emergence of a continuum hydrodynamic description, or (more precisely) the timescale beyond which non-hydrodynamic information such as discrete particle worldlines can no longer be inferred. We support this picture by analyzing a model in which we exploit SW-SSB to derive a classical stochastic hydrodynamic description from the underlying quantum dynamics.

\end{abstract}

\maketitle

\tableofcontents

\section{Introduction}\label{sec:introduction}
Systems coupled to the outside world can undergo information-theoretic transitions, which differ in many crucial ways from equilibrium phase transitions. The quantum error correction threshold \cite{shor1996fault, aharonov1997fault, knill1998resilient, kitaev2003fault, dennis2002topological} is the most studied example of such a transition, but recent work has revealed a much broader class of transitions occurring in systems coupled to external environments. Beyond quantum error correction, known information-theoretic transitions include measurement-induced criticality \cite{Li_2018,Li_2019,Skinner_2019}, separability transitions \cite{chen2024separability, chen2024symmetry}, teleportation transitions~\cite{PRLfinitet,Googleteleportation}, complexity transitions~\cite{RTN,NappFiniteTtransition}, topological phase transitions driven by decoherence \cite{Ma_2023, lu2023mixed, wang2025intrinsic, ma2025symmetry, sohal2025noisy, ellison2025toward, guo2025strong, schafer2025symtft}, and strong-to-weak spontaneous symmetry breaking (SW-SSB) \cite{Lee_2023, lessa2025strong, gu2025spontaneous,  sala2024spontaneous, huang2025hydrodynamics, zhang2025strong, kim2024error, chen2025strong, sa2025exactly, ziereis2025strong, zerba2025strong, hauser2025information}. The latter two transitions are also known in the literature as mixed-state phase transitions. 
A crucial distinction between information-theoretic (or mixed-state) transitions and conventional phase transitions is that the former can occur at finite time: for example, an error-correcting code subjected to errors at rate $\gamma$ (absent error correction) will typically accumulate enough errors by time $\sim 1/\gamma$ that the logical information is irretrievable. Under spatially local dynamics, whether unitary or dissipative, standard observables such as correlation functions---which are linear in the system's density matrix---can only spread inside a causal light cone, so the associated correlation lengths cannot diverge at finite time. Accordingly, such observables are not singular at information-theoretic transitions. Instead, information-theoretic phase transitions can be diagnosed in one of two ways: either by the performance of information-processing tasks (such as error correction~\cite{dennis2002topological}), or by nonlinear functions of the system's density matrix. In the latter case, various Rényi correlation measures \cite{Lee_2023, zhang2024fluctuation, liu2025diagnosing, sun2025scheme, weinstein2025efficient}, as well as the so-called ``Markov length'' which measures the decay scale of conditional mutual information \cite{sang2025stability, sang2025mixed, negari2024spacetime,  zhang2025conditional, lloyd2025diverging, chen2025local}, have emerged as powerful diagnostics of mixed state phase transitions. 

SW-SSB is an information-theoretic analogue of spontaneous symmetry breaking that occurs in mixed states \cite{lessa2025strong}. Unlike pure states, mixed states can be invariant under symmetries in two ways---strong and weak. In this work we will focus on the case of $U(1)$ symmetry, associated with a conventional scalar charge. In this context, a (globally) strongly symmetric state is one with a definite value of the total charge in the system, while a weakly symmetric state can be any incoherent mixture of distinct charge sectors. Intuitively, SW-SSB occurs when a state has a sharp value of the global charge, but this information is not stored (or retrievable) locally. An instructive example of a state with SW-SSB is the maximally mixed state in a fixed charge sector. This is an ensemble of all bit-strings with a global charge constraint. Locally, this constraint is invisible and the state looks maximally mixed; however, if part of the state is lost, one needs \emph{global} data (namely, the total charge of the rest of the system) to correctly restore it. By contrast, a highly-structured bit string like ``$010101\ldots$" can be restored by a local dissipation which forces each bit to take its assigned value: in states like this, we say the strong symmetry is ``intact'' and not spontaneously broken.

As the example above might suggest, SW-SSB has profound connections to thermalization and (in the $U(1)$ case) hydrodynamics \cite{huang2025hydrodynamics}. Under generic nonintegrable dynamics on the lattice, a conserved charge will spread diffusively\footnote{One exception to this is the case of traditional $U(1)$ SSB, where a superfluid with ballistic charge modes can arise. However, such transitions (where a weak symmetry is spontaneously broken down to nothing) are not the focus of this work.}. 
The local density matrix looks thermal on length scales that are limited by diffusion, i.e., $x \sim \sqrt{t}$, and conventional correlation functions also only spread on this scale. It is tempting to conjecture that since the system looks maximally mixed on the diffusive scale, probes of SW-SSB will also seem to develop order on this scale.
We argue that, in fact, such probes expose interesting dynamics that are entirely distinct from the diffusion of charge.
In one dimension, \mbox{SW-SSB} does not occur at any finite time, but the length scale associated with SW-SSB correlators grows parametrically faster than charge spreads (specifically, it grows linearly in time).
In two or more dimensions, this length scale grows still faster, diverging at a finite $t_c$, past which the SW-SSB correlators are quasi-long-range ordered (in 2d) or long-range ordered (in higher d). 
These results are closely parallel to those for conventional superfluidity in $d$ dimensions; we will make this analogy explicit below.

We will provide evidence for these claims in three models, which provide complementary perspectives on the universal features of $U(1)$-symmetric open quantum dynamics. Two of these models are purely dissipative: the first is a microscopic spin-$1/2$ Lindblad dynamics and the second is a phenomenological decohered $U(1)$ rotor model. The spin-$1/2$ model maps to a well-known classical model (the ``simple symmetric exclusion process'', or SSEP), and may be efficiently simulated, as we do in one and two spatial dimensions. The rotor model admits an analytical replica field theory and renormalization-group (RG) treatment, allowing a direct comparison to the numerics in 1d and 2d. The third model adds coherent Hamiltonian evolution on top of the $U(1)$-symmetric decoherence. This extension lets us discuss how finite-time SW-SSB transitions in higher dimensions can be accompanied by an emergent long-wavelength classical hydrodynamics.

In one dimension, a continuous $U(1)$ symmetry forbids spontaneous symmetry breaking at any finite evolution time. Accordingly, the SW-SSB order parameters we study do not develop true long-range order. Instead, both the mixed-state correlators and the information-theoretic diagnostics decay exponentially at long distances. While their short-distance behavior can be nonuniversal, the asymptotic decay is characterized by a correlation length that grows linearly in time. This linear growth is reminiscent of one-dimensional finite-temperature bosonic systems, where $\beta$-periodic worldlines of a conserved $U(1)$ charge lead to an exponential correlation length set by the inverse temperature~$\beta$. For the mixed-state correlators, the relevant length scale is controlled by boundary-condition constraints on these charge worldlines, rather than by ordinary diffusive charge spreading. Finally, from a decoding perspective on charge fluctuations, we explain why the information-theoretic diagnostics are governed by the same ballistic length scale rather than by diffusion.

In higher dimensions we expect qualitatively different behavior, including a finite-time SW-SSB transition. We develop an analytical framework based on a replica rotor field theory that applies in arbitrary spatial dimension.
In one dimension, we benchmark this theory against numerics and find that it captures the correct physics, including the predicted factor-of-two relation between the correlation lengths extracted from the R\'enyi-$1$ and R\'enyi-$2$ correlators.
In two dimensions, the same framework predicts a Berezinskii--Kosterlitz--Thouless (BKT)-type transition \cite{berezinskii1971destruction, kosterlitz1972long, kosterlitz1973ordering}. 
A key feature is that the replica constraint changes the relevant topological defects from single vortices to vortex dipoles across replicas. 

By including coherent Hamiltonian evolution together with decoherence in the rotor setting, we can address the higher-dimensional regime where a finite-time SW-SSB transition is expected. 
We consider a generic $U(1)$-symmetric open-system dynamics built from simple Lindblad generators, including on-site dephasing and decoherence of charge-conserving hopping terms. While at early times the density matrix can remain highly nontrivial due to coherent evolution, we argue that once SW-SSB occurs at a finite time, the long-wavelength dynamics is effectively described by a classical stochastic hydrodynamic description. In our model, the resulting long-wavelength theory is the standard Hohenberg-Halperin Model~F, a well-known stochastic classical hydrodynamic description of a $U(1)$ order parameter coupled to an additional conserved density \cite{RevModPhys.49.435}. We conjecture that, for generic $U(1)$-symmetric systems with decoherence in sufficiently high spatial dimension, a finite-time SW-SSB drives a quantum-to-classical transition, with a long-wavelength hydrodynamic description emerging.

We finally study the possibility of SW-SSB starting directly from classical continuum hydrodynamic equations. We verify the scalings of the R\'enyi correlators computed in the previous models in $d=1$ to $3$; however, we observe that there is no possibility for a finite-time transition in $d>1$ (the SW-SSB occurs at infinitesimal times). Moreover, we find that the Markov length diverges at all times in the 1d hydrodynamic theory. These findings point to discreteness of charge as being an essential feature for the SW-SSB \emph{transition} to occur, at finite times. 

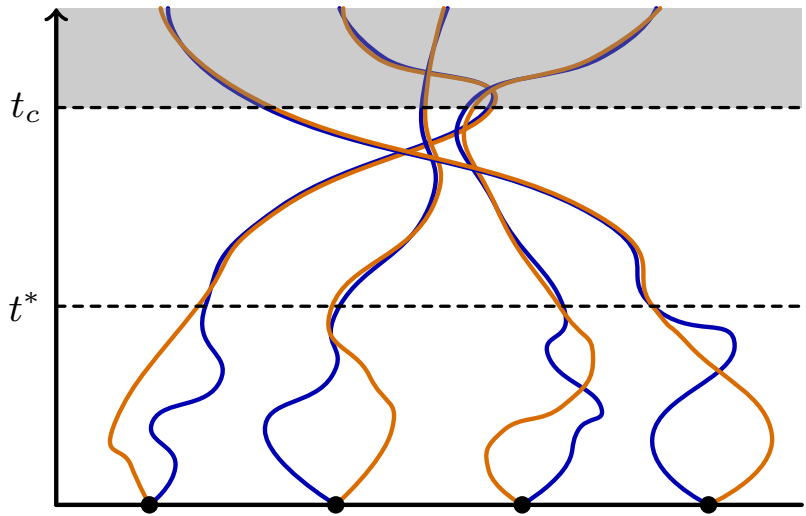
\begin{figure}[!t]
\centering   
\resizebox{0.6\linewidth}{!}{
\def\FIGW{6cm}
\def\FIGH{4cm}
\def\RAILLW{1pt}
\def\DASHW{0.75pt}
\def\WORLDLW{1pt}
\def\DOTR{2pt}
\def\ARROWY{0.125}

\def\FIGFONTSIZE{\small}%

\colorlet{WORLDCOL1}{blue!70!black}%
\colorlet{WORLDCOL2}{orange!85!black}%
\colorlet{ARROWCOL}{black}%

\def\PARTICLES{0.125,0.375,0.625,0.875}%

\def\ARROWXS{0.125,0.25,0.375,0.5,0.625,0.75,0.875}%

\begin{tikzpicture}[
  x=\FIGW, y=\FIGH,
  line cap=round, line join=round,
  every node/.style={font=\FIGFONTSIZE} 
]



\draw[line width=\WORLDLW, draw=WORLDCOL1]
  plot[smooth, tension=0.9] coordinates {
    (0.38,1.0)
    (0.44,0.9)
    (0.58,0.8)
    (0.30,0.6)
    (0.20,0.4)
    (0.22,0.25)
    (0.13,0.17)
    (0.16,0.08)
    (0.125,0)
  };

\draw[line width=\WORLDLW, draw=WORLDCOL2]
  plot[smooth, tension=0.9] coordinates {
    (0.385,1.0)
    (0.43,0.9)
    (0.585,0.8)
    (0.305,0.6)
    (0.19,0.4)
    (0.08,0.17)
    (0.10,0.08)
    (0.125,0)
  };



\draw[line width=\WORLDLW, draw=WORLDCOL1]
  plot[smooth, tension=0.9] coordinates {
    (0.525,1.0)
    (0.49,0.8)
    (0.50,0.6)
    (0.38,0.4)
    (0.37,0.26)
    (0.28,0.15)
    (0.375,0)
  };

  \draw[line width=\WORLDLW, draw=WORLDCOL2]
  plot[smooth, tension=0.9] coordinates {
    (0.52,1.0)
    (0.495,0.8)
    (0.505,0.6)
    (0.37,0.4)
    (0.42,0.26)
    (0.45,0.15)
    (0.375,0)
  };



\draw[line width=\WORLDLW, draw=WORLDCOL1]
  plot[smooth, tension=0.9] coordinates {
    (0.805,1.0)
    (0.73,0.9)
    (0.55,0.8)
    (0.59,0.6)
    (0.68,0.4)
    (0.66,0.3)
    (0.73,0.2)
    (0.70,0.15)
    (0.68,0.07)
    (0.625,0)
  };

\draw[line width=\WORLDLW, draw=WORLDCOL2]
  plot[smooth, tension=0.9] coordinates {
    (0.81,1.0)
    (0.72,0.9)
    (0.56,0.8)
    (0.59,0.6)
    (0.675,0.4)
    (0.72,0.3)
    (0.68,0.2)
    (0.60,0.15)
    (0.58,0.07)
    (0.625,0)
  };



  \draw[line width=\WORLDLW, draw=WORLDCOL1]
  plot[smooth, tension=0.9] coordinates {
    (0.15,1)
    (0.28,0.80)
    (0.7,0.6)
    (0.8,0.4)
    (0.91,0.32)
    (0.8,0.14)
    (0.875,0)
  };

\draw[line width=\WORLDLW, draw=WORLDCOL2]
  plot[smooth, tension=0.9] coordinates {
    (0.14,1)
    (0.29,0.80)
    (0.705,0.6)
    (0.8,0.4)
    (0.85,0.32)
    (0.96,0.14)
    (0.875,0)
  };

\fill[gray, opacity=0.4]
  (0,0.8) rectangle (1,1);


\draw[line width=\DASHW, dash pattern=on 2pt off 2pt] (0,0.4) -- (1,0.4);
\draw[line width=\DASHW, dash pattern=on 2pt off 2pt] (0,0.8) -- (1,0.8);
\draw[line width=\RAILLW] (0,0) -- (1,0);
\draw[line width=\RAILLW, ->] (0,0) -- (0,1);

\node[anchor=east] at (0,0.8) {$t_c$};

\node[anchor=east] at (0,0.4) {$t^*$};

\foreach \u in \PARTICLES {
  \fill (\u,0) circle[radius=\DOTR];
}





\end{tikzpicture}%
}
\caption{Illustration of the ``two-step'' dynamical process underlying SW-SSB in an open quantum system. In the density matrix representation of a quantum system, each particle has two worldlines, corresponding to the ket and bra space of the density matrix, or the forward and backward paths in the Keldysh formalism. At a crossover time scale~$t^\ast$, the two worldlines merge together, meaning the density matrix becomes approximately diagonal. In two and three dimensions, at a later time, $t_c \gtrapprox t^\ast$, there is a finite-time {\it phase transition} whereupon the merged worldlines intertwine, exchange and condense, losing track of their original positions, marking the onset of SW-SSB and emergent classical hydrodynamics. }
\label{fig:worldlines_for_back}

\end{figure}

We propose that an intrinsic open quantum system undergoes a ``two-step'' process during decoherence. Unlike a ground-state closed quantum system, each particle in an open system must be represented by two worldlines, one in the ket and one in the bra space, or equivalently, the forward and backward paths in the Keldysh formalism. As decoherence proceeds, the system first reaches a crossover time scale $t^\ast$, at which the two worldlines effectively merge into a bound state. In this sense, the density matrix becomes diagonal in the configuration basis. In two and three dimensions, at a later critical time $t_c$ there is a SW-SSB phase transition, where the merged worldlines become sufficiently scrambled that information about the initial positions is irreversibly lost: the particles' initial positions become undecodable, and a classical hydrodynamic description emerges. This two-step process is depicted schematically in Fig.~\ref{fig:worldlines_for_back}. 

The two time scales \(t^\ast\) and \(t_c\) are not always well separated in a generic open quantum system. Within the scope of this paper, two of the three quantum models we study (the decohered spin-\(1/2\) model and the rotor model) effectively begin at times \(t \gtrsim t^\ast\), since their density matrices are always diagonal in the occupation basis; equivalently, their dynamics can be mapped to a classical model. The third model discussed in Sec.~\ref{sec:ModelF} includes the regime \(t<t^\ast\) in Fig.~\ref{fig:worldlines_for_back}, and in that case \(t^\ast\) and \(t_c\) are not clearly separated. 

We note that an ordinary zero-temperature superfluid (a phase that spontaneously breaks the strong symmetry to nothing) corresponds to a different situation: the ket and bra worldlines scramble and condense separately, without first merging.

We begin in Sec.~\ref{sec:diagnostics} with an overview of relevant observables and the worldline picture for charge-conserving dynamics. Next, in Sec.~\ref{sec:spin-1/2}, we discuss our decohered spin-$1/2$ model and provide numerical results for linear and nonlinear observables in 1d and 2d. In Sec.~\ref{sec:rotor}, we discuss our decohered rotor model and provide analytical results in 1d and 2d. In Sec.~\ref{sec:ModelF}, we define a richer $U(1)$-symmetric model and show how Model-F dynamics arises after a finite-time phase transition. In Sec.~\ref{sec:classical} we directly analyze the hydrodynamics as an effective classical description of a system with SW-SSB. Finally, in Sec.~\ref{sec:discussion}, we provide a discussion and outlook, and discuss consequences of our results for the onset of a continuum hydrodynamic description in $U(1)$-symmetric many-body systems.

\section{Diagnostics and the worldline picture}\label{sec:diagnostics}

\subsection{Observables}
\label{sec:diagnostics:observables}
\subsubsection{SW-SSB measures}\label{sec:diagnostics:observables:SWSSB}

\begin{table}[t!]
\centering
\renewcommand{\arraystretch}{1.25}
\setlength{\tabcolsep}{10pt}

\begin{tabular}{|c|c|c|c|}
\hline
\textbf{Model/Theory} & \textbf{Dim.} & \textbf{
Key results} & \textbf{Section} \\
\hline

\multirow{4}{*}[+1.8ex]{spin-$1/2$}
& \multirow{2}{*}{$1\mathrm{d}$}
& Diffusion, No SW-SSB & \multirow{2}{*}{Sec.~\ref{sec:spin-1/2:1d}} \\
& & short-range CMI, efficient charge decoding & \\
\cline{2-4}
& $2\mathrm{d}$
& SW-SSB, BKT-like transition & Sec.~\ref{sec:spin-1/2:2d} \\
\hline

\multirow[c]{2}{*}[-1.8ex]{rotor}
& $1\mathrm{d}$
& No SW-SSB, $\xi^{(1)}/\xi^{(2)}=2$ & Sec.~\ref{sec:rotor:1d} \\
\cline{2-4}
& $2\mathrm{d}$
& \begin{tabular}[c]{@{}c@{}}
    SW-SSB, BKT-like transition,\\
    $\Delta^{(2)} / \Delta^{(1)} = 2$
  \end{tabular}
& Sec.~\ref{sec:rotor:2d} \\
\hline

including Hamiltonian
& $\ge 2\mathrm{d}$
& emergent Model~F & Sec.~\ref{sec:ModelF} \\
\hline

\multirow{2}{*}{hydrodynamics}
& $1\mathrm{d}$
& No SW-SSB, long-range CMI
& \multirow{2}{*}{Sec.~\ref{sec:classical}} \\
\cline{2-3}
& $\ge 2\mathrm{d}$
& SW-SSB
& \\
\hline

\end{tabular}

\caption{Summary of models, dimensions, results, and the corresponding sections.}
\label{tab:roadmap}
\end{table}

In this work we will focus exclusively on systems with $U(1)$ symmetry, or equivalently with a single conserved charge~$\hat Q$. A state is strongly symmetric if it satisfies $e^{ \ii \theta \hat Q} \h{\rho} = e^{ \ii \theta Q_0} \hat{\rho}$; i.e., if $\h{\rho}$ has a definite charge value~$Q_0$. A state is weakly symmetric if $e^{ \ii \theta \hat Q} \h{\rho} e^{- \ii \theta \hat Q} = \h{\rho}$; i.e., if $\h{\rho}$ contains no off-diagonal matrix elements between states of different charge. Dynamics preserves strong (weak) symmetry if a strongly (weakly) symmetric density matrix retains this symmetry at all times. Under strongly symmetric quantum channels, the system does not exchange charge with its environment. If one decomposes these channels in Kraus form, each Kraus operator separately commutes with $\hat Q$. We will take the charge $\hat Q$ to be a sum of local terms, $\hat Q = \sum_r \hat{q}_r$, and restrict ourselves to spatially local dynamics. Under spatially local dynamics with a strong $U(1)$ symmetry, charge typically spreads diffusively: $\langle \h{q}_r(t) \h{q}_{r'}(0) \rangle  = t^{-d/2} \phi(|r - r'|/\sqrt{t})$, where $\phi(\cdot)$ is a scaling function that is asymptotically Gaussian. As we will argue, this diffusive hydrodynamics does not control the length- or timescales associated with SW-SSB.

Physically, $U(1)$ SW-SSB is a statement about the \emph{distinguishability} of quantum states, i.e., the number of measurements that one needs to tell two states apart. The distinguishability of two general quantum states $\h{\rho}$ and $\h{\sigma}$ is canonically measured by the fidelity, $\mathcal{F}(\h{\rho}, \h{\sigma}) = \left(\mathrm{Tr}\sqrt{\!\sqrt{\h{\rho}}\h{\sigma} \sqrt{\h{\rho}}}\right)^2$. The ``fidelity correlator'' $C^{(F)}_{ij} \equiv \mathcal{F}(\h{\rho}, \h{O}_{ij} \h{\rho} \h{O}^\dagger_{ij})$ \cite{lessa2025strong} measures the fidelity between the reference state $\h{\rho}$ and a state $\h{O}_{ij} \h{\rho} \h{O}^\dagger_{ij}$ derived from it by moving a charge from site $i$ to site $j$. Thus, if $\h{S}^+_i$ and $\h{S}^-_i$ are charge raising and lowering operators, one can choose $\h{O}_{ij} = \h{S}^+_i \h{S}^-_j + \h{S}^-_i \h{S}^+_j$. A state is said to exhibit SW-SSB if $\lim_{|i - j| \to \infty} C^{(F)}_{ij} \neq 0$, i.e., if moving a particle across the system does not lead to a sharply distinguishable state.
Interestingly, we will find that states are indistinguishable even after moving charges across distances much larger than the diffusive length scale. This indicates that SW-SSB is not simply a measure of localization of charge information.

The fidelity correlator is of particular interest because long-range order in the fidelity correlator is stable under strongly symmetric finite-depth local channels~\cite{sang2025mixed}. Thus, it is a suitable diagnostic of mixed-state phases, which may be organized in terms of two-way local, short-depth channel connectivity~\cite{sang2025mixed} (analogous to the classification of pure-state phases in terms of short-depth unitary connectivity~\cite{PhysRevB.72.045141}). 
In practice, fidelities are hard to compute, so one works with proxies for the fidelity correlator. The most robust of these diagnostics is the R\'enyi-$1$ correlator,
\begin{align}
    C^{(1)}(i,j)
    \equiv
    \Tr\left(\sqrt{\hat{\rho}}\hat{O}_{ij}\sqrt{\hat{\rho}}\hat{O}_{ij}^\dagger\right).
\end{align}
{which is tightly related to $C^{(F)}$ but is easier to compute. Since $[C^{(F)}(i,j)]^2 \leq C^{(1)}(i,j) \leq C^{(F)}(i,j)$, long-range
order in $C^{(1)}(i,j)$ provides a stable diagnosis of SW-SSB, and we adopt it throughout this work~\cite{weinstein2025efficient}. When $C^{(1)}$ decays exponentially, we denote its correlation length by $\xi^{(1)}(t)$, where $t$ is the evolution time.

Another commonly-used diagnostic is the R\'enyi-$2$ correlator. For our operator $\hat{O}_{ij}$ acting on sites $i$ and $j$, it is defined as
\begin{align}
    C^{(2)} (i,j)
    \equiv \frac{\Tr\left(\hat{O}_{ij}\hat{\rho}\hat{O}_{ij}^\dagger\hat{\rho}\right)}
    {\Tr\left(\hat{\rho}^2\right)}.
\end{align}
This quantity is convenient to calculate, but has no direct information-theoretic interpretation and does not satisfy the same stability properties as the fidelity and R\'enyi-$1$ correlators under strongly symmetric finite-depth channels. As a diagnostic of a mixed-state phase, therefore, R\'enyi-2 correlators are merely suggestive.  The \cx{R\'enyi-2 correlator is also the simplest example of the replica generalization of the R\'enyi-1 correlator, a formalism often used in analytical studies of SW-SSB, as discussed in Sec.~\ref{sec:rotor}.}
When $C^{(2)}$ decays exponentially, we denote its correlation length by $\xi^{(2)}(t)$.

Much of our numerical analysis for the spin-$1/2$ model  concerns density matrices that are diagonal in the $S^z$ basis, $\hat{\rho} = \sum_{s} P(s) |s\rangle\langle s|$, where $s$ denotes a spin configuration and $P(s)$ is a classical probability distribution. It is convenient to introduce the associated vectors
\begin{align}
\label{eq:charge_basis}
    |\rho\rangle \equiv \sum_{s} P(s) |s\rangle,
    \qquad
    |\sqrt{\rho}\rangle \equiv \sum_{s} \sqrt{P(s)} |s\rangle.
\end{align}
The vector $|\rho\rangle$ has a nonstandard normalization, since $\langle \rho|\rho\rangle=\sum_s P(s)^2$ is generally not equal to one. Instead, $\langle \mathbb{I}|\rho\rangle=1$ (with $\langle \mathbb{I}|=\sum_s \langle s|$).
In contrast, $\langle \sqrt{\rho}|\sqrt{\rho}\rangle=\sum_s P(s)=1$.

For the operator insertion $\hat{O}_{ij}=\hat{S}_i^+\hat{S}_j^-$, define $s'$ as the configuration obtained from $s$ by shifting only the two spins at $i$ and $j$ by $s_i' = s_i+1$ and $s_j' = s_j-1$ while all other sites are unchanged. Then the diagonal-sector correlators reduce to
\begin{align}
    C^{(2)}(i,j)
    &= \frac{\langle \rho| \hat{S}_i^+ \hat{S}_j^- |\rho\rangle}{\langle \rho|\rho\rangle}
    = \frac{\sum_{s}' P(s)P(s')}{\sum_s P(s)^2},
\end{align}
and
\begin{align}
    C^{(1)}(i,j)
    = \langle \sqrt{\rho}| \hat{S}_i^+ \hat{S}_j^- |\sqrt{\rho}\rangle
    = \sum\nolimits_{s}' \sqrt{P(s)P(s')}.
\end{align}
Here $\sum_s'$ denotes a restricted sum over configurations $s$ for $s_i=-\tfrac12$ and $s_j=+\tfrac12$ so that $s'$ is well-defined. Note that in this diagonal (classical) sector, the fidelity correlator is identical to the R\'enyi-$1$ correlator. Furthermore, these definitions coincide precisely with the Bhattacharyya coefficient $B(P(s), P'(s))$, where we have defined $P'(s)= P(s')$, which is a standard metric for the similarity of classical probability distributions~\cite{dodge2003oxford}. The Bhattacharyya coefficient is related to a more familiar metric of similarity, the Kullback-Leibler divergence $D_{\mathrm{KL}}(P\Vert P') \equiv \mathrm{Tr}(P (\log P - \log P'))$, by the inequality $D_{\mathrm{KL}}(P\Vert Q) \geq - \log B(P, Q)$. The latter quantity is an estimate of the number of samples one would need from distribution $P'$ to determine if it were distinct from $P$. For Gaussian distributions, the inequality above is saturated, so the two measures carry the same information.

\subsubsection{Conditional mutual information}\label{sec:diagnostics:observables:CMI}

So far, we have focused on direct order parameters for SW-SSB: these incorporate our knowledge of the expected pattern of symmetry breaking via the choice of the operator $\h{O}_{ij}$. For equilibrium phase transitions, the mutual information between two distant regions offers an observable-agnostic way of probing a diverging length-scale. The mutual information can be defined as follows. For a region $\mathcal{R}$, we define the reduced density matrix $\hat{\rho}_{\mathcal{R}} \equiv \Tr_{\mathcal{R}^c}(\hat{\rho})$ and its von Neumann entropy
\begin{align}
S_{\mathcal{R}} \equiv S(\hat{\rho}_{\mathcal{R}}) = -\Tr\left(\hat{\rho}_{\mathcal{R}} \log \hat{\rho}_{\mathcal{R}}\right).
\end{align}
The mutual information is defined as $I(A:B) \equiv S_A + S_B - S_{AB}$. It upper-bounds all connected correlations between operators supported in regions $A$ and $B$~\cite{wolf2008area}. Like individual correlation functions, the mutual information is bounded by the Lieb-Robinson light cone~\cite{lieb1972finite}, and thus cannot diverge at finite time. 
In recent work Ref.~\cite{sang2025stability} it was argued that an analog of the mutual information for information-theoretic transitions is the conditional mutual information (CMI).
The CMI for three regions $A$, $B$, and $C$ is
\begin{align}
I(A:C|B) = I(A:BC) - I(A:B) =  S_{AB} + S_{BC} - S_B - S_{ABC},
\end{align}
where $A$ and $C$ are separated by the intermediate region $B$. {In many cases the CMI is expected to be short ranged, decaying exponentially in the separation between $A$ and $C$}:
\begin{align}
I(A:C|B) \propto f(|A|,|C|)\,e^{-R_B/\xi^M}.
\end{align}
{Here $R_B$ is the distance separating $A$ from $C$, which in 1d is $R_B=|B|$.
The length scale $\xi^M$ is called the Markov length \cite{sang2025stability, sang2025mixed}. Provided the Markov length stays finite as a state is evolved, \emph{no} mixed-state phase transition can occur: any two mixed states that can be connected via a finite-Markov-length (FML) path are in the same mixed-state phase.

The Markov length has an operational meaning in terms of recoverability~\cite{sang2025mixed}. States within the same mixed-state phase are connected by local quantum channels, so one can map between them in both directions. Across a phase transition, this two-way connectivity is lost. In particular, an initial state evolved under a non-FML path cannot be recovered using a short depth circuit comprised of only local channels. This can be viewed as an inference or decoding problem relevant to error correction~\cite{sang2024mixed}. Starting from a state with no errors, decoherence evolves it to a mixed state, and recoverability asks whether one can decode back. The amount of information required for successful recovery is controlled by the Markov length $\xi^M$. 

Mixed state phase transitions associated with diverging Markov length might encompass a broader class of transitions than those associated with SW-SSB. However, the Markov length upper-bounds the length-scale associated with the decay of the fidelity correlator, by the following argument (made precise in Theorem 4 of Ref.~\cite{lessa2025strong}). Suppose the Markov length is $\xi^M$. We now consider moving a particle from site $i$ to site $j$ with $\abs{i-j} \gg \xi^M$, and consider the fidelity between the initial state $\rho$ and the state with the particle moved, $\rho'$. Recall that fidelity increases monotonically under quantum channels. Consider erasing site $i$, and trying to recover it using all information available within a distance $\xi^M$ of site $i$. Since $\xi^M \ll |j - i|$, the recovery map does not have access to the charge on site $j$, so it must act the same way regardless of the charge on that site. Therefore it will recover $\rho$ and $\rho'$ to states of distinct total charge. The fidelity between these is zero, so (by monotonicity) the fidelity between the initial states leading to these is also zero (to within exponential error). Thus the SW-SSB correlators all decay on the scale of $\xi^M$ or faster---as we will see below, in many cases they in fact decay strictly faster.}

In our numerics we will study the time evolution of the Markov length as a probe of SW-SSB. The standard geometry used in discussions of mixed-state phases and recoverability takes $A\cup B\cup C$ to cover the entire system. In one dimension, this corresponds to three consecutive intervals $A$, $B$, and $C$ that partition the full chain [Fig.~\hyperref[fig:CMI_geometries]{\ref*{fig:CMI_geometries}(a)}]. Here, instead, we choose $A$, $B$, and $C$ to be consecutive intervals that occupy only part of the chain, with $|A|$ and $|C|$ short (typically $O(1)$, often a single site), and we vary the length $R_B$ [Fig.~\hyperref[fig:CMI_geometries]{\ref*{fig:CMI_geometries}(b)}]. Equivalently, one may view this as a two-point correlator in which $A$ and $C$ are separated by $B$ while the rest of the system is traced out. 
Although the two geometries lead to conceptually different quantities, once there is no true long-range order they are expected to track the same characteristic length scale. 

To build intuition for these nonlinear diagnostics, we next re-express them in a worldline representation. This viewpoint also clarifies how nonlinear observables differ from linear observables, even though both are evaluated from the same underlying density matrix.

\begin{figure}
    \centering

    \begin{minipage}{0.48\textwidth}
        \centering
        \begin{tikzpicture}
            \node (fig) {
\begin{tikzpicture}[x=1cm,y=1cm, line cap=butt]
  \def\xL{0}
  \def\xA{1.5}
  \def\xB{4.0}
  \def\xC{1.5}

  \def\y{0}
  \def\tick{0.22}
  \def\hw{14pt}     

  \draw[blue!30,  line width=\hw] (\xL,\y) -- ++(\xA,0);
  \draw[red!30,   line width=\hw] (\xL+\xA,\y) -- ++(\xB,0);
  \draw[green!30, line width=\hw] (\xL+\xA+\xB,\y) -- ++(\xC,0);

  \draw[thick] (\xL,\y) -- ++(\xA+\xB+\xC,0);

  \foreach \x in {0,\xA,\xA+\xB,\xA+\xB+\xC} {
    \draw[thick] (\x,\y-\tick) -- (\x,\y+\tick);
  }

  \node[above=6pt] at (\xL+\xA/2,\y) {$A$};
  \node[above=6pt] at (\xL+\xA+\xB/2,\y) {$B$};
  \node[above=6pt] at (\xL+\xA+\xB+\xC/2,\y) {$C$};
\end{tikzpicture}};
            \node[left=-6pt of fig.west] {(a)};
        \end{tikzpicture}
    \end{minipage}
    \hfill
    \begin{minipage}{0.48\textwidth}
        \centering
        \begin{tikzpicture}
            \node (fig) {
\begin{tikzpicture}[x=1cm,y=1cm, line cap=butt]
  \def\xL{0}
  \def\xG{1.3}
  \def\xA{0.7}
  \def\xB{3.0}
  \def\xC{0.7}

  \def\y{0}
  \def\tick{0.22}
  \def\hw{14pt}

  \draw[gray!30,  line width=\hw] (\xL,\y) -- ++(\xG,0);
  \draw[blue!30,  line width=\hw] (\xL+\xG,\y) -- ++(\xA,0);
  \draw[red!30,   line width=\hw] (\xL+\xG+\xA,\y) -- ++(\xB,0);
  \draw[green!30, line width=\hw] (\xL+\xG+\xA+\xB,\y) -- ++(\xC,0);
  \draw[gray!30,  line width=\hw] (\xL+\xG+\xA+\xB+\xC,\y) -- ++(\xG,0);

  \draw[thick] (\xL,\y) -- ++(2*\xG+\xA+\xB+\xC,0);

  \foreach \x in {0,\xG,\xG+\xA,\xG+\xA+\xB,\xG+\xA+\xB+\xC,2*\xG+\xA+\xB+\xC} {
    \draw[thick] (\x,\y-\tick) -- (\x,\y+\tick);
  }

  \node[above=6pt] at (\xL+\xG+\xA/2,\y) {$A$};
  \node[above=6pt] at (\xL+\xG+\xA+\xB/2,\y) {$B$};
  \node[above=6pt] at (\xL+\xG+\xA+\xB+\xC/2,\y) {$C$};
\end{tikzpicture}};
            \node[left=-6pt of fig.west] {(b)};
        \end{tikzpicture}
    \end{minipage}

    \caption{CMI geometries in 1d. (a) In the first case, regions $A$, $B$, and $C$ cover the entire system. This is the standard geometry for CMI in the context of 1d mixed state phases and recoverability. (b) In the second case, the three regions do not cover the entire system, with regions $A$ and $C$ typically of size $O(1)$. In both cases, we denote the separation between regions $A$ and $C$ as $R_B = \abs{B}$.}
    \label{fig:CMI_geometries}
\end{figure}
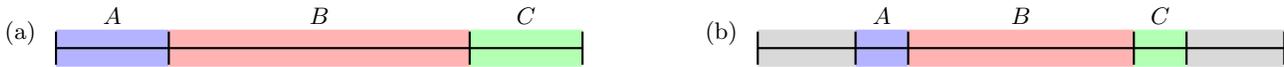

\subsection{The worldline picture}\label{sec:diagnostics:worldline}

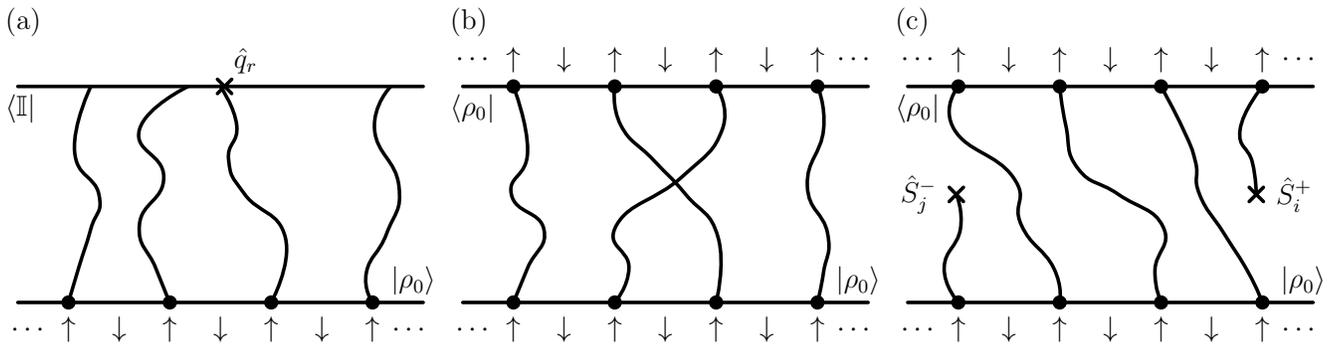
\begin{figure}[!t]
\centering   

\begin{minipage}[t]{0.33\linewidth}
  \centering
  \resizebox{\linewidth}{!}{
\def\FIGW{6cm}
\def\FIGH{3.2cm}
\def\RAILLW{1.5pt}
\def\WORLDLW{1.5pt}
\def\DOTR{3pt}
\def\ARROWY{0.125}

\def\FIGFONTSIZE{\large}%

\colorlet{WORLDCOL}{black}%
\colorlet{ARROWCOL}{black}%

\def\PARTICLES{0.125,0.375,0.625,0.875}%

\def\ARROWXS{0.125,0.25,0.375,0.5,0.625,0.75,0.875}%

\begin{tikzpicture}[
  x=\FIGW, y=\FIGH,
  line cap=round, line join=round,
  every node/.style={font=\FIGFONTSIZE} 
]


\draw[line width=\WORLDLW, draw=WORLDCOL]
  plot[smooth, tension=0.9] coordinates {
    (0.18,1)
    (0.14,0.70)
    (0.20,0.5)
    (0.18,0.34)
    (0.15,0.16)
    (0.125,0)
  };

\draw[line width=\WORLDLW, draw=WORLDCOL]
  plot[smooth, tension=0.9] coordinates {
    (0.42,1)
    (0.30,0.80)
    (0.36,0.58)
    (0.30,0.36)
    (0.34,0.16)
    (0.375,0)
  };

\draw[line width=\WORLDLW, draw=WORLDCOL]
  plot[smooth, tension=0.9] coordinates {
    (0.50,1)
    (0.54,0.82)
    (0.52,0.66)
    (0.56,0.50)
    (0.66,0.30)
    (0.625,0)
  };

\draw[line width=\WORLDLW, draw=WORLDCOL]
  plot[smooth, tension=0.9] coordinates {
    (0.92,1)
    (0.88,0.80)
    (0.94,0.54)
    (0.91,0.32)
    (0.86,0.14)
    (0.875,0)
  };

\foreach[count=\k] \u in \ARROWXS {
  \pgfmathtruncatemacro{\odd}{mod(\k,2)}
  \ifnum\odd=1
    \node[text=ARROWCOL] at (\u,-\ARROWY) {$\uparrow$};
  \else
    \node[text=ARROWCOL] at (\u,-\ARROWY) {$\downarrow$};
  \fi
}

\draw[line width=\RAILLW] (0,1) -- (1,1);
\draw[line width=\RAILLW] (0,0) -- (1,0);

\foreach \u in \PARTICLES {
  \fill (\u,0) circle[radius=\DOTR];
}

\node[text=ARROWCOL] at (0.03,-\ARROWY) {$\cdots$};
\node[text=ARROWCOL] at (0.97,-\ARROWY) {$\cdots$};

\node[anchor=west] at (-0.05,0.9) {$\bra{\mathbb{I}}$};
\node[anchor=east] at (1.05,0.1) {$\ket{\rho_0}$};

\node[anchor=west] at (-0.05,1.3) {(a)};


\node[anchor=east] at (0.61, 1.12) {$\hat{q}_r$};
\def\Xpt{3pt}%
\draw[line width=\WORLDLW, draw=WORLDCOL]
  ($(0.51,1)+(-\Xpt,-\Xpt)$) -- ($(0.51,1)+(\Xpt,\Xpt)$);
\draw[line width=\WORLDLW, draw=WORLDCOL]
  ($(0.51,1)+(-\Xpt,\Xpt)$) -- ($(0.51,1)+(\Xpt,-\Xpt)$);

\end{tikzpicture}%
}
\end{minipage}%
\begin{minipage}[t]{0.33\linewidth}
  \centering
  \resizebox{\linewidth}{!}{
\def\FIGW{6cm}
\def\FIGH{3.2cm}
\def\RAILLW{1.5pt}
\def\WORLDLW{1.5pt}
\def\DOTR{3pt}
\def\ARROWY{0.125}

\def\FIGFONTSIZE{\large}%

\colorlet{WORLDCOL}{black}%
\colorlet{ARROWCOL}{black}%

\def\PARTICLES{0.125,0.375,0.625,0.875}%

\def\ARROWXS{0.125,0.25,0.375,0.5,0.625,0.75,0.875}%

\ignorespaces%
\begin{tikzpicture}[
  x=\FIGW, y=\FIGH,
  line cap=round, line join=round,
  every node/.style={font=\FIGFONTSIZE}%
]

\draw[line width=\WORLDLW, draw=WORLDCOL]
  plot[smooth, tension=0.9] coordinates {
    (0.125,1)
    (0.16,0.70)
    (0.12,0.50)
    (0.20,0.34)
    (0.16,0.16)
    (0.125,0)
  };

\draw[line width=\WORLDLW, draw=WORLDCOL]
  plot[smooth, tension=0.9] coordinates {
    (0.625,1)
    (0.64,0.80)
    (0.54,0.58)
    (0.38,0.36)
    (0.41,0.16)
    (0.375,0)
  };

\draw[line width=\WORLDLW, draw=WORLDCOL]
  plot[smooth, tension=0.9] coordinates {
    (0.375,1)
    (0.39,0.82)
    (0.47,0.66)
    (0.55,0.50)
    (0.63,0.30)
    (0.625,0)
  };

\draw[line width=\WORLDLW, draw=WORLDCOL]
  plot[smooth, tension=0.9] coordinates {
    (0.875,1)
    (0.89,0.80)
    (0.85,0.54)
    (0.90,0.32)
    (0.89,0.14)
    (0.875,0)
  };

\foreach[count=\k] \u in \ARROWXS {%
  \pgfmathtruncatemacro{\odd}{mod(\k,2)}%
  \ifnum\odd=1
    \node[text=ARROWCOL] at (\u,-\ARROWY) {$\uparrow$};%
  \else
    \node[text=ARROWCOL] at (\u,-\ARROWY) {$\downarrow$};%
  \fi
}

\foreach[count=\k] \u in \ARROWXS {%
  \pgfmathtruncatemacro{\odd}{mod(\k,2)}%
  \ifnum\odd=1
    \node[text=ARROWCOL] at (\u,1+\ARROWY) {$\uparrow$};%
  \else
    \node[text=ARROWCOL] at (\u,1+\ARROWY) {$\downarrow$};%
  \fi
}

\draw[line width=\RAILLW] (0,1) -- (1,1);
\draw[line width=\RAILLW] (0,0) -- (1,0);

\foreach \u in \PARTICLES {%
  \fill (\u,0) circle[radius=\DOTR];%
}

\foreach \u in \PARTICLES {%
  \fill (\u,1) circle[radius=\DOTR];%
}

\node[text=ARROWCOL] at (0.03,-\ARROWY) {$\cdots$};
\node[text=ARROWCOL] at (0.97,-\ARROWY) {$\cdots$};
\node[text=ARROWCOL] at (0.03,1+\ARROWY) {$\cdots$};
\node[text=ARROWCOL] at (0.97,1+\ARROWY) {$\cdots$};

\node[anchor=west] at (-0.05,0.9) {$\bra{\rho_0}$};
\node[anchor=east] at (1.05,0.1) {$\ket{\rho_0}$};
\node[anchor=west] at (-0.05,1.3) {(b)};


\end{tikzpicture}
}
\end{minipage}%
\begin{minipage}[t]{0.33\linewidth}
  \centering
  \resizebox{\linewidth}{!}{
\def\FIGW{6cm}
\def\FIGH{3.2cm}
\def\RAILLW{1.5pt}
\def\WORLDLW{1.5pt}
\def\DOTR{3pt}
\def\ARROWY{0.125}

\def\Xpt{3pt}%

\def\FIGFONTSIZE{\large}%

\colorlet{WORLDCOL}{black}%
\colorlet{ARROWCOL}{black}%

\def\PARTICLES{0.125,0.375,0.625,0.875}%

\def\ARROWXS{0.125,0.25,0.375,0.5,0.625,0.75,0.875}%

\ignorespaces%
\begin{tikzpicture}[
  x=\FIGW, y=\FIGH,
  line cap=round, line join=round,
  every node/.style={font=\FIGFONTSIZE}%
]

\draw[line width=\WORLDLW, draw=WORLDCOL]
  plot[smooth, tension=0.9] coordinates {
    (0.12,0.50)
    (0.13,0.34)
    (0.09,0.16)
    (0.125,0)
  };

\draw[line width=\WORLDLW, draw=WORLDCOL]
  plot[smooth, tension=0.9] coordinates {
    (0.125,1)
    (0.115,0.80)
    (0.26,0.58)
    (0.28,0.36)  
    (0.36,0.16)
    (0.375,0)
  };

\draw[line width=\WORLDLW, draw=WORLDCOL]
  plot[smooth, tension=0.9] coordinates {
    (0.375,1)
    (0.39,0.82)
    (0.42,0.62)
    (0.52,0.50)  
    (0.63,0.35)
    (0.61,0.15)
    (0.625,0)
  };

\draw[line width=\WORLDLW, draw=WORLDCOL]
  plot[smooth, tension=0.9] coordinates {
    (0.625,1)
    (0.68,0.80)
    (0.71,0.65)
    (0.72,0.50)
    (0.78,0.32)
    (0.84,0.14)
    (0.875,0)
  };

\draw[line width=\WORLDLW, draw=WORLDCOL]
  plot[smooth, tension=0.9] coordinates {
    (0.875,1)
    (0.82,0.80)
    (0.85,0.65)
    (0.86,0.50)
  };

\foreach[count=\k] \u in \ARROWXS {%
  \pgfmathtruncatemacro{\odd}{mod(\k,2)}%
  \ifnum\odd=1
    \node[text=ARROWCOL] at (\u,-\ARROWY) {$\uparrow$};%
  \else
    \node[text=ARROWCOL] at (\u,-\ARROWY) {$\downarrow$};%
  \fi
}

\foreach[count=\k] \u in \ARROWXS {%
  \pgfmathtruncatemacro{\odd}{mod(\k,2)}%
  \ifnum\odd=1
    \node[text=ARROWCOL] at (\u,1+\ARROWY) {$\uparrow$};%
  \else
    \node[text=ARROWCOL] at (\u,1+\ARROWY) {$\downarrow$};%
  \fi
}

\draw[line width=\RAILLW] (0,1) -- (1,1);
\draw[line width=\RAILLW] (0,0) -- (1,0);


\draw[line width=\WORLDLW, draw=WORLDCOL]
  ($(0.12,0.5)+(-\Xpt,-\Xpt)$) -- ($(0.12,0.5)+(\Xpt,\Xpt)$);
\draw[line width=\WORLDLW, draw=WORLDCOL]
  ($(0.12,0.5)+(-\Xpt,\Xpt)$) -- ($(0.12,0.5)+(\Xpt,-\Xpt)$);

\node[anchor=east] at (0.12-0.03,0.5) {$\hat{S}_j^-$};

\draw[line width=\WORLDLW, draw=WORLDCOL]
  ($(0.86,0.5)+(-\Xpt,-\Xpt)$) -- ($(0.86,0.5)+(\Xpt,\Xpt)$);
\draw[line width=\WORLDLW, draw=WORLDCOL]
  ($(0.86,0.5)+(-\Xpt,\Xpt)$) -- ($(0.86,0.5)+(\Xpt,-\Xpt)$);

\node[anchor=west] at (0.86+0.03,0.5) {$\hat{S}_i^+$};

\foreach \u in \PARTICLES {%
  \fill (\u,0) circle[radius=\DOTR];%
}

\foreach \u in \PARTICLES {%
  \fill (\u,1) circle[radius=\DOTR];%
}

\node[text=ARROWCOL] at (0.03,-\ARROWY) {$\cdots$};
\node[text=ARROWCOL] at (0.97,-\ARROWY) {$\cdots$};
\node[text=ARROWCOL] at (0.03,1+\ARROWY) {$\cdots$};
\node[text=ARROWCOL] at (0.97,1+\ARROWY) {$\cdots$};

\node[anchor=west] at (-0.05,0.9) {$\bra{\rho_0}$};
\node[anchor=east] at (1.05,0.1) {$\ket{\rho_0}$};
\node[anchor=west] at (-0.05,1.3) {(c)};


\end{tikzpicture}
}
\end{minipage}

\caption{Worldline representation of the local $U(1)$ charge dynamics (with time running upward), illustrated for the spin-$1/2$ model. The gray arrows indicate the N\'{e}el initial state $\hat{\rho}_0$. Panel (a) shows the local charge expectation value, where the final-time boundary is free 
and is probed by the insertion of a charge operator $\hat q_r$. 
Panels (b) and (c) show the corresponding representation for the R\'enyi-$2$ correlator $C^{(2)}(i,j)$, where the temporal boundaries are both $\hat{\rho}_0$ at the bottom and the top. They correspond to the denominator and numerator of $C^{(2)}(i,j)$, respectively; in (c) the operator insertions $\hat S_i^+$ and $\hat S_j^-$ create and annihilate a defect worldline. }
\label{fig:worldlines}
\end{figure}

It is useful to distinguish observables that are \emph{linear} in the density matrix, such as the local charge density $\langle \hat{q}_r(t)\rangle$, from the \emph{nonlinear} observables discussed above. 
In $U(1)$-symmetric open dynamics, the conserved charge typically exhibits diffusive hydrodynamics~\cite{huang2025hydrodynamics}. This behavior can be understood from the structure of the $U(1)$-symmetric Lindbladian, which supports a quadratic long-wavelength Goldstone mode~\cite{Ogunnaike2023hydro}, and it can sometimes also admit an explicit mapping to an effective random-walking classical dynamics in the diagonal sector as described in the next section for our first model.

Given the charge dynamics, one might wonder whether mixed-state correlators and information-theoretic quantities such as the CMI are also controlled by the same diffusion with dynamical exponent $z=2$. The worldline representation (Fig.~\ref{fig:worldlines}) provides a simple explanation for why this need not be the case. Here the worldlines represent a bound-state of the forward and backward paths, that were illustrated in Fig.~\ref{fig:worldlines_for_back} for times $t> t^*$. The dynamics is a path-integral sum over all trajectories of the local $U(1)$ charge (for a diagonal density matrix), with weights set by the local update rules and boundary conditions fixed by the observable.

Much of our analysis focuses on the diagonal sector, where $\hat{\rho}$ is diagonal in the $\hat{q}_r$ basis and can be viewed as a classical probability distribution over charge configurations. In this setting, the charge expectation value can be written in the vectorized notation as
\begin{align}
    \langle \hat{q}_r(t)\rangle
    = \Tr\left(\hat{q}_r\hat{\rho}(t)\right)
    = \langle \mathbb{I}|\hat q_r e^{\mathcal{L}^{\textsf{diag}}t}|{\rho}_0\rangle,
\end{align}
where $\langle \mathbb{I}|$ is the vector with all components equal to one in the configuration basis which implements the trace, and $\mathcal{L}^{\textsf{diag}}$ is the effective generator governing the diagonal-sector evolution. In the worldline picture, this belongs to a \emph{free} boundary condition at the final time. Worldlines are not required to match prescribed endpoints, and the resulting behavior is therefore governed by ordinary diffusion. Fig.~\hyperref[fig:worldlines]{\ref*{fig:worldlines}(a)} illustrates this structure for the spin-$1/2$ model,
where we are taking an initial pure state $\hat\rho_0$ corresponding to a N\'{e}el arrangement diagonal in the $\hat S^z_j$ basis.

By contrast, mixed-state correlators impose a different boundary condition. In particular for the R\'enyi correlators, even though the \emph{bulk} evolution is generated by the same diagonal-sector updates from $\mathcal{L}^{\textsf{diag}}$, the observable fixes both the initial and the final boundary conditions to be $\hat\rho_0$. As a consequence, worldlines cannot simply wander freely. They must be paired up consistently between the two boundaries, and exchange processes become well defined. This change in boundary conditions is the basic reason why these quantities are not described by simple diffusion.

Fig.~\hyperref[fig:worldlines]{\ref*{fig:worldlines}(b,c)} illustrates the worldline picture for the spin-$1/2$ R\'enyi-2 correlator $C^{(2)}(i,j)$. In the denominator [Fig.~\hyperref[fig:worldlines]{\ref*{fig:worldlines}(b)}], the initial and final boundary conditions are both set by $\hat\rho_0$. The doubled construction produces a bulk of duration $2t$, and worldlines are constrained to connect the bottom boundary to the top boundary rather than terminate freely. In the numerator [Fig.~\hyperref[fig:worldlines]{\ref*{fig:worldlines}(c)}], there is an additional insertion at time $t$, inside the bulk: $\hat S_i^+$ creates a defect worldline at site $i$, while $\hat S_j^-$ annihilates it at site $j$. Because of this additional creation/annihilation, other worldlines must rearrange to accommodate the defect, and the correlator measures the cost of this rearrangement. The R\'enyi-1 correlator has the same boundary-conditions, but with a different probabilistic weight.

This structure is reminiscent of the worldline representation of quantum fluids. In that setting, superfluidity is tied to the appearance of large cycles, where many worldlines permute and become indistinguishable. In a normal fluid, such exchanges are typically rare and worldlines remain mostly separated. The key difference for us is the boundary condition and the ensemble. In the usual finite-temperature path integral one computes $\mathrm{tr}(e^{-\beta \hat H})$, so the trace imposes $\beta$-periodic boundary conditions and automatically sums over all worldline permutations. Also, one often works in a grand-canonical ensemble, so the particle number can fluctuate. In our case, the boundaries are fixed by $\hat\rho_0$ at both ends. This effectively selects a canonical sector in which the number of worldlines is fixed by the boundary conditions rather than being summed over, but the boundary conditions being the same at the initial and final times automatically sums over all worldline permutations.

For periodic boundary conditions in space, the worldline ensemble splits into sectors labeled by an integer winding number $W$. Here $W$ counts the net number of worldlines that wind around the periodic direction during the evolution. 
In equilibrium path integrals, the second moment of winding-number fluctuations, $\langle W^2\rangle$, is proportional to the superfluid stiffness~\cite{PhysRevB.36.8343}, and a nonzero $\langle W^2\rangle$ in the thermodynamics limit quantifies the world-line indistinguishability. In our setting, instead of a $\beta$-periodic temporal boundary condition, the bottom and top boundaries are fixed to the same state $\hat{\rho}_0$. Nevertheless, one can still organize worldline trajectories by their spatial winding sectors and study the statistics of $W$. This motivates us to consider winding-number observables for both the R\'enyi-$2$ and R\'enyi-$1$ constructions. The distinction between them is the probabilistic weight assigned to trajectories, which we spell out in the next section. These winding observables are especially important in two dimensions, where the stiffness inferred from winding-number fluctuations captures the BKT-like transition.

\section{Decohered spin-1/2 model}\label{sec:spin-1/2}

The first model we consider is a minimal interacting $U(1)$-symmetric Lindblad dynamics acting on a $d$-dimensional spin-$1/2$ system. 
Decoherence is generated by the nearest-neighbor jump operators $\h{L}_{ij}=\h{S}_i^+\h{S}_j^-$, with the Lindbladian taking the form
\begin{align}\label{eq:spin-half-lindblad}
\mathcal{L}(\rho) = \gamma \sum_{\langle i,j \rangle} \Big[
\h{L}_{ij}\h{\rho} \h{L}_{ij}^\dagger
- \frac{1}{2}\big\{\h{L}_{ij}^\dagger \h{L}_{ij}, \h{\rho}\} + (i \leftrightarrow j)
\Big],
\end{align}
where $\langle i,j\rangle$ denotes a pair of adjacent sites.
This dynamics has a strong $U(1)$ symmetry because each jump operator commutes with $\h{U}_\theta = e^{\ii\theta \sum_j \h{S}^z_j}$. The strong symmetry also implies a weak symmetry, enabling the possibility of SW-SSB during time evolution starting from a $U(1)$-symmetric initial state (i.e., a state with fixed initial charge).

A simplifying feature of this model is that the diagonal and off-diagonal sectors (in the charge basis) of the density matrix evolve independently. A short calculation, provided in Appendix~\ref{app:diagonal_lindbladian}, shows that the diagonal density matrix $\ket{\rho}$ evolves under the effective Lindblad operator 
\begin{align}\label{eq:imaginaryevolution}
\mathcal{L}^{\textsf{diag}} = 2\gamma \sum_{\langle ij\rangle}
\left(\h{\mathbf{S}}_i\cdot\h{\mathbf{S}}_j - \frac{1}{4}\right), \hspace{1cm} \ket{\rho_t}= e^{\mathcal{L}^{\textsf{diag}} t}\ket{\rho_0},
\end{align}
where $\langle ij \rangle$ denotes nearest-neighbor sites $i$ and $j$. Furthermore, as discussed in Appendix~\ref{app:diagonal_lindbladian}, all off-diagonal sectors feature a finite dissipative gap relative to the diagonal sector. Therefore the late-time dynamics is governed only by the diagonal evolution. In the remainder of this section, we restrict our attention to initial states diagonal in the charge basis. 

Within the diagonal sector, the Lindblad dynamics is therefore equivalent to imaginary-time evolution under the \emph{ferromagnetic} Heisenberg model \cite{gwa1992bethe}. Note that $-\mathcal{L}^{\textsf{diag}}$ is positive semi-definite and has a lowest eigenvalue zero, so gives a well-defined Markovian evolution on diagonal density matrices. The ferromagnetic Heisenberg model supports quadratic Goldstone modes, which is even true for higher spin $S$~\cite{Ogunnaike2023hydro,watanabe2024oneD}, implying diffusive scaling for linear observables. 

The diagonal dynamics can be cast in another form which is particularly instructive. Using the identity for two spins,
$
    \mathbf{\h{S}}_i\cdot \mathbf{\h{S}}_j
    =\frac{1}{2}\,\mathrm{SWAP}_{ij}-\frac{1}{4}\,\mathbb{I},
$
where $\mathrm{SWAP}_{i,j}$ exchanges the neighboring spins, we obtain 
\begin{gather}\label{eq:SSEP}
    e^{\mathcal{L}^{\textsf{diag}}t} 
    = \exp\!\left(\gamma t\sum_{\langle i,j\rangle} (\mathrm{SWAP}_{i,j}-1)\right)
    =  \lim_{\delta t\rightarrow 0}\exp\!\left(\gamma \delta t\sum_{\langle i,j\rangle} (\mathrm{SWAP}_{i,j}-1)\right)^{t/\delta t} \nonumber \\
    \lim_{\delta t\rightarrow 0}\bigg(\prod_{\langle i,j\rangle} \Big(1-\gamma \delta t+\gamma \delta t \mathrm{SWAP}_{i,j}\Big)\bigg)^{t/\delta t} \nonumber \\
    = \int \rho_{[0,t]}(d\omega)\, \prod_{({\langle i,j\rangle},t)\in\omega} \mathrm{SWAP}_{i,j}.
\end{gather}

In other words, within a short time window $\delta t$ on a single bond $(i,i+1)$, the operator $\mathrm{SWAP}_{i,i+1}$ is applied with average rate $\gamma$. The resulting history can be visualized as a ladder where swap events are rungs that are placed randomly in time and space according to a Poisson process. We denote a realization of this stochastic update pattern by $\omega$. Identifying an occupied site $\bullet$ with $\uparrow$ and an empty site $\circ$ with $\downarrow$, the dynamics leads to hopping of particles with a rate $\gamma$, modulo the exclusion constraint that at most one particle can occupy a given site. Since the left and right hopping rates are symmetric, this results in the classical symmetric simple exclusion process \cite{spohn2012large} (SSEP) in the diagonal charge sector. This dynamics is illustrated in Fig.~\ref{fig:worldlines}. In the remainder of this section, we will study the dynamics of SW-SSB in this model in one and two dimensions. We impose the strong symmetry starting from an initial state with fixed charge; specifically, we consider dynamics of the initial state $|\rho_0\rangle = |\text{N\'eel}\rangle$, in both one and two dimensions. In the particle language, this is the half-filled state with particles on every other site.

Our results are summarized as follows: 

\begin{itemize}
    
    \item In 1d, SW-SSB does not occur at finite time, but the dynamics of linear and nonlinear observables are parametrically different. Charge spreads diffusively, but length scales associated with nonlinear observables ($\xi^{(1)}$ and $\xi^{(2)}$ for R\'enyi correlations, the Markov length $\xi^M$ for the CMI, and the region size $\xi^{\text{dec}}$ required for accurate charge decoding) all grow ballistically. Intuitively, successful recovery of information becomes increasingly difficult at late times, but is always possible given access to a sufficiently large snapshot of the system.
    
    \item In 2d, we find signatures of a BKT-like SW-SSB transition at finite time by evaluating R\'enyi susceptibilities and stiffness.  We expect the associated state recovery and decoding tasks to fail in 2d past this transition.
        
\end{itemize}

\subsubsection{Numerical method: sampling trick}\label{sec:spin-1/2:numerics}

In the following sections, we use the $U(1)$-symmetric spin-$1/2$ model as a numerically accessible setting. Throughout, we take the initial state $\hat \rho_0$ to be a N\'eel state. In most of the cases we consider, the dynamics is effectively classical in the sense that the density matrix remains diagonal in the charge basis [see Eq.~\eqref{eq:charge_basis}]. We therefore treat $\hat{\rho}_t$ as a classical probability distribution over charge configurations, evolved by a diagonal-sector generator, $|{\rho}_t\rangle = e^{\mathcal{L}^{\textsf{diag}} t}|{\rho}_0\rangle$.
We simulate this evolution using time-evolving block decimation (TEBD) with matrix product states (MPS) in one dimension, and quantum Monte Carlo (QMC) in two dimensions. In particular, for the 2d case we use a worm algorithm to sample and average over spacetime trajectories with the specified boundary conditions.

R\'enyi-$2$ observables are straightforward in both approaches. In 1d they reduce to standard MPS contractions. In 2d they correspond to a spacetime trajectory ensemble with the temporal boundary condition fixed as $\hat{\rho}_0$ at both the bottom and the top, over a total temporal extent $2t$. Operator insertions are implemented directly in both cases, so R\'enyi-$2$ correlators are easy to evaluate. In two dimensions, winding-number observables are likewise straightforward to measure within the same trajectory ensemble.

R\'enyi-$1$ observables are less direct because they involve $\ket{\sqrt{\rho}}$ and therefore do not reduce to a single standard contraction. There are two ways to overcome this issue. First, one can use tensor cross interpolation (TCI) to infer $\ket{\sqrt{\rho}}$ by sampling $\ket{\rho}$, after which  R\'enyi-$1$ correlations may be computed in the standard way~\cite{OSELEDETS201070}. Second, one can sample R\'enyi-$1$ correlations directly from $\ket{\rho}$ by the following approach. In 1d, this can be done by rewriting the configuration sum as an average over the spins away from $i$ and $j$. Denoting by $\bar{s}_{ij}$ the configuration on all sites except $i$ and $j$, one can write
\begin{align}
C^{(1)}(i,j)
= \sum_{\bar{s}_{ij}} P(\bar{s}_{ij})\sqrt{P(\downarrow_i \uparrow_j \mid \bar{s}_{ij})P(\uparrow_i \downarrow_j \mid \bar{s}_{ij})}.
\end{align}
Then, $\bar{s}_{ij}$ can be sampled from its marginal $P(\bar{s}_{ij})$ by sequential sampling of the MPS~\cite{lloyd2025diverging}. For each sampled $\bar{s}_{ij}$, one fixes all spins except those at $i$ and $j$, and evaluates the two conditional probabilities, and then averages the resulting estimator over samples. In this work, we primarily use TCI, but also verify the results using direct sampling of $\ket{\rho}$.

We evaluate the CMI using the same sampling method~\cite{lloyd2025diverging}. In particular, we sample a configuration $s_B$ on region $B$ from its marginal, and for each sample we compute the conditional distribution on $A$ and $C$ needed to estimate the CMI using $I(A:C|B) = \sum_{s_B} P(s_B)I(A:C|B=s_B)$.

In 2d, we evaluate R\'enyi-$1$ and related disorder-averaged quantities using a two-step sampling procedure. We first generate an outcome configuration $\{s\}$ by evolving the diagonal-sector classical dynamics defined by $\mathcal{L}^{\textsf{diag}}$ up to time~$t$, which samples $\{s\}$ with the correct weight. Conditioned on this sampled $\{s\}$, we then run a second QMC with fixed temporal boundaries given by $|\rho_0\rangle$ at the bottom and $|s\rangle$ at the top, and we measure the desired conditional quantities within this fixed-boundary ensemble. This gives the probability ratio $P(s')/P(s)$ entering the estimator for $C^{(1)}(i,j)$. We also compute a stiffness-like quantity by measuring the integrated current (total displacement) in a chosen direction within the fixed-boundary QMC run and then averaging its fluctuations over the outer samples of~$\{s\}$. In the classical limit, this reduces to the winding-number definition used in Ref.~\cite{temkin2025charge}.

\subsection{1d results}\label{sec:spin-1/2:1d}

\subsubsection{Diffusion and R\'enyi correlations}\label{sec:1d:correlators} 

In this section, we present numerical data for the evolution over time of local charge, R\'enyi-$2$ correlations, R\'enyi-$1$ correlations, and the infinite MPS R\'enyi-$2$ transfer matrix gap. As is well known, the spreading of local charge displays a diffusive length scale $\xi^D(t) \sim \sqrt{t}$. Nevertheless, the other observables all indicate the presence of ballistic length scales $\xi^{(1)}(t) \sim t$ and $\xi^{(2)}(t) \sim t$ for R\'enyi-$1$ and R\'enyi-$2$ correlations respectively. Notably, we find numerically that $\xi^{(1)}(t) = 2\,\xi^{(2)}(t)$ in the spin-$1/2$ model, matching an analytical result obtained for the 1d decohered rotor model in Sec.~\ref{sec:rotor:1d}.

As expected from the Mermin--Wagner theorem, in one dimension a continuous $U(1)$ symmetry does not exhibit symmetry breaking after any finite time evolution, neither in the strong nor in the weak sense. Therefore, SW-SSB is not observed at finite times, and long-range order can emerge only in the asymptotic limit of infinite time. This is reflected in our numerical results for the nonlinear observables, whose correlation lengths diverge only after extensive time.

Although there is no SW-SSB at finite times, we still observe the expected diffusive spreading of the conserved charge. This is directly seen from domain-wall melting starting from a sharp domain-wall initial state $\ket{{\uparrow}\dots{\uparrow}{\downarrow}\dots{\downarrow}}$. The magnetization profiles $\langle S_x^z\rangle$ at different times collapse under the diffusive rescaling $x/\sqrt{t}$, indicating that ${\langle S_x^z\rangle = f(x/\sqrt{t})}$ for suitable times (after a constant initial onset time of diffusion, and before the boundaries become relevant). These data are given in Fig.~\hyperref[fig:diffusive_spreading]{\ref*{fig:diffusive_spreading}} with $L=100$ and $\gamma = 0.1$.

\begin{figure}[h]
    \centering
    \includegraphics[width=0.5\linewidth]{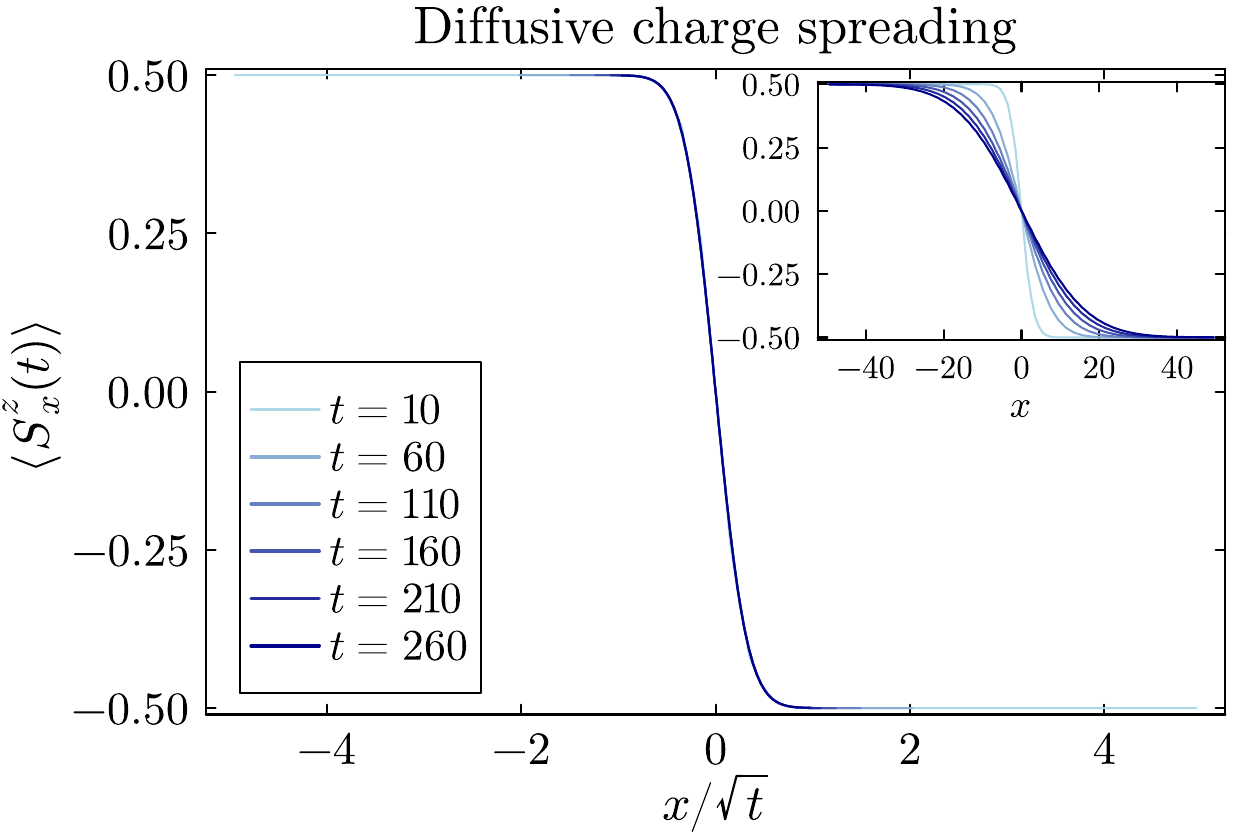}
    \caption{Spatial charge profiles over time from a domain wall initial state, with $L = 100$ and $\gamma = 0.1$ from $t = 10$ to $t = 260$. In the inset, the horizontal axis is not scaled and it is apparent that the initial domain wall melts over time, with charge spreading from the left to the right. When the horizontal axis is rescaled by $\sqrt{t}$, the charge profiles at different times all collapse to one curve. This indicates a scaling form $\expval{S^z_i(t)} = f(x/\sqrt{t})$ over this time range, indicative of diffusion. In both cases, $x=0$ is taken to be halfway between site $50$ and site $51$.}
\label{fig:diffusive_spreading}
\end{figure}

The behavior of nonlinear observables is markedly different. In particular, we find that the R\'enyi-$1$ and \mbox{R\'enyi-$2$} correlators decay exponentially in space, with a decay length that grows \emph{linearly} with the evolution time. This behavior is reminiscent of one-dimensional free bosonic systems at finite temperature. There, thermal fluctuations introduce a finite thermal length scale, so correlators such as $\Tr(b_0 b^\dagger_r e^{-\beta H})$ decay exponentially at large $r$, with a correlation length $\xi^\beta$ proportional to $\beta$. As discussed in the Sec.~\ref{sec:diagnostics:worldline}, the R\'enyi-$2$ correlation length is 
akin to the finite-temperature correlation length of the Heisenberg model. We remark that if the environment were non-Markovian such that symmetric feedback could be applied to the system, the system could develop standard SSB instead of SW-SSB in the steady state~\cite{hauser2024continuous}.

Numerical data for R\'enyi-$1$ and R\'enyi-$2$ correlators in our spin-$1/2$ model are given in Fig.~\hyperref[fig:1d-renyi]{\ref*{fig:1d-renyi}(a)} for $L = 128$ and $\gamma = 0.1$. We observe that $C^{(1)}(x;t)$ and $[C^{(2)}(x;t)]^{1/2}$ are linear on a semi-log plot when $x$ is sufficiently large (but not so large that boundary effects arise), indicating that $C^{(1)}(x;t) \propto e^{-x/\xi^{(1)}(t)}$ and $C^{(2)}(x;t) \propto e^{-x/\xi^{(2)}(t)}$. More specifically,  we observe that the data collapse reasonably well under the ballistic rescaling $x/t$. Both $C^{(1)}(x;t)$ and $[C^{(2)}(x;t)]^{1/2}$ are proportional to a scaling form $f(x/t)$, which implies that $\xi^{(1)}(t)$ and $\xi^{(2)}(t)$ both grow ballistically. Furthermore, since $C^{(1)}(x;t)$  and $[C^{(2)}(x;t)]^{1/2}$ have the same slope, it follows that $\xi^{(1)}(t) = 2 \,\xi^{(2)}(t)$. The R\'enyi-$1$ data in Fig.~\hyperref[fig:1d-renyi]{\ref*{fig:1d-renyi}(a)} are obtained using TCI, and we show in Appendix~\ref{app:TCI_vs_sampling} that the results are consistent with data obtained from direct sampling. 

The long-distance decay of R\'enyi-$2$ correlators is directly related to the gap of the MPS transfer matrix, similar to the case of standard observables in quantum ground states \cite{bridgeman2017hand}. We can work directly with infinite MPS (iMPS) so that the transfer matrix is translationally invariant.
The largest eigenvalue of the transfer matrix, if appropriately normalized, will be $1$; let the next-largest eigenvalue be $\lambda_2$. Then, (R\'enyi-2) correlations at sufficiently large distance $r$ will be suppressed by $\lambda_2^r$, from which we can infer a correlation length $\xi = -(\log \lambda_2)^{-1}$. The growth of this correlation length over time is given in Fig.~\hyperref[fig:1d-renyi]{\ref*{fig:1d-renyi}(b)}, and is clearly ballistic. 

In Sec.~\ref{sec:rotor:1d}, we study both the R\'enyi-$1$ and R\'enyi-$2$ observables analytically in a phenomenological rotor model. This theory predicts that the correlation lengths extracted from the R\'enyi-$1$ and R\'enyi-$2$ correlators are related by a factor of two, as we observed in Fig.~\hyperref[fig:1d-renyi]{\ref*{fig:1d-renyi}(a)}.

\begin{figure}
    \centering
    \begin{overpic}[width=0.49\linewidth]{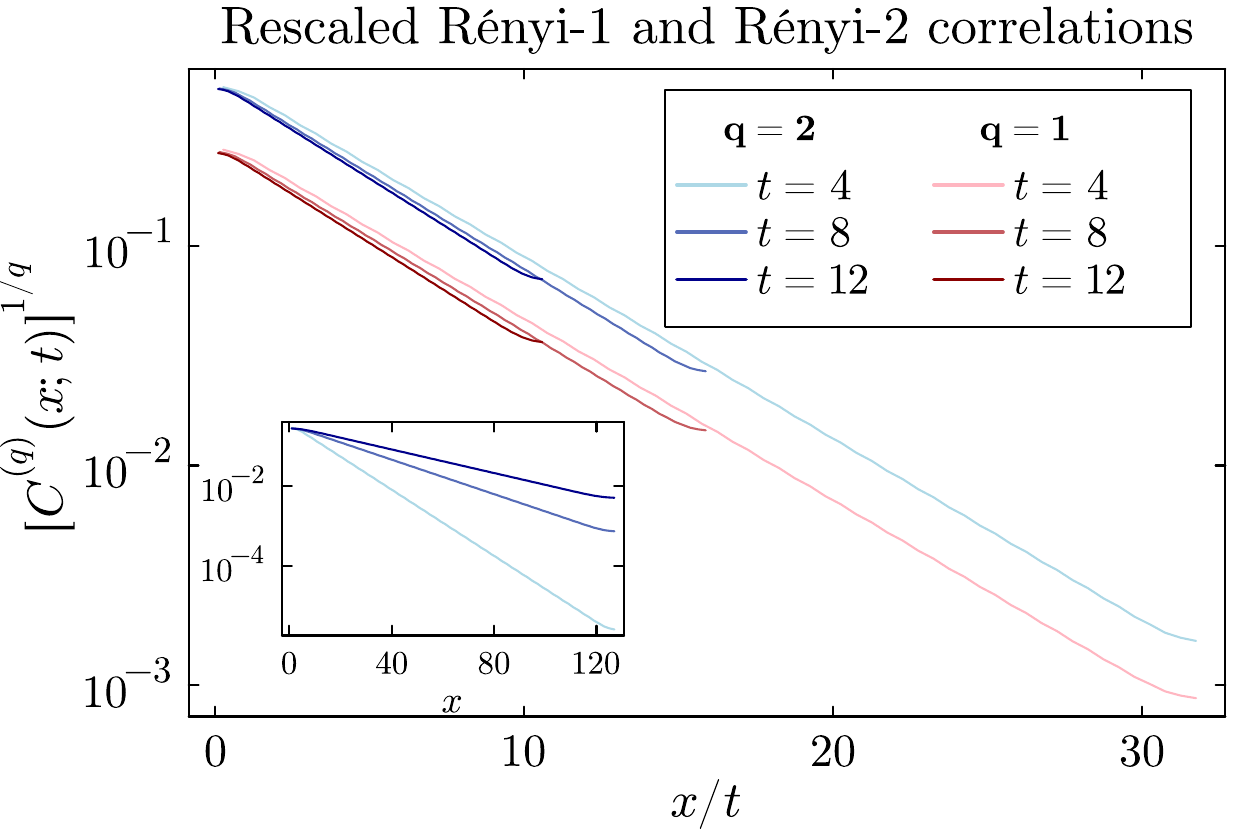}
      \put(5,60){(a)}
    \end{overpic}\hfill
    \begin{overpic}[width=0.49\linewidth]{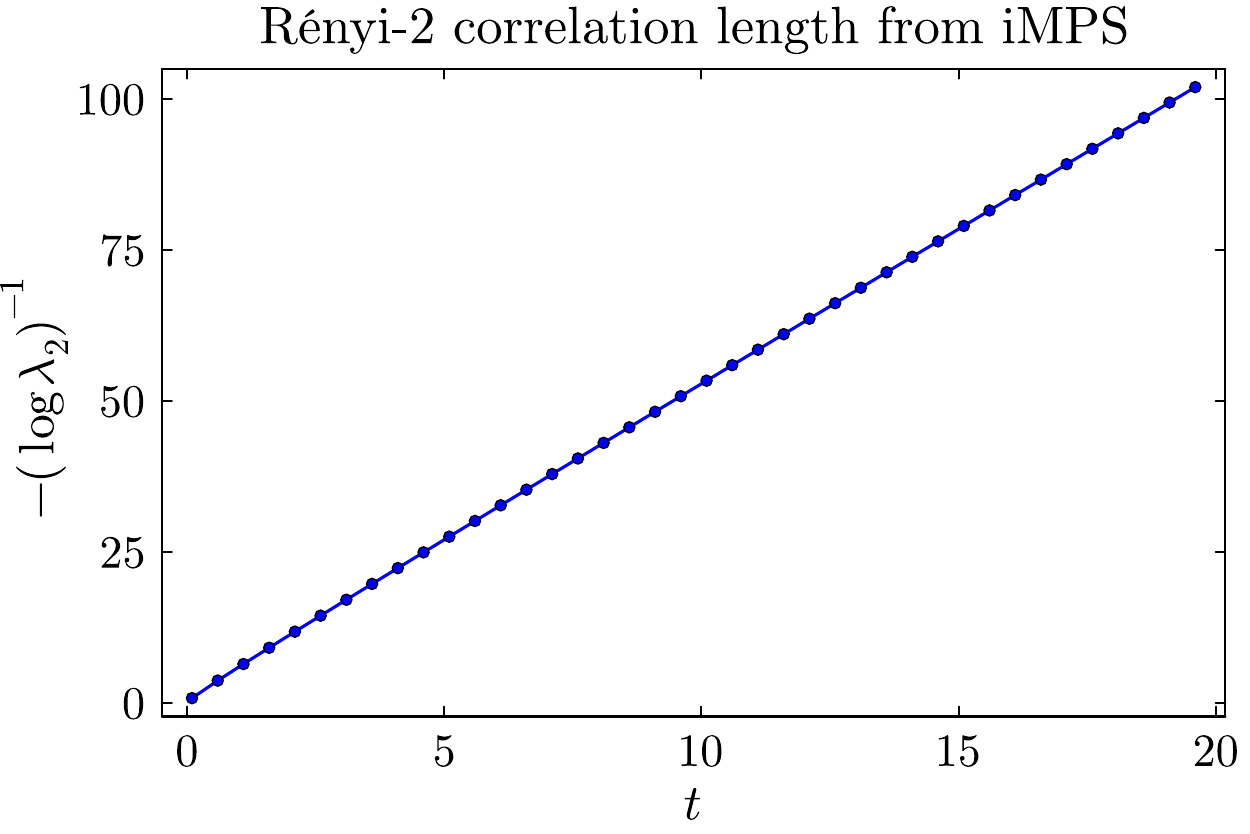}
      \put(0,60){(b)}
    \end{overpic}
    \caption{(a) Evolution of R\'enyi-$1$ and R\'enyi-$2$ correlations over time, with $L=128$ and with $\gamma = 0.1$. In the inset, $[C^{(2)}(x;t)]^{1/2}$ is given on a semi-log plot vs. $x$ for a range of times, demonstrating that $[C^{(2)}(x;t)]^{1/2} \propto e^{-x/(2\,\xi^{(2)}(t))}$. Upon rescaling the horizontal axis, we see that both $C^{(1)}(x;t)$ and $[C^{(2)}(x;t)]^{1/2}$ are functions of $x/t$, and specifically that $C^{(1)}(x;t) \propto [C^{(2)}(x;t)]^{1/2} \propto e^{-x/(2\,\xi^{(2)}(t))}$. This implies that $\xi^{(1)}(t) = 2\,\xi^{(2)}(t) \propto t$. (b) The infinite MPS R\'enyi-$2$ transfer matrix gap is given on a log-log scale. The data indicates saturation to $t^{-1}$ scaling. Specifically, applying a linear fit to the data from $t=2$ to $t=20$ yields a slope of $0.97 \approx 1$.}
    \label{fig:1d-renyi}
\end{figure}

\subsubsection{Conditional mutual information (CMI)}\label{sec:1d:information}

We now turn to a numerical study of the system's conditional correlations. The CMI is expected to decay according to the Markov length, as reviewed in Section~\ref{sec:1d:correlators}. The SW-SSB of a zero-form symmetry (such as $U(1)$) is known to imply the divergence of the Markov length \cite{lessa2025strong}. We expect the Markov length to follow a similar scaling to the R\'enyi correlation lengths, growing as $\xi^M \propto t$. 

We use the method of Ref.~\cite{lloyd2025diverging} to numerically sample the CMI, representing the time-evolved state as an MPS. We work directly in the thermodynamic limit using the framework of uniform MPS \cite{vanderstraeten2019tangent}. Results below are for a maximum bond dimension $\chi=48$, which we have checked is well-converged for the CMI data and density correlators, and a hopping rate $\gamma=0.1$. The results for the CMI are shown with a contour plot in Fig.~\hyperref[fig:1d-CMI]{\ref*{fig:1d-CMI}(a)}, as a function of evolution time $t$ and size of the buffer region $R_B=|B|$. The data are plotted on a logarithmic color scale and normalized at each timestep by $I_{0,t} = \text{max}_{R_B}\ I_t(A:C|B)$. It is clear that the Markov length increases with time, consistent with a divergence at infinite time suggested by our results of the previous section. To demonstrate the Markov length scaling, we show a scaling collapse of the CMI data at different times with respect to the ballistic scaling variable $x_b=t$, in Fig.~\hyperref[fig:1d-CMI]{\ref*{fig:1d-CMI}(b)}. The CMI data are rescaled by a factor of $t^2$, to better highlight the collapse. The data shows a good collapse to an exponential decay for $R_B \gg x_b$, consistent with a Markov length $\xi^M \propto t$. Interestingly, a collapse with respect to the diffusive scaling variable $x_d = \sqrt{t}$, shown in the inset to Fig.~\hyperref[fig:1d-CMI]{\ref*{fig:1d-CMI}(b)}, reveals an intermediate regime $x_d \lesssim R_B \ll x_b$ where the CMI decays \emph{algebraically}, $I_t(A:C|B)\propto R_B^{-\alpha}$. Numerically we find $\alpha \approx 1.2-1.4$, but the exponent drifts continuously toward larger values over the accessible times. We will connect this surprising behavior to the emergence of a Gaussian hydrodynamic regime in Section \ref{sec:classical}, which predicts $\alpha = 2$.

Our data suggests that there is no mixed-state phase transition at finite time in the 1d model, but instead the Markov length demonstrates the same ballistic scaling $\xi^M \propto t$ as the R\'enyi correlation lengths studied in the previous section. The possibility remains for a finite-time transition in higher dimensions. We provide numerical evidence for this scenario by extracting the Markov length for the spin-1/2 model in quasi-1d ladders of increasing width $W$. We compute the Markov length using the Lyapunov spectrum of the MPS (see \cite{lloyd2025diverging}), which provides a cleaner extrapolation than fitting the CMI data. We show in Fig.~\hyperref[fig:1d-CMI]{\ref*{fig:1d-CMI}(c)} the Markov length scaling for the 1d chain ($W=1$) and several ladders of width $W=2$ to $W=4$. The Markov lengths display clear linear-time growth, with the larger-width ladders growing as $\xi^M_W \propto W\xi^M_1$. Extrapolating to the 2d limit, we expect that a mixed-state phase transition occurs at finite time in $d>1$, consistent with our finding in Sec.~\ref{sec:spin-1/2:2d}.

\begin{figure}
  \centering
  \includegraphics[width=0.96\linewidth]{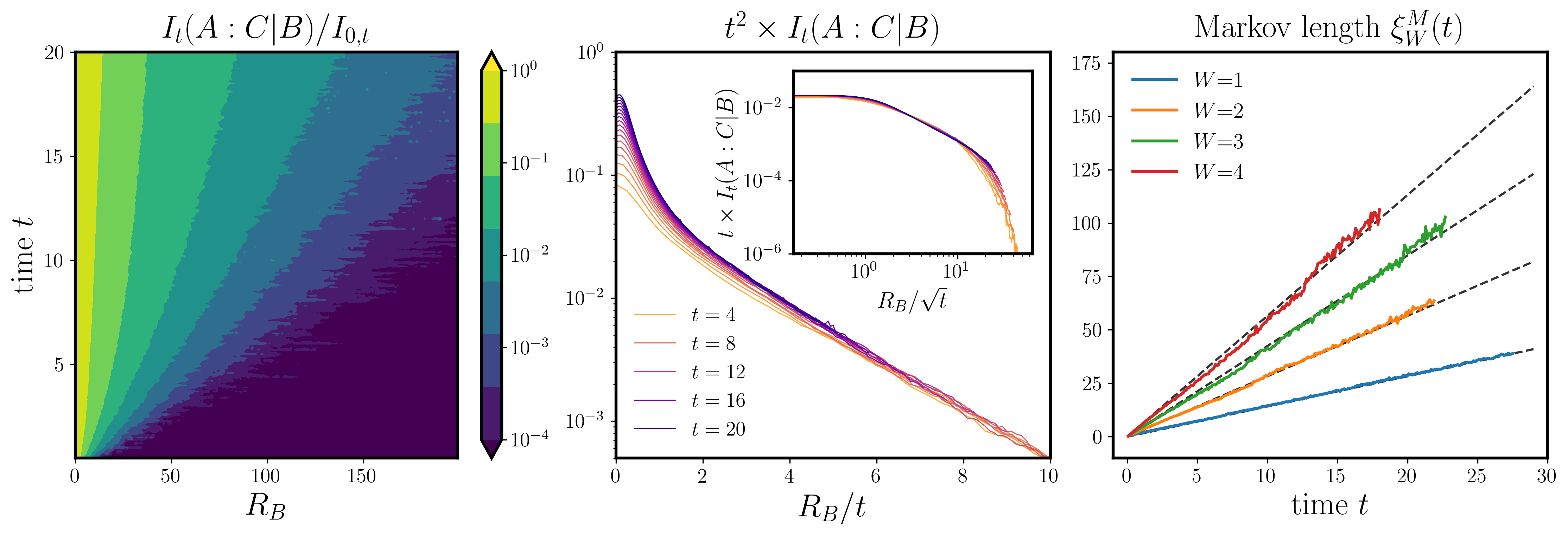}
  
  \caption{(a) CMI $I_t(A:C|B)$ for the spin-$1/2$ model, as a function of evolution time $t$ and subsystem size $R_B= |B|$. The colorbar is log scaled. The lengthscale of the CMI (Markov length) diverges as a function of the evolution time, $\xi^M(t)\propto t$. (b) Scaling collapse of CMI vs ballistic lengthscale $x_b=t$, demonstrating the linearly growing Markov length. \emph{Inset:} Scaling collapse of CMI vs diffusive lengthscale $x_d=\sqrt{t}$. The data shows a pre-Markov regime where conditional correlations decay algebraically (see Sec.~\ref{sec:classical}). (c) Markov length $\xi^M_W(t)$ as a function of time for the chain ($W=1$) and ladders of increasing width ($W=2-4$). The Markov length follows the scaling $\xi^M_W(t) \sim W t$ -- dashed lines display the lines $W\xi^M_1(t)$.}
  \label{fig:1d-CMI}
\end{figure}

\subsubsection{Decoding charge}\label{sec:1d:decoding}

\begin{figure}
    \centering
    \includegraphics[width=0.7\textwidth]{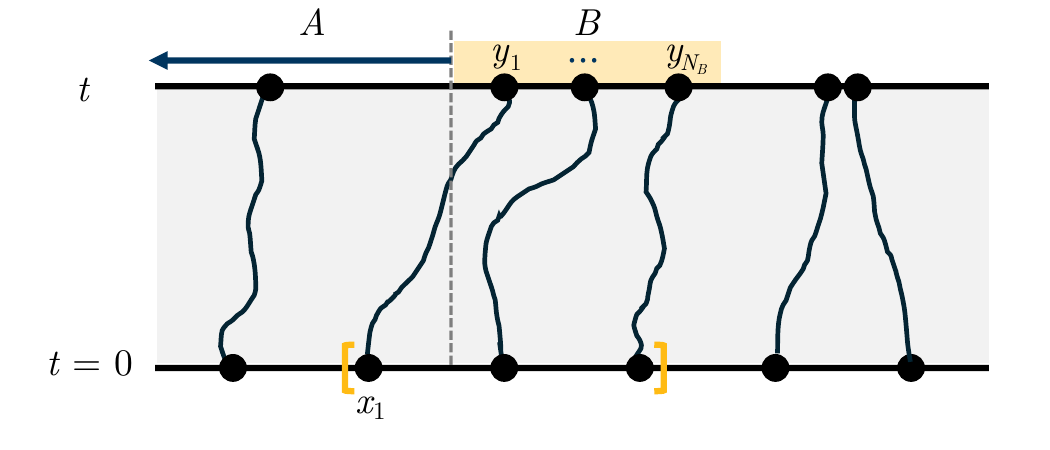}
    \caption{Cartoon of the decoding task. At a fixed time $t$ we observe occupations in the window $B$ (yellow) and aim to infer the number of particles remaining in $A$. The observed particle positions are $y_1,\dots,y_{R_B}$. From these, we infer the originating block of $N_B$ consecutive odd sites with leftmost site $x_1$ and obtain the total particle number passed through the boundary between $A$ and $B$.}
    \label{fig:decoding1d}
\end{figure}

To complete our 1d results, we consider several decoding protocols which connect the SW-SSB transition to the difficulty of learning the charge of a subregion, which we will denote $A$. We note that the Markov length studied in the previous section sets the size of subregion $B$ needed in order to accurately recover the \emph{entire} state in region $A$, given access to the state on $B$ \cite{sang2025stability, fawzi2015quantum}. Here we consider the simpler task of decoding the total charge $N_A$ of the region, ignoring information about the spatial distribution of charges. As we will see, the accuracy of charge decoding increases with the size $R_B$ of the accessible region $B$, enabling us to pull out a decoding length scale, $\xi^{\rm dec}(t)$. We find that $\xi^{\rm dec}(t)$ diverges linearly with time, matching our results for the R\'enyi and Markov length scalings. Our results provide an intuitive picture for the ballistic growth of these length-scales in 1d. 

We take a system of $2L$ spin-$1/2$ degrees of freedom initialized in the N\'eel state, with open boundary conditions. In the particle language, this corresponds to a total of $L$ particles occupying the odd sites and the dynamics is given by the SSEP. We work in the regime $L\gg t^2$ so boundary effects are negligible. We take the region $A$ to consist of the first $L$ sites, and region $B$ as the sites $[L+1,\dots,L+R_B]$, so $ R_B=|B|$. 
The decoding task is to correctly infer the number of particles in $A$, given the charge configuration of region $B$. Note that for $R_B=L-|A|$, the answer follows immediately from number conservation, $N_A = L-N_B$.

Since the dynamics is an exclusion process, particle worldlines cannot cross. Therefore, the $N_B$ particles observed in $B$ at the final time $t$ must originate from $N_B$ consecutive particles in the initial N\'eel pattern. Let $x_1$ denote the leftmost site of this candidate block. Given an estimate of $x_1$, we obtain the inferred number of particles in $A$ by counting how many positive odd integers lie strictly to the left of $x_1$. The decoder succeeds if this inferred value equals the true count: see Fig.~\ref{fig:decoding1d}. An optimal decoder would evaluate the likelihood for every possible $x_1$ of producing the observed configuration in $B$ and then choose the maximum-likelihood estimate. Below, we analyze three transparent non-optimal decoders that may be tested with very large system sizes. Although they are suboptimal, all three capture the correct scaling of the minimal size of $B$ needed for successful decoding, namely $\xi^{\rm dec}\sim t$. Lastly, we implement the optimal decoder using MPS simulations of the dynamics and analyze its performance, again finding scaling consistent with $\xi^{\rm dec} \sim t$.

\noindent\rule{\linewidth}{0.4pt}

\paragraph{{Center-of-mass (COM) decoder.}}
The COM decoder infers the leftmost site $x_1$ from the center of mass of the $N_B$ occupied sites observed in $B$ at time $t$. Denote these occupied positions by $\{y_1,\dots,y_{N_B}\}$ and define $\bar{y}\equiv \frac{1}{N_B}\sum_{j=1}^{N_B} y_j$. If $N_B=0$, the decoder trivially fails. If the inference length $R_B$ is large, then typically $N_B$ is large, so $\bar{y}$ is self-averaging and fluctuates weakly around its initial position.

The decoder models the occupied sites as originating from a length-$N_B$ block of consecutive odd sites $\{x_1,\,x_1+2,\,\dots,\,x_1+2(N_B-1)\}$. The center of mass of this block is $\bar{x}\equiv x_1+(N_B-1)$. Since $x_1$ is odd, $\bar{x}$ is an integer with parity $\bar{x}\equiv N_B\ (\mathrm{mod}\ 2)$. We therefore round the observed $\bar{y}$ to the nearest integer $x^\star$ with the same parity as $N_B$ as an estimator of $\bar{x}$ and infer
\begin{align}
x_1 = x^\star - (N_B-1).
\end{align}

The decoder fails when $\bar{y}$ crosses a decision boundary, halfway between adjacent admissible integers, which are spaced by $2$. On hydrodynamic scales SSEP is diffusive, implying $\mathrm{Var}(\bar{y})\sim t/N_B$. Thus the success probability is controlled by $t/N_B$. Since $N_B\sim R_B$ at finite density, this suggests the decoding success rate is controlled by $R_B/t$, consistent with Fig.~\ref{fig:decoding_result}(a).

\noindent\rule{\linewidth}{0.4pt}

\paragraph{{Minimal-weight perfect matching (MWPM) decoder.}}
Let the occupied positions in $B$ be $y_1\le \cdots \le y_{N_{B}}$. For each candidate odd integer $x_1$, we define the $\ell_1$ matching cost
\begin{align}
    \sum_{i=1}^{N_{B}}\big|y_i-(x_1+2(i-1))\big|.
\end{align}
The MWPM decoder chooses an $x_1$ that minimizes this cost. If there are multiple minimizers, one is selected uniformly at random.

\noindent\rule{\linewidth}{0.4pt}

\paragraph{Height decoder.}
Instead of matching individual particles, we summarize the configuration in $B$ by an integrated \textit{height} profile. We infer how many particles have crossed the cut by estimating the height offset, which we fix by requiring that the average height in the window stays close to its initial value $1/2$ from the N\'eel state. A useful intuition is that the height is a cumulative sum. A fluctuation entering the window from the left boundary shifts the entire profile inside $B$, whereas a fluctuation near the right edge affects the profile only locally. This asymmetry makes the height profile sensitive to the net particle transfer across the cut.\\

Let $n_j\in\{0,1\}$ denote the occupation at site $j$ at time $t$ ($n_j=1$ if occupied and $n_j=0$ if empty). We define the height field by
\begin{align}
    h(0)=0,\qquad
    h(j)=h(j-1)+\big(2n_j-1\big),\quad j=1,\dots,2L,
\end{align}
so the height increases (decreases) by one on occupied (empty) sites. We anchor a relative height at the cut and build it across the window,
\begin{align}
    h_{\mathrm{rel}}(j)=\sum_{x=L+1}^{j}\big(2n_x-1\big),
    \quad j\in B.
\end{align}
This determines the profile inside $B$ up to an unknown additive constant, $h(L)$. We estimate this constant by shifting the profile so that its window average is closest to $1/2$. Once $h(L)$ is inferred, we obtain the number of particles remaining in $A$.

\noindent\rule{\linewidth}{0.4pt}

\paragraph{Optimal decoder.} At time $t$ the density matrix of the system will be some distribution
\[\label{eq:optimal_decoder_1}
\ket{\rho_t} = \sum_{s_A, s_B, s_C} P(s_A, s_B, s_C; t)\ket{s_A}\ket{s_B}\ket{s_C}.
\]
Thus for each configuration $s_B$, there is a distribution of possible total charges $N_A$ in region $A$:
\[\label{eq:optimal_decoder_2}
P(N_A | s_B; t) = \sum_{s_A,s_C}P(s_A,s_B,s_C;t)\,\delta_{\abs{s_A},N_A}\delta_{\abs{s_C},N-N_A-\abs{s_B}} 
\]
where $\abs{s}$ counts the total charge in configuration $s$ and $N$ is the total charge of the whole system. The optimal decoder is a \emph{maximum likelihood} decoder which guesses that the total charge in region $A$ is
\[\label{eq:optimal_decoder_3}
N_A^* = \argmax_{N_A} P(N_A | s_B; t).
\]
This decoder will succeed with probability $P(N_A^*)$, which is the quantity plotted in Fig.~\hyperref[fig:decoding_result]{\ref*{fig:decoding_result}(d)}.

The sequence of Eqs.~\eqref{eq:optimal_decoder_1}-\eqref{eq:optimal_decoder_3} essentially describes the protocol for efficiently computing $N_A^*$. In 1d, we may efficiently represent $\ket{\rho_t}$ as a state. Within this framework, we can erase all information in region $A$ except for $N_A$ by applying a sequence of gates that map $\ket{n_1}\ket{n_2} \to \ket{n_1+n_2}$ until the sites in region $A$ have been replaced by a single qudit with dimension $R_A+1$. Then, when region $B$ is projected down to configuration $s_B$ and region $C$ is traced out, the result is a vector of dimension $R_A+1$ encoding $P(N_A | s_B; t)$. Consequently, the optimal decoder may be efficiently benchmarked and efficiently applied in practice.

\noindent\rule{\linewidth}{0.4pt}
\\

Fig.~\ref{fig:decoding_result} shows the performance of the four decoders. The success probabilities collapse well when plotted as a function of $R_B/t$, indicating the scaling $\xi^{\rm dec}(t)\sim t$. This highlights that although local charge correlations develop diffusively under the SSEP dynamics, successfully inferring the charge fluctuation at a given cut requires access to a window whose size grows proportionally to the evolution time. We note that the task of estimating $N_A$ is closely related to the task of estimating the (``disorder''-dependent) winding number $W$ defined in Eq.~\ref{disorderstiff} (Section \ref{sec:spin-1/2:2d}). In other words, $N_A(t)-N_A(0)$ counts the total flux of particle worldlines crossing the cut shown by the dashed line in Fig.~\ref{fig:decoding1d}; in the Figure, $W=+1$. The possibility of decoding accurately ($p\approx 1$) for $R_B\gg \xi^{\rm dec}$ implies the variance of $W$ vanishes for system sizes $L \gg \xi^{\rm dec}$, consistent with there being no sharp SW-SSB transition in 1d. We will show next that the situation is very different in 2d, and the winding number fluctuations do not decay with the system size.

\begin{figure}[t]
    \centering
    \includegraphics[width=0.85\textwidth]{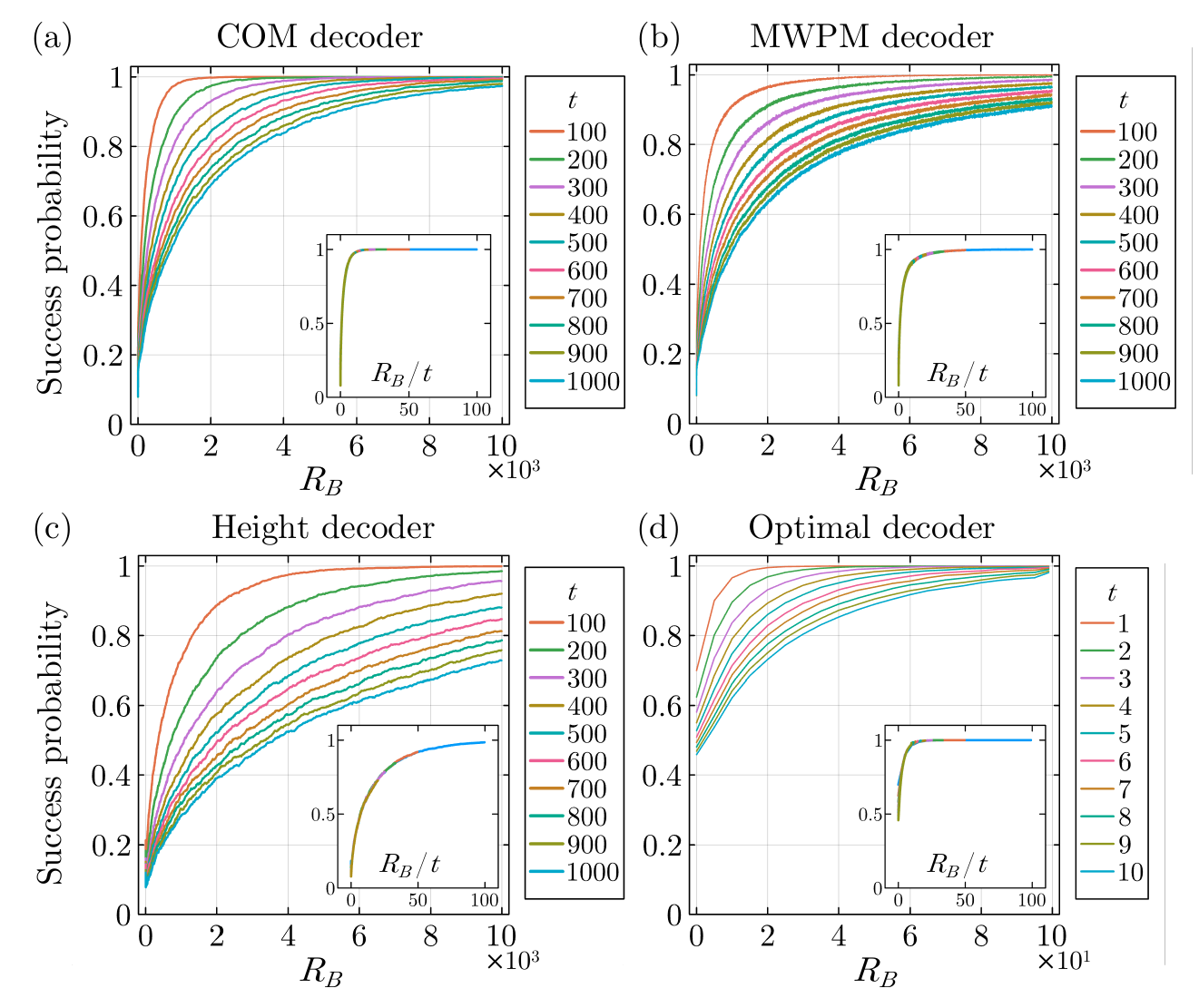}
    \caption{Success probability versus size of $B$ region $R_B$, for several times $t$ and for (a) the COM decoder, (b) the MWPM decoder, (c) the height decoder, and (d) the optimal decoder. Insets show the same data plotted against the rescaled width $R_B/t$, exhibiting an excellent collapse. Simulations for (a,b,c) use $L=2\times10^4$ and each curve is averaged over $\sim 10^4$ samples. Simulations for (d) use $L=2\times 10^2$ and each curve is averaged over $2000$ samples. We note that the large probabilities of success at $R_B = 0$ in (d) would be suppressed at larger system sizes and later times.}
    \label{fig:decoding_result}
\end{figure}

\subsection{2d results}\label{sec:spin-1/2:2d}
According to the Mermin-Wagner theorem, in a Gibbs state at equilibrium there cannot be a sharp transition that spontaneously breaks the $U(1)$ symmetry in 1d. In 2d there can be a quasi-long-range order in the low temperature phase below the BKT phase transition. From the formalism we developed we have seen the time $t$ of the Lindbladian plays the role of the inverse temperature in the imaginary time evolution. The goal of this subsection is to pursue evidence for a BKT-like transition at finite time during the evolution, resulting in a BKT algebraic mixed state phase. 

Even though the evolution Eq.~\eqref{eq:imaginaryevolution} can be regarded as an imaginary time evolution with $t$ identified as inverse temperature, the physics can be quite different from an equilibrium problem defined by the effective Hamiltonian $-\mathcal{L}^{\text{diag}}$. Indeed various quantities we will be considering show properties different from their equilibrium counterparts. Moreover, due to the $SU(2)$ invariance, at equilibrium the equal time correlation function will never be long range ordered in 2d (We note at this point that the $SU(2)$ symmetry only occurs for spin-1/2 systems. If we instead consider a general decohered spin-$S$ system, there is no $SU(2)$ symmetry even in the diagonal subspace). However our numerics suggest that $\hat \rho_t$ does undergo a finite time transition, and develop a quasi-long-range order, for the nonlinear observables we consider.

We consider a 2d square lattice of size $L\times L$. To numerically access the various nonlinear observables related to SW-SSB, we use a quantum Monte Carlo method for results obtained in this subsection. The QMC method samples spacetime trajectories from which an estimator of the various observables can be obtained. To detect SW-SSB of $U(1)$ symmetry, we first examine the R\'enyi-2 and R\'enyi-1 correlation functions (defined in Section~\ref{sec:diagnostics:observables}). We define the R\'enyi susceptibility as
\beqn    \chi^{(Q)}(t) = \frac{1}{L^2}\sum_{i,j}C^{(Q)}(i,j;t),  \label{chi}
\eeqn
for $Q=1,2$. The R\'enyi susceptibilities are expected to saturate to a constant in the disordered phase, while inside the BKT algebraic phase it grows with system size as a power-law. We plot the computed susceptibilities in Fig.~\ref{fig:susceptibility}. The expected behavior is verified for the R\'enyi-2 susceptibility. In particular, $\chi^{(2)}/L^2$ decays with $L$ as a power law in the BKT algebraic SW-SSB phase. We provide further evidence with a rotor-model calculation performed in Appendix~\ref{sec:winding}. The R\'enyi-1 susceptibility is less definitive, as $\chi^{(1)}/L^2$ does not show a clear power-law scaling with $L$. We comment on this below.

\begin{figure}
    \centering    \includegraphics[width=0.32\linewidth]
    {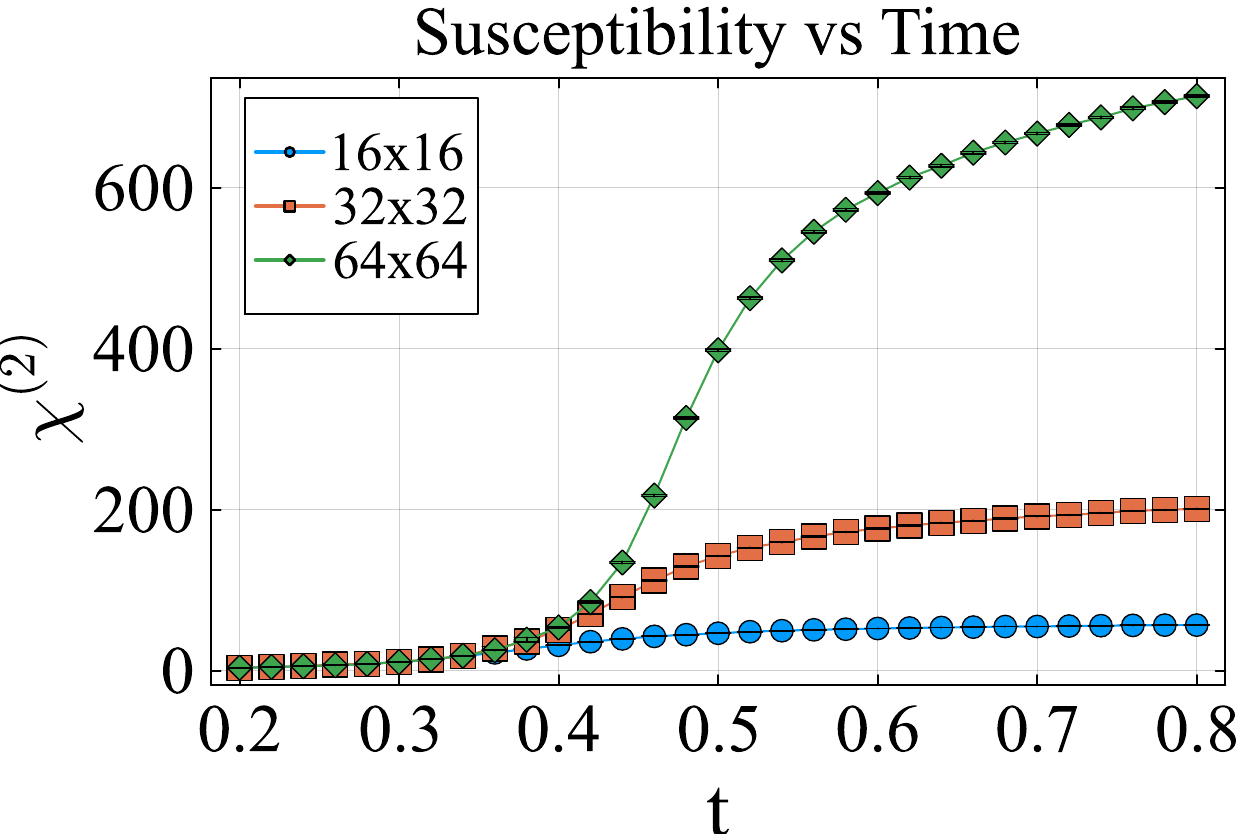}
    \includegraphics[width=0.32\linewidth]{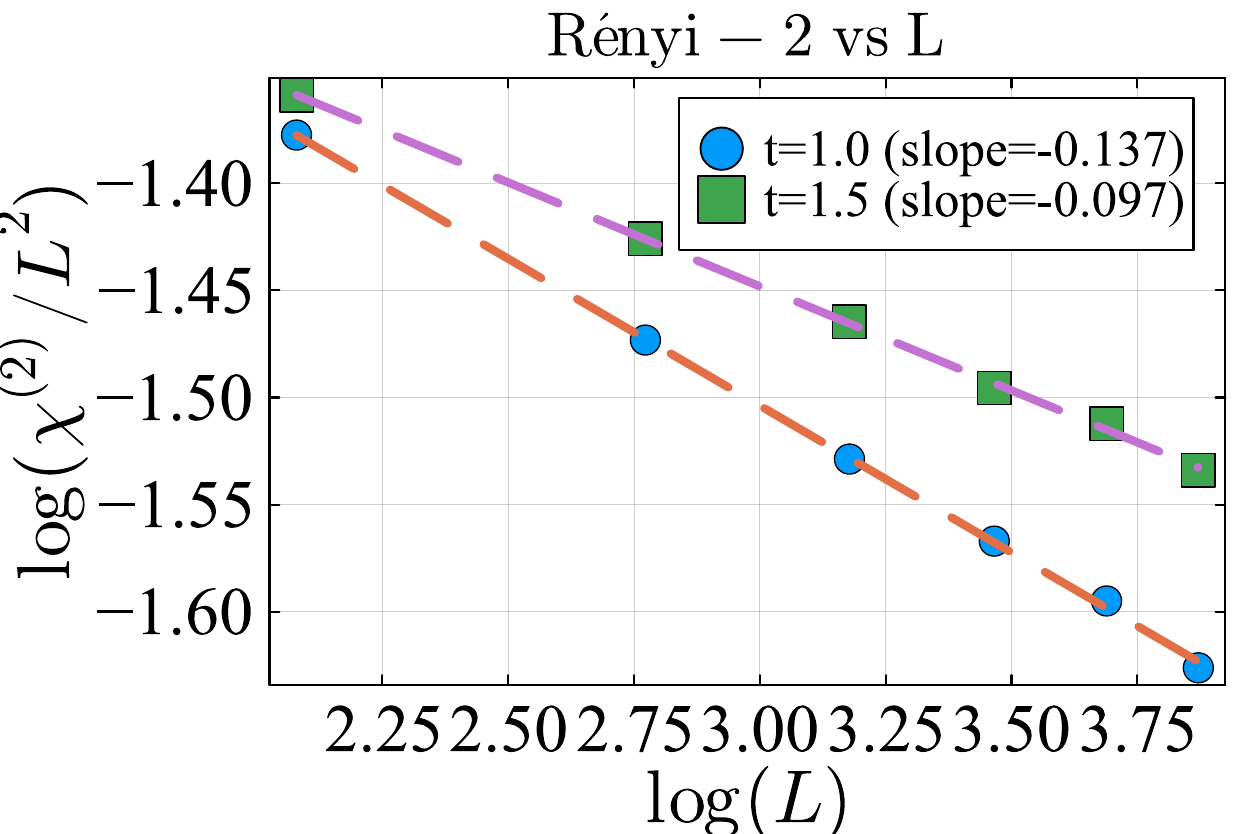}
    \includegraphics[width=0.32\linewidth]{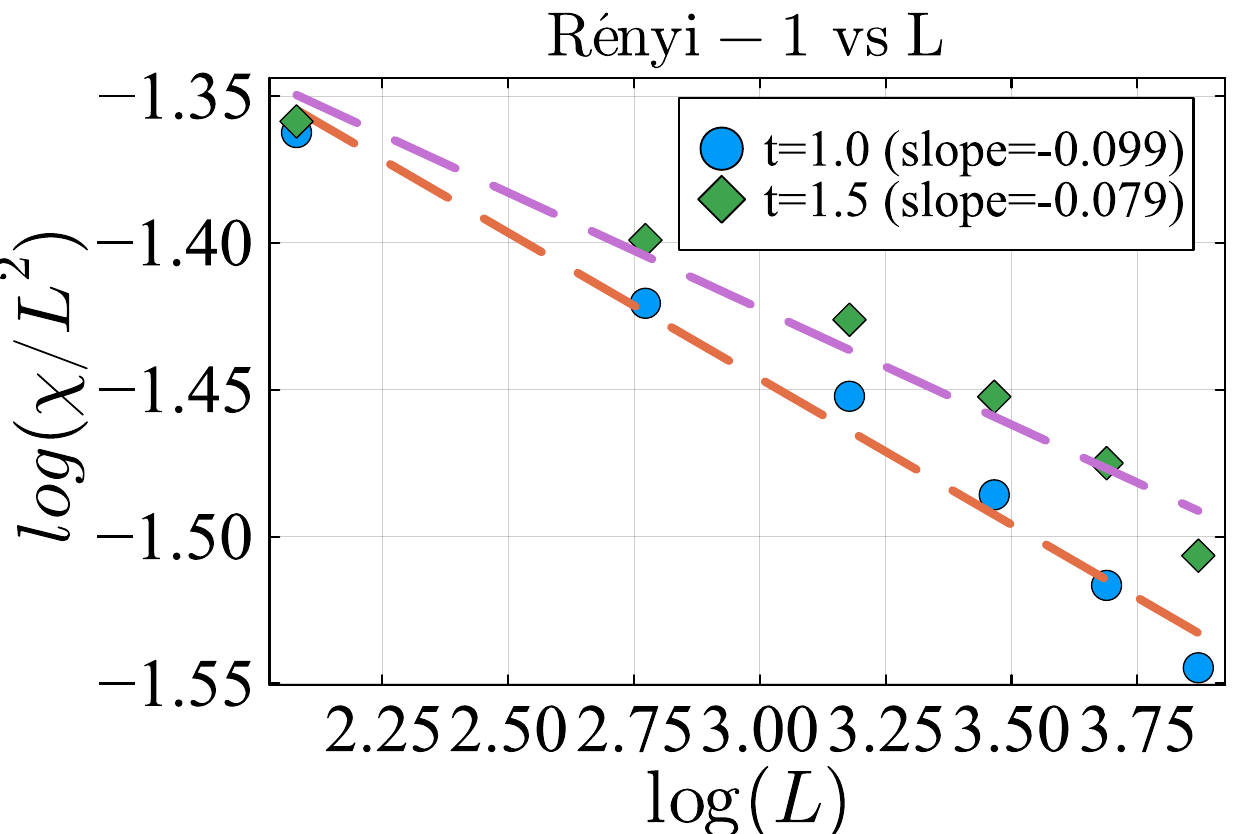}
    
    \caption{R\'{e}nyi susceptibilities defined in Eq.~\eqref{chi} for our spin-1/2 model under Lindblad evolution.} 
    \label{fig:susceptibility}
\end{figure}

\begin{figure}
    \centering
    \includegraphics[width = 0.43\linewidth]{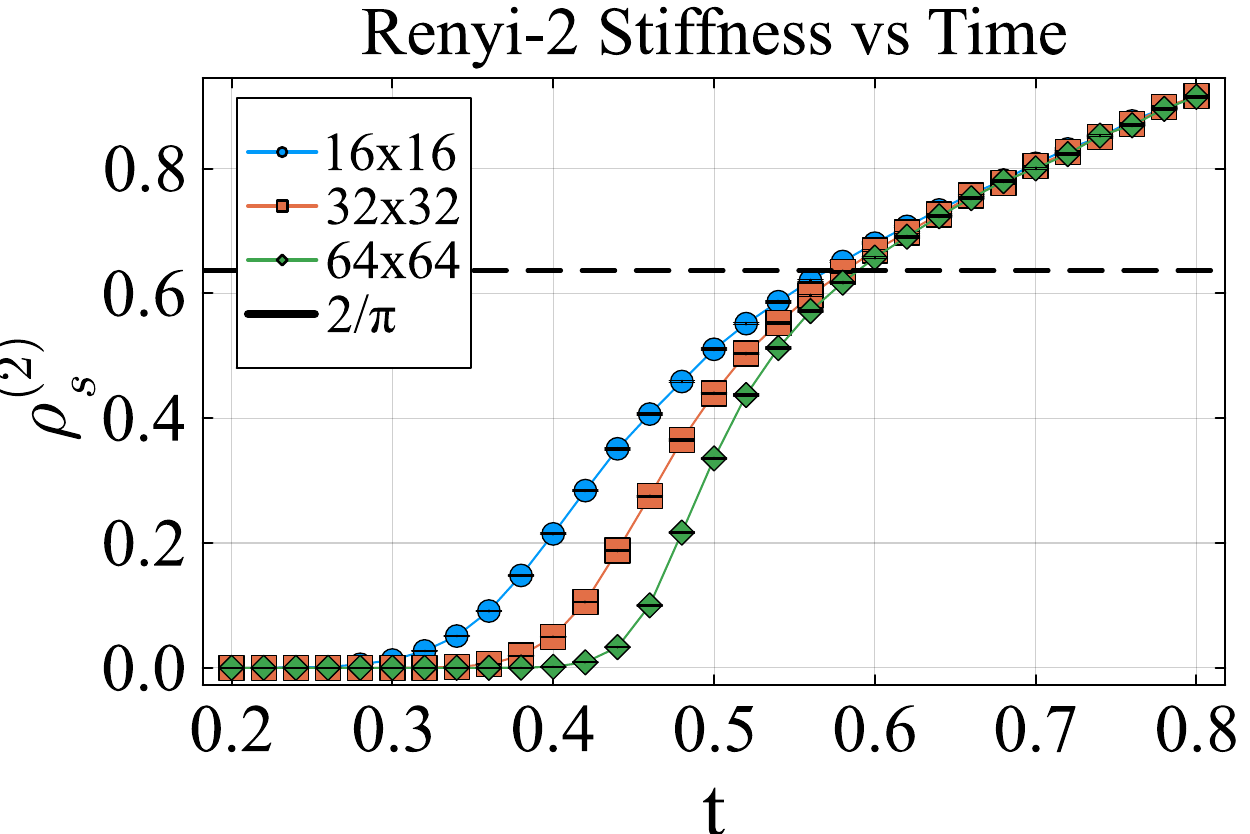}\includegraphics[width=0.43\linewidth]{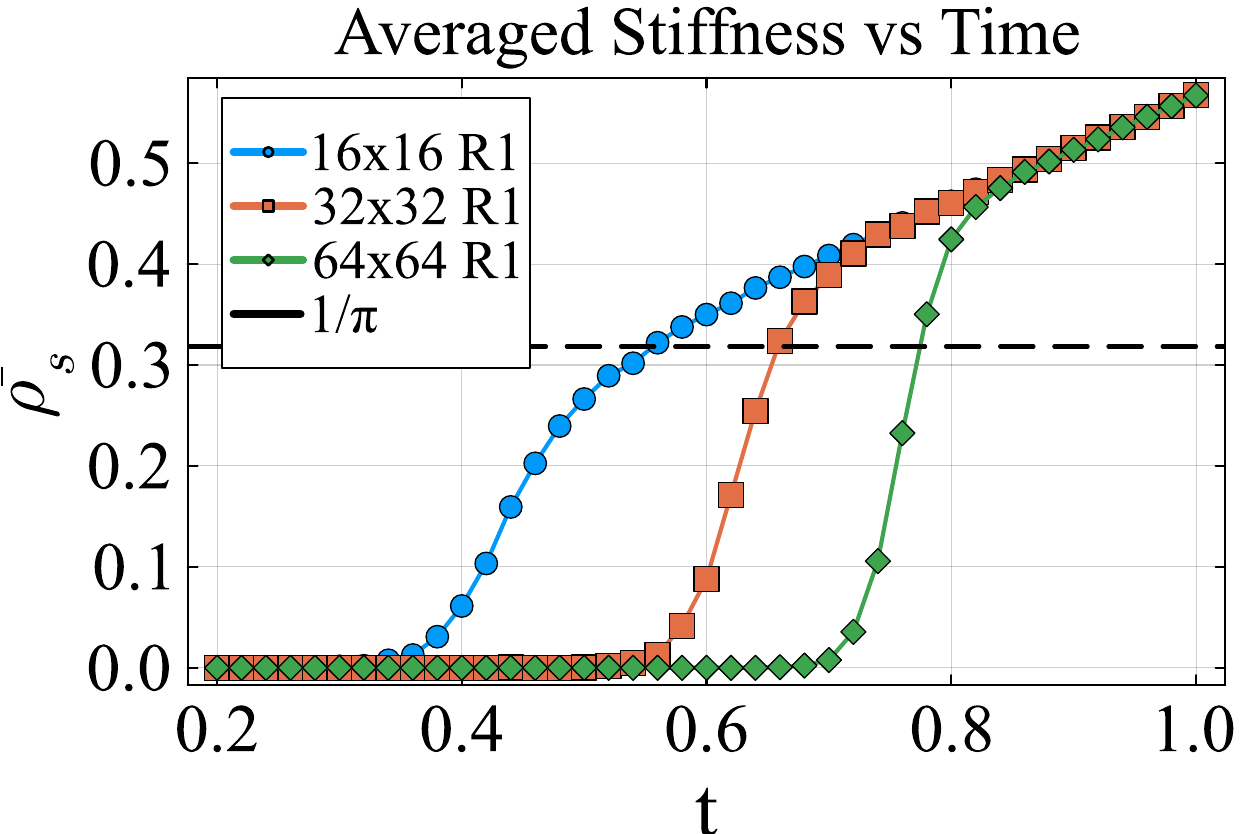}
    \caption{ The R\'enyi-2 stiffness defined in Eq.~\eqref{renyi2stiff} (left) and the ``disorder-averaged" stiffness defined in Eq.~\eqref{disorderstiff} (right). The R\'enyi-2 stiffness behaves consistently with our theoretical expectation, with a jump at the transition with amplitude $2/\pi$. The disorder-averaged stiffness exhibits a stronger finite-size effect, which we attribute to the modified RG equation to be presented in the next section. }
    \label{fig:stiffness}
\end{figure}

Another sharp diagnosis for the BKT-like transition is the jump of the dimensionless stiffness. The stiffness can be defined as the winding number fluctuation 
in the space-time path integral. The winding number can be accessed more concretely in Stochastic Series Expansion, which samples the operator strings \begin{equation}
    \frac{t^n}{n!}\langle \rho_0| \cL_1 \cdots \cL_n |\rho_0\rangle,\label{eq:opstring}
    \end{equation}
obtained from Taylor expansion of $e^{t \cL}$. Here $\cL_{i}$ labels a local term in the Lindbladian. The weight of the operator string sampled is determined both by the matrix element in Eq.~\eqref{eq:opstring} and the factor $\frac{t^n}{n!}$. This can be viewed as a way to sample configurations in spacetime. For each configuration consisting of a list of operators, the winding number in $\mu$ direction is defined as~\cite{Sandvik_2010}
\begin{equation}
    W_\mu = \frac{N_+ - N_-}{L},
\end{equation}
where $N_+$ denotes the amount of $S_{i+\mu}^+S_i^-$ in the operator string (\ref{eq:opstring}) and $N_-$ denotes the amount of $S_i^+S_{i+\mu}^-$ in the operator string. By averaging over the configurations sampled, we can obtain the stiffness from the statistics of the winding number. Using other discretized representations of path integrals (e.g. the one used in continuous time worm algorithm) would give equivalent results. 

We define two different measures of the stiffness depending on the final boundary condition. The R\'enyi-2 stiffness is defined as
\begin{equation}
    \rho^{(2)}_s = \langle W^2\rangle_{R2}\label{renyi2stiff}
\end{equation}
with the initial configuration evolved for time $2\,t$ and the boundary conditions at time $0$ and $2\,t$ fixed to be the N\'{e}el state --- the expectation with the R\'enyi-2 boundary condition is denoted as $\langle\,\cdot\,\rangle_{R2}$.
This is in slight contrast to the equilibrium stiffness which imposes periodic boundary conditions in imaginary time. In the equilibrium setting, this quantity is known to exhibit a universal jump at the transition, and therefore serves as a clear diagnostic for BKT-like physics. 

Similarly, we define a ``disorder-averaged" stiffness with free final boundary condition, evolving the state for time $t$ and averaging over the final spin configuration $s$ at time $t$:
\begin{equation}
    \bar{\rho}_s = \sum_s P(s) \left(\langle W^2\rangle_s - \langle W\rangle_s^2 \right).\label{disorderstiff}
\end{equation}
The average for fixed disorder configuration $s$ is denoted $\langle\, \cdot\,\rangle_s$. The procedure to evaluate this quantity is again to sample $s$, and then to use another QMC sampling to estimate the winding number fluctuation. 
We display results for both the R\'enyi-2 and disorder-averaged stiffness in Fig.~\ref{fig:stiffness}. We observe a clear jump in the R\'enyi-2 stiffness with magnitude $2/\pi$ as $t$ is increased. The disorder-averaged stiffness is less conclusive: there is a strong finite size effect and the apparent transition point $t_c$ drifts to larger values as the system size increases.
Our finding for the R\'enyi-2 observables are consistent with rotor model calculations carried out in the next section. The strong finite-size effect of the disorder-averaged stiffness is also suggested within the RG analysis in the replica limit discussed in the next section, as we find that an extra term in the RG equation would lead to even longer correlation length near the critical point, compared with the BKT transition. 

\section{Decohered rotor model}\label{sec:rotor}

In this section we turn to our second model which consists of a lattice of $U(1)$ quantum rotors, with number operators $\hat{n}_j$ and phases $\hat{\phi}_j$ which satisfy commutation relations,
\begin{equation}
    [\hat{n}_i, e^{ \ii \hat{\phi}_j}] = \delta_{ij} e^{ \ii \hat{\phi}_i}.
\end{equation}
We will again explore Lindbladian dynamics with an initial state of the evolution to be $|0\rangle = \prod_{i}|\hat{n}_i = 0\rangle$, and the jump operator $L_{ij} = e^{\ii \hat{\phi}_i} e^{ - \ii \hat{\phi}_j}$. In the absence of a coherent Hamiltonian (which we shall explore in the following Section) the master equation is $ \partial_t \h{\rho} = \frac{1}{2}\sum_{i, j} \h{L}_{ij} \h{\rho} \h{L}_{ij}^\dagger - \hat{\rho} $, here the sum $i,j$ includes both orientations. At time $t$ the density matrix can be represented as, 
\beqn \hat{\rho}(t) \sim \int D\phi_L D\phi_R \ e^{ \sum_{ij} t \left( \cos(\phi_{L,i} - \phi_{R,i} - \phi_{L,j} + \phi_{R,j} ) - 1 \right) } |\vect{\phi}_L \rangle \langle \vect{\phi}_R |. \label{O2}\eeqn

We first note that just like the spin-1/2 models studied in the previous sections, the system has a ``weak-local-$U(1)$'' symmetry $\phi_{L,i} \ra \phi_{L,i} + \alpha_i$, $\phi_{R,i} \ra \phi_{R,i} + \alpha_i$, therefore the field $\phi \sim \phi_{L,i} + \phi_{R,i}$ is a redundant ``gauge mode" and can be integrated out. In the next section we will see that $\phi$ corresponds to the ``classical field", while $\tphi = \phi_L - \phi_R$ is the ``quantum field". 
A more generic channel or Lindbladian would break the weak-local-$U(1)$ symmetry, but we later will argue that breaking this symmetry will not lead to qualitative change of the discussion, at least at the perturbative level. 

Due to the weak-local-$U(1)$ symmetry, the density matrix is always diagonal in the occupational basis $| \mm \rangle =  \prod_i | \hat{n}_i = m_i \rangle$, and the probability of finding configuration $\mm$ at time $t$ is 
\beqn P(\mm)_t = \langle \mm | \hat \rho(t) | \mm \rangle \sim \int D \tphi \ e^{ \sum_{ij} t \cos( \tphi_i - \tphi_j) } e^{ \ii \sum_i m_i \tphi_i } = \langle e^{\ii \sum_i m_i \hat{\tphi}_i} \rangle_t, \label{P}\eeqn where $\tphi = \phi_L - \phi_R$. 

In this section we will often use the Gaussian theory after a spin wave expansion of the rotor model, i.e.~$
\cos (\tilde{\phi}_i  - \tilde{\phi}_j ) \approx 1 - \frac{1}{2} (\tilde{\phi}_i  - \tilde{\phi}_j)^2$, as an approximation of Eq.~\eqref{O2} at large-$t$. The Gaussian field theory in d-dim space reads: \beqn S = \int d^dx \ \frac{\td}{2} ( \partial_\mu \phi)^2. \label{S} \eeqn $\td$ is a parameter in the continuum field theory, we expect it to be largely proportional to $t$ albeit renormalization. In this Gaussian field theory we are interested in the correlation function $C(\vect{x}) = \langle e^{-\ii \phi_0} \ e^{\ii \phi_{\vect{x}}} \rangle $, whose behaviors in 1d and 2d are evaluated as \beqn {\rm d = 1} &:& C(x) \sim e^{- |x|/\xi}, \hspace{.5cm} \xi = 2 \, \td; \cr\cr {\rm d = 2} &:& C( \vect{x}) \sim \left( \frac{1}{|\vect{x}|} \right)^{1/\left( 2 \pi \td \right) }. \eeqn

Our results are summarized as follows: 

\begin{itemize}
    
    \item  For the 1d decohered rotor model, though the system remains strongly-symmetric at any finite $t$, the correlation length of the R\'enyi-1 and R\'enyi-2 correlators grows linearly in time, $\xi \propto t$. Moreover, they satisfy a universal relation at large $t$: $\xi^{(1)}/\xi^{(2)} = 2$, which is confirmed numerically in the previous section. 
    
    \item For the 2d decohered rotor model, there is a phase transition between the strongly symmetric phase and a quasi-long-range SW-SSB phase. We perform two calculations for these correlators, with and without using the replica formalism, and they lead to the same conclusion. We also perform the RG calculation to demonstrate that the phase transition belongs to a modified BKT-like universality class, with an unusual scaling at the critical point.

\end{itemize}

\subsection{1d rotor model}\label{sec:rotor:1d}

We first discuss the $U(1)$ rotor model in 1d. This model should not have a $U(1)$-SW-SSB phase, not even a quasi-long-range order except for $t \ra \infty$. However, we will show that with large $t$, this model still exhibits universal behaviors: the correlation length of the R\'enyi-1 and R\'enyi-2 correlators satisfy a simple relation $ \xi^{(1)} = 2 \, \xi^{(2)}$, as was reported in our numerical study of the $s=1/2$ model.

\subsubsection{R\'enyi-2 correlator}

We first evaluate the R\'enyi-2 correlator of the 1d rotor model: \beqn C^{(2)}(x)_t = \int D\phi_L D\phi_R \ e^{ \sum_{ij} 2 t \left( \cos(\phi_{L,i} - \phi_{R,i} - \phi_{L,j} + \phi_{R,j} ) - 1 \right) }  \ e^{- \ii (\phi_{L,0} - \phi_{R,0}) + \ii (\phi_{L,x} - \phi_{R,x}) }. \eeqn As we discussed already, the system has a weak-local-$U(1)$ symmetry and the field $\phi = \phi_L + \phi_R$ is redundant. After integrating out $\phi$, the R\'enyi-2 correlator becomes \beqn C^{(2)}(x)_t \sim \int D\tphi \ e^{ \sum_{ij} 2 t \left( \cos(\tphi_i - \tphi_j ) - 1 \right) }  \ e^{- \ii \tphi_0 + \ii \tphi_x }, \label{1dRenyi2} \eeqn

In 1d we focus on the regime with large $t$, and perform a spin-wave expansion of Eq.~\eqref{1dRenyi2}. The R\'enyi-2 correlator reduces to the correlation of the Gaussian theory Eq.~\eqref{S}, with $\td$ replaced by $2 \, \td$. Therefore the R\'enyi-2 correlator decays exponentially as \beqn {\rm d = 1} &:& C^{(2)}(x) \sim e^{- |x|/\xi^{(2)}}, \hspace{.5cm} \xi^{(2)} = 4 \, \td. \eeqn

The Gaussian theory after spin-wave expansion is a very convenient way of evaluating the R\'{e}nyi-2 correlator, but here the spin-wave expansion is not necessary. One can take the standard Villain form analysis of the rotor model, and the R\'enyi-2 correlator becomes \beqn C^{(2)}(x) \sim \sum_{E_{\bar{i}}} \delta_{\nabla E, \rho} \exp\left( \sum_{\bar{i}} - \frac{1}{4t} E_{\bar{i}}^2 \right). \eeqn Here $E_{\bar{i}}$ can be viewed as the discrete electric field on the links $\bar{i}$, and it is constrained by the Gauss law $\nabla E = \bd_{0,x}$, where $\bd_{0,x} = \{ (-1)_0; (+1)_x \}$ is a configuration of dipole with $\pm 1$ charges residing at $0$ and $x$. Therefore $E_{\bar{i}}$ is a step function, and $C^{(2)}(x)$ reads \beqn C^{(2)}(x) \sim \sum_{E} \exp\left( - \frac{L - x}{4t} E^2 - \frac{x}{4t} (E+1)^2 \right) \sim e^{-\frac{x}{4t}}\theta_3 \left( \frac{ \ii x}{4t},\frac{\ii L}{4\pi t} \right), \eeqn where $L$ is the length of the chain, and $\theta_3$ is the third Jacobi theta function. In the limit of large $L$, the sum of $E$ is evidently dominated by $E = 0$, and the R\'{e}nyi-2 correlator reduces to the same form as the Gaussian theory.

\subsubsection{R\'enyi-1 correlator}

To evaluate the R\'enyi-1 correlator at large $t$, we first note that the probability of finding configuration $\mm = ( \cdots m_i \cdots)$ at time $t$ is given by \beqn P(\mm)_t &=& \langle \prod_{i} e^{\ii m_i \hat{\tilde{\phi}}_i }\rangle_t \sim \delta_{\sum_i m_i, 0} \exp\left( - \cE_1[ \mm ] \right) , \hspace{1cm} \cE_1[ \mm ] = - \frac{1}{2 \td} \sum_{j< k} m_j \, m_k \, |x_j-x_k|, \label{1dcoulomb} \eeqn 
where we have used the spin-wave approximation.  Here, $\cE_1[ \mm ]$ is the energy of 1d Coulomb gas with charge configuration $\mm$, which interact with each other through a linear Coulomb potential energy.

The R\'enyi-1 correlator is defined as: \beqn C^{(1)}(x)_t = \Tr \left( \sqrt{\rho(t)} \ e^{- \ii \hphi_0 + \ii \hphi_x} \sqrt{\rho(t)} \ e^{\ii \hphi_0 - \ii \hphi_x} \right). \eeqn Since the density matrix is always diagonal in the occupational basis, the R\'enyi-1 correlator can also be written as \beqn C^{(1)}(x,t) = \sum_{\mm} \sqrt{ P(\mm) P(\mm') }, \eeqn 
where $\mm'$ is the charge configuration $\mm$ after moving one particle from $0$ to $x$: $ \mm' = \mm + \bd_{0,x}$, with dipole $\bd_{0,x}$. Using Eq.~\eqref{1dcoulomb}, the R\'enyi-1 correlator can be written as: \beqn C^{(1)}(x)_t &=& \sum_{\mm } \sqrt{ \langle \prod_i e^{\ii m_i \hphi_i} \rangle \langle \prod_i e^{\ii m_i \hphi_i} \ e^{- \ii \hphi_0 + \ii \hphi_x } \rangle } \cr\cr &=& \sum_{\mm} \sqrt{ \exp(- \cE_1[ \mm ] ) 
\exp(- \cE_1 [ \mm + \bd_{0,x} ] ) }  \cr\cr &=& \sum_{\mm} 
\exp( - \cE_1 [ \mm + \bd_{0,x} ] ) \times 
\exp\left( - \cE_1 [ \, \frac{1}{2} \bd_{0,x} ] \right) \cr\cr &=& A(x)_t \times B(x)_t. 
\eeqn 
The two factors in the last equations above are: \beqn && A(x)_t = \sum_{\{m\}} \exp\left( \frac{1}{2 \td}\sum_{j<k} m^A_j \, m^A_k \, |x_j-x_k| \right), \hspace{1.2cm} B(x)_t \sim e^{- |x|/\xi^{(1)}}, \ \ \xi^{(1)} =  8 \, \td. \label{AB1d} \eeqn Here $\mm^A$ is a charge configuration that is obtained from configuration $\mm$ by inserting ``test-charges" $\pm 1/2$ at position $0$ and $x$: \beqn \mm^A =  \mm + \frac{1}{2} \bd_{0,x}. \eeqn

The factor $A(x)_t$ is the partition function of 1d Coulomb gas, with extra test charges. We will show in Appendix \ref{app:rotor} that the decay length of $A(x)_t$ increases very rapidly with $\tilde{t}$, so the R\'enyi-1 correlator is dominated by $B(x)_t$. The physical picture is that the test charges of a 1d Coulomb gas are screened, therefore the partition function (and $A(x)_t$) does not vary strongly with $x$. Then the dominant factor of $C^{(1)}(x)_t$ is $B(x)_t$, and it decays exponentially with correlation length $\xi^{(1)} = 8 \, \td$, which is twice of that of $\xi^{(2)}$, i.e. $\xi^{(1)} = 2 \, \xi^{(2)}$, exactly what was observed numerically for our $s=1/2$ model.

\subsection{2d rotor model}\label{sec:rotor:2d}

Under the Lindbladian evolution Eq.~\eqref{O2}, the 2d rotor model will have a quasi-long-range $U(1)$-SW-SSB at large $t$. We will perform a renormalization group analysis later to discuss the phase transition, but let us assume that we are already in the quasi-long-range SW-SSB phase, and evaluate the R\'enyi-1 and R\'enyi-2 correlators. We still first ignore the compactness of $\tilde{\phi}$, and take a spin-wave expansion. When evaluating the R\'enyi-2 correlator, the effective theory is given by Eq.~\eqref{S}, with $\td$ replaced by $2\,\td$. Therefore the R\'enyi-2 correlator in the SW-SSB phase reads \beqn C^{(2)}(\bm x)_t \sim \frac{1}{|\bm x|^{1/(4 \pi \td)}}, \eeqn and at the critical point $\td_c = 1/\pi$ predicted through an RG calculation (to be presented, below), the R\'enyi-2 correlator has a universal scaling $C^{(2)}(\bm x)_t \sim 1/|\bm x|^{1/4}$.

\subsubsection{R\'enyi-1 correlator: direct calculation}

Now let us evaluate the R\'enyi-1 correlator. In the large-$t$ SW-SSB phase, the formalism we used for the 1d rotor model still applies, 
provided we replace the probability of finding configuration $\mm$ at time $t$ by 
\beqn P(\mm)_t &=& \langle \prod_{i} e^{\ii m_i \hphi_i }\rangle \sim \delta_{\sum_i m_i, 0} \exp\left( - \cE_2[ \mm ] \right) , \hspace{1cm} \cE_2[ \mm ] = - \frac{1}{2 \pi \td} \sum_{j< k} m_j \, m_k \, \ln|x_j-x_k|. \label{2dcoulomb} \eeqn Here, $\cE_2[ \mm ]$ is the energy of 2d Coulomb gas with charge configuration $\mm$, which interact with each other through a logarithmic Coulomb potential energy.
We can then write,
\beqn C^{(1)}(\vect{x})_t &=& \sum_{\mm} \sqrt{ P(\mm) P(\mm') } \cr\cr &=& \sum_{\mm } \exp( - \cE_2 [\mm + \frac{1}{2} \bd_{0,x} ] ) \times \exp\left( - \cE_2 [ \, \frac{1}{2} \bd_{0,x} ] \right) \cr\cr &=& A(\vect{x})_t \times B( \vect{x})_t. \eeqn 

The factor $B( \vect{x})_t
$ gives a power-law decay, 
\beqn B( \vect{x})_t = \exp\left( - \cE_2 [ \, \frac{1}{2} \bd_{0,x} ] \right) = \left( \frac{1}{|\vect{x}|} \right)^{1/(8\pi \td)}, \label{B2d} \eeqn
while the factor $A(\vect{x})_t$ can be evaluated as \beqn A(\vect{x})_t \sim \int D\chi (\vect{x}) \exp\left( - \int d^2x \ \frac{\td}{2} ( \partial_\mu \chi)^2 - \lambda \cos(\chi) \right) e^{ - \frac{\ii}{2} \chi_0 + \frac{\ii}{2} \chi_x }. \label{2dA}\eeqn Here the sum over $\mm$ will be recovered once we expand in powers of the fugacity $\lambda$. Since the sum over $\mm$ has no bound, $\lambda \ra \infty$. 
With infinite $\lambda$, the fluctuation of $\chi$ in Eq.~\eqref{2dA} is pinned completely, and $A(\vect{x})_t$ does not decay with $\vect{x}$. 

Thus in the Gaussian phase for $t > t_c$ , the long distance behavior of the R\'enyi-1 correlator is given by $B( \vect{x} )_t$ only, and it decays as \beqn C^{(1)}( \vect{x})_t \sim \left( \frac{1}{| \vect{x}|} \right)^{1/(8\pi \td)}. \label{2dRenyi1} \eeqn We can see that the relation $ C^{(2)}( \vect{x})_t \sim \left( C^{(1)}( \vect{x})_t \right)^2 $ still holds in 2d, or in other words, inside a Gaussian phase (the phase with $U(1)$ SW-SSB), the scaling dimension of the R\'{e}nyi-1 and R\'{e}nyi-2 correlator satisfy $ \Delta^{(2)} = 2 \, \Delta^{(1)} $.  At the critical point $\td_c = 1/\pi$, as predicted through the RG calculation to be presented below, the R\'enyi-1 correlator has scaling $C^{(1)}(x)_t \sim 1/|x|^{1/8}$.

\subsubsection{R\'enyi-1 correlator: replica formula}

In this subsection we reevaluate the R\'enyi-1 correlator of the 2d rotor model with the replica formula.
We first generalize the R\'enyi-1 correlator to the following R\'enyi-$Q$ correlator: \beqn C^{(Q)}( \vect{x})_t &=& \Tr \left( \rho(t)^{Q/2} \ e^{- \ii \hphi_0 + \ii \hphi_\vect{x}} \rho(t)^{Q/2} \ e^{\ii \hphi_0 - \ii \hphi_\vect{x}} \right). \eeqn This R\'enyi-Q correlator quantity can be conveniently evaluated with even integer $Q$. We will eventually take the replica limit $Q \ra 1$.

The generating theory for the R\'enyi-$Q$ correlator is the following partition function: \beqn Z^{(Q)}_t &=& \frac{1}{({\cal N}_t)^Q} \int \prod_{a = 1}^Q D \tilde{\phi}_a \ \exp\left( - \Big[ \sum_{a = 1}^Q \sum_{ij} - t \cos(\nabla_{ij} (\tilde{\phi}_a - \tilde{\phi}_{a+1})) \Big] \right), \cr\cr\cr &=& \frac{1}{({\cal N}_t)^Q} \int \prod_{a = 1}^Q D \varphi_a \ \delta\left( \sum_{a = 1}^Q \varphi_a \right) \ \exp\left( - \Big[ \sum_{a = 1}^Q \sum_{ij} - t \cos(\nabla_{ij} \varphi_a)) \Big] \right). \label{ZQ2} \eeqn In the second line above we have defined $\varphi_a = \tilde{\phi}_a - \tilde{\phi}_{a+1}$, and the new fields $\varphi_a$ are subject to a hard constraint $\sum_{a = 1}^Q \varphi_a(\vect{x}) = 0$.

After a spin wave expansion of Eq.~\eqref{ZQ2}, the R\'enyi-$Q$ correlator is now evaluated as: \beqn C^{(Q)}(\vect{x})_t \sim \int \prod_{a = 1}^Q D \varphi_a \ \delta\left( \sum_{a = 1}^Q \varphi_a \right) \ \exp\left( \ii \sum_{a =1}^{Q/2} ( \varphi_{a,0} - \varphi_{a, \vect{x}} ) \right) \exp\left( - \sum_{a = 1}^Q \int d^2x \ \frac{\td}{2} (\partial_\mu \varphi_a)^2 \right). \eeqn Due to the hard constraint $\sum_{a = 1}^Q \varphi_a = 0$, the correlation of the Gaussian fields $\varphi_a$ is \beqn && \langle \varphi_a(0) \varphi_b(\vect{x}) \rangle = - \left( \delta_{ab} - \frac{1}{Q} \right) \frac{1}{2\pi \td} \ln |\vect{x}|, \hspace{1cm} \sum_{a,b = 1}^{Q/2} \langle \varphi_a(0) \varphi_b(\vect{x}) \rangle = - \frac{Q}{4} \frac{1}{2\pi \td} \ln |\vect{x}|.  \eeqn This eventually leads to \beqn C^{(Q)}(x)_t \sim \left( \frac{1}{|\vect{x}|} \right)^{Q/(8 \pi \td)}. \eeqn In the replica limit $Q \ra 1$, it recovers exactly the same result Eq.~\eqref{2dRenyi1} obtained without using the replica formula.

\subsubsection{Renormalization group}

Now let us approach the SW-SSB phase transition from the quasi-long-range $U(1)$-SW-SSB phase. In an ordinary 2d superfluid, the transition is a BKT transition driven by proliferation of vortices, with critical $\td_c = 2/\pi$. The critical $\td_c$ corresponds to the well-known universal jump of the dimensionless stiffness (also called the helicity modulus) at the BKT transition. The replica-$Q$ partition function Eq.~\eqref{ZQ2} are $Q$-copies of Gaussian fields with a constraint $\sum_{a} \varphi_a = 0$. This constraint enforces that the total vorticity of system is zero everywhere, i.e. $ \sum_{a = 1}^Q v_{a, \bar{i}} = 0$. Therefore the minimal vortices are ``replica-dipoles" such as $(+1, -1, 0, \cdots)$. Since a replica-dipole costs twice the energy of a single vortex, the standard ``energy-entropy" argument concludes that the transition associated with the vortex proliferation should be at $\td_c = 1/\pi$, half of that of a single component superfluid. A similar structure of inter-replica dipoles was observed in field theories describing monitored $U(1)$ quantum~\cite{SharpeningEFT} and classical systems~\cite{NahumJacobsenBayesian,gopalakrishnan2026monitoredfluctuatinghydrodynamics}, where the corresponding ``charge-sharpening'' transition~\cite{ChargeSharpening} can also be associated with SW-SSB~\cite{singh2025mixedstatelearnabilitytransitionsmonitored,vijay2025holographicallyemergentgaugetheory}. Our finite time transition is a 2+0d version of this monitored 1+1d problem, where conditioning occurs only at the final time step. 

For a more systematic RG, let us return to Eq.~\eqref{ZQ2}. The most standard RG for a BKT transition is performed in the dual theory. We first consider the following modified theory: \beqn S = \int d^2x \ \sum_{a} \frac{\td}{2} \,(\partial_\mu \varphi_a)^2 + \frac{m}{2} (\partial_\mu  \sum_a \varphi_a)^2 = \int d^2x \ \frac{K_{ab}}{2} \, \partial_\mu  \varphi_a \partial_\mu  \varphi_b, \label{SQ}\eeqn where $K_{ab} = \tilde{t} \, \delta_{ab} + m$ is a matrix in the replica space. We eventually will need to take $m \ra \infty$, to freeze the mode $\sum_a \varphi_a$, and effectively enforce the constraint $\sum_a \varphi_a = 0$. If we keep $m$ finite, the dual theory of Eq.~\eqref{SQ} is \beqn S_d = \int d^2x \ \left( \frac{K^d_{ab}}{2} \, \partial_\mu  \theta_a \partial_\mu  \theta_b - \sum_a v \cos(2\pi \theta_a)\right), \eeqn here $K^d = K^{-1}$, and $e^{\ii 2\pi \theta_a}$ creates a vortex of boson species $\varphi_a$ .

How do we see that the vortices need to form a replica-dipole? Suppose we evaluate the correlation function of $\cos(2\pi \theta_a)$, i.e. the single vortex operator, we will see that its scaling dimension is \beqn \Delta_a = \pi \, \td + \pi m, \eeqn and when we take large-$m$, the single vortex operator is highly irrelevant. The most relevant vortex configuration turns out to be exactly a replica-dipole: \beqn V_{ab} = e^{\ii 2\pi \theta_a - \ii 2 \pi \theta_b}, \ \ \ \Delta_{ab} = 2\pi \, \td. \eeqn If we only include the most relevant vortices, the dual theory should be \beqn S_d = \int d^2x \ \left( \frac{K^d_{ab}}{2} \, \partial_\mu \theta_a \partial_\mu \theta_b - \sum_{a \neq b} \tilde{v} \cos(2\pi \theta_a - 2\pi \theta_b) \right). \eeqn
Using $\lim_{m \to \infty } K^{d}_{ab} = \hat{\Pi}_{ab}/\tilde{t}$ with $  \hat{\Pi}_{ab} = \delta_{ab} - \frac{1}{Q}$ the projector onto inter-replica modes, we can also write the action directly enforcing the constraint
\begin{equation} 
S = \int d^2 x \left( \frac{1}{2 \tilde{t}} \sum_a  \left(\partial_\mu \hat{\theta}_a \right)^2  -  \sum_{a \neq b} \tilde{v} \cos(2 \pi \hat{\theta}_a - 2\pi \hat{\theta}_b) \right) ,
\end{equation} 
where we have introduced the fields $\hat{\theta}_a = \hat{\Pi}_{ab} \theta_b$, which satisfy $\sum_a \hat{\theta}_a=0$. This form matches the field theory of monitored 1+1d systems with $U(1)$ symmetry discussed in Refs.~\cite{SharpeningEFT,NahumJacobsenBayesian,gopalakrishnan2026monitoredfluctuatinghydrodynamics}.

The RG flow for $\tilde{v}$ near the critical point includes two contributions: \beqn \frac{d \tilde{v}}{d\ln l} = (2 - 2\pi \, \td) \tilde{v} + 2\pi (Q - 2) \tilde{v}^2. \label{RGv} \eeqn The first term is the standard BKT beta function, and it confirms that the critical point for the transition happens at $\td_c = 1/\pi$, rather than $2/\pi$ as is the case for an ordinary rotor model. The second term stems from the OPE between $V_{ab} \times V_{bc} \ra V_{ac}$, where $V_{ab}$ is the vortex dipole, and the factor $Q-2$ comes from choosing $b \neq a,c$ with $a\neq c$.

\begin{figure}
    \centering
\includegraphics[width=0.75\linewidth]{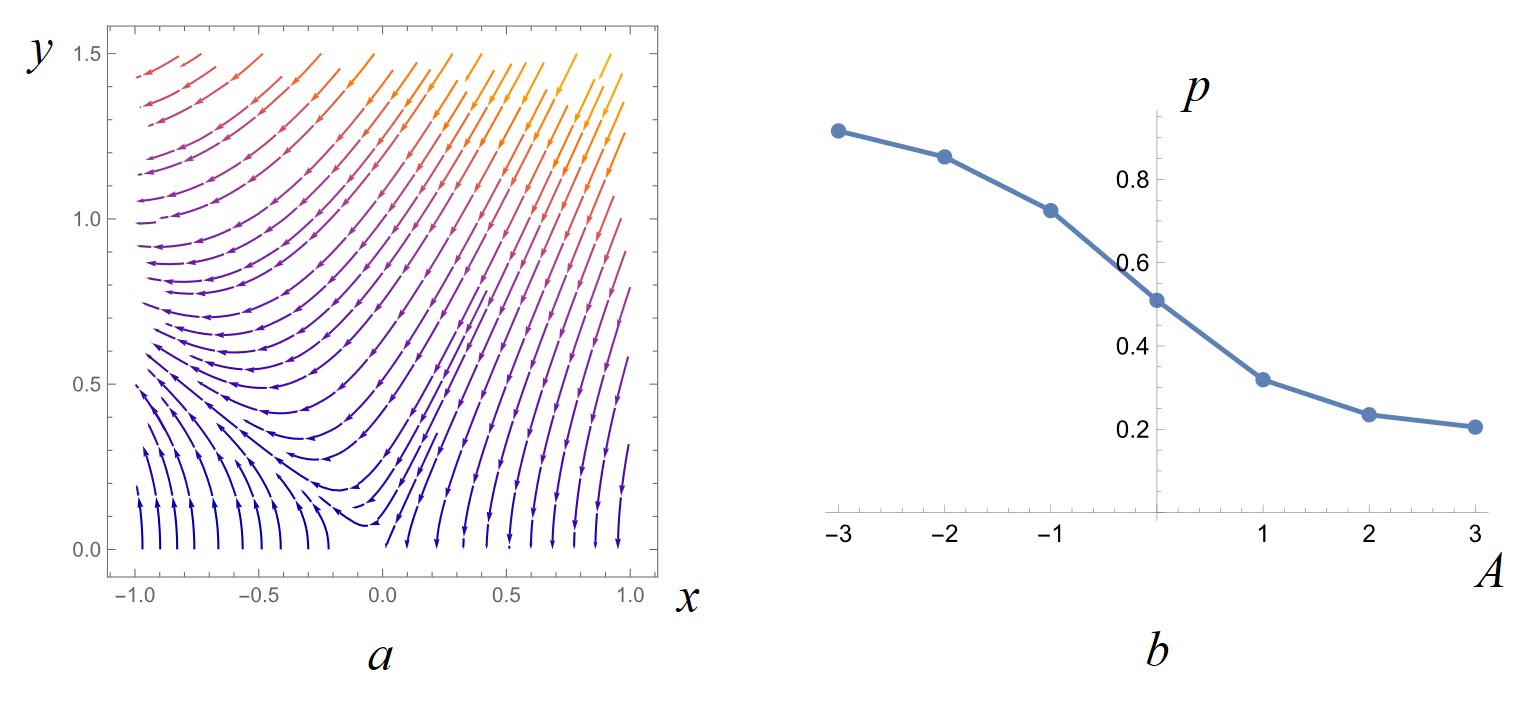}
    \caption{($a$) The RG flow of Eq.~\eqref{RGxy} with $A < 0$. The separatrix and line of fixed points still exist. ($b$) The exponent $p$ defined in Eq.~\eqref{exponent} as a function of $A$ in Eq.~\eqref{RGxy}, which is extracted from numerical solution of the RG equations Eq.~\eqref{RGxy}. }
    \label{RG}
\end{figure}

To derive the RG equation for $\td$, we follow the standard RG steps in the dual theory, and use the OPE: \beqn \cos(2\pi \theta_a - 2\pi\theta_b)_{\vect{x}}\cos(2\pi\theta_a - 2\pi\theta_b)_0 
\sim - \frac{\pi^2}{|\vect{x}|^{4}} \left( x^\mu \partial_\mu ( \theta_a - \theta_b) \right)^2 + \cdots \eeqn Under rescaling $x \ra x \, l$ in real space we generate a new term \beqn \delta S &=& \int d^2x \ \frac{\pi^3}{2} \tilde{v}^2 \ln l \left( \sum_{a,b} (\partial_\mu \theta_a - \partial_\mu \theta_b)^2 \right) = \int d^2 x \ \frac{\pi^3}{2} \tilde{v}^2 \ln l \left( Q \sum_a (\partial_\mu \theta_a)^2 - (\partial_\mu  \sum_a \theta_a)^2 \right). \eeqn
This means that at the new scale $l$, $K^d$ becomes \beqn K^{d}(l) = K^{d}(0) + \delta_{ab} \pi^3 Q \tilde{v}^2 \ln l - \pi^3 \tilde{v}^2 \ln l. \eeqn Then at the new scale, $K(l)$ in the original formalism is $K(l) = K^d(l)^{-1}$. After expanded to the leading order of $\ln l$, $K(l)$ takes the same form as the original $K$, with separately renormalized $\td(l)$ and $m(l)$: \beqn && K(l)_{ab} = (K^d(l))^{-1}_{ab} = \td(l) \, \delta_{ab} + m(l) \cr\cr \ra && \frac{d \, \td}{ d\ln l} = - Q \pi^3 \tilde{v}^2 \td^2, \ \ \ \frac{d \, m }{d\ln l} = \pi^3 \tilde{v}^2 \td^2. \label{RGt}\eeqn Though $m$ will be renormalized under RG, it won't affect the physics, as we eventually take $m$ to infinity to enforce the constraint. Eq.~\eqref{RGv} and Eq.~\eqref{RGt} together constitute the RG flow, and after rescaling $\tilde{v}$ and $\tilde{t}$, the RG equations near the transition point $\td \sim 1/\pi$ reduce to \beqn \frac{d y }{d\ln l} = - s y + A y^2, \quad \frac{d s}{d\ln l} = - y^2, \label{RGxy} \eeqn $A = (Q - 2) \sqrt{\frac{2}{Q}}$. The same form of RG equations were obtained in previous literature~\cite{cardyostlund,NahumJacobsenBayesian}.

At $Q = 2$, the coefficient $A$ vanishes, therefore the R\'{e}nyi-2 quantities should undergo a standard BKT transition. But for other $Q$ these RG equations are expected to lead to different solutions. Like the BKT transition, there is still a line of fixed points at $\tilde{v} \sim y = 0$ and $\td > t_c = 1/\pi$, therefore in the thermodynamic limit we still expect a stiffness jump with the size $1/\pi$. However, the modified RG equations lead to different scaling of the correlation length: \beqn \xi \sim \exp\left( \frac{c}{|\tilde{t} - \td_c|^p} \right), \label{exponent} \eeqn where the exponent $p$ is a continuous function of $A$ (Fig.~\ref{RG}$b$), with $p = 1/2$ for $A = 0$.

The BKT transition is associated with a jump of the dimensionless stiffness. Our theory predicts that the jump of the R\'{e}nyi-2 stiffness defined in Eq.~\eqref{renyi2stiff} should be $2 \, \td_c = 2/\pi$, while the jump of the ``disorder-averaged" stiffness in Eq.~\eqref{disorderstiff} is expected to be $\td_c = 1/\pi$, i.e. the stiffness in the replica limit $Q \ra 1$. Our simulation of the spin-1/2 model in the previous section confirms the jump of the R\'{e}nyi-2 stiffness (Fig.~\ref{fig:stiffness}), but exhibits more complicated behavior of the disorder-averaged stiffness. There is evidently a strong finite-size effect for the disorder averaged stiffness, qualitatively consistent with prediction from the RG equation with the extra $A y^2$ term. 

As we discussed previously, our Lindbladian for rotor model has a weak-local-$U(1)$ symmetry $\phi_{L,i} \ra \phi_{L,i} + \alpha_i$, $\phi_{R,i} \ra \phi_{R,i} + \alpha_i$. A generic channel or Lindbladian should break this local symmetry. For example, one can consider an extra jump operator $L'_{ij} = \cos(\hphi_i - \hphi_j)$, which adds another perturbation to the action Eq.~\eqref{O2}: \beqn \delta S \sim (\cos(\nabla_{ij} \phi_L) - \cos(\nabla_{ij} \phi_R))^2 \sim \sin \left(\frac{1}{2} \nabla_{ij} \tphi \right)^2 \sin \left( \frac{1}{2} \nabla_{ij} \phi \right)^2, \eeqn where $\phi = \phi_L + \phi_R$. When $\delta S$ is weak, $\phi$ remains massive (i.e. there is no spontaneous breaking of the global weak symmetry), and we can safely replace $\sin \left( \frac{1}{2} \nabla_{ij} \phi \right)^2$ by its expectation value, and the role of $\delta S$ is to renormalize $\td$ in our rotor theory. All our results remain qualitatively unchanged. 

\section{Model F rotor dynamics -- emergent classicality}\label{sec:ModelF}

In this section we consider our third model, consisting of Lindblad evolution for the rotor model with open system $U(1)$ dynamics that includes coherent Hamiltonian dynamics, and also targets several classical long-time steady states.  As we shall see, if we start in a pure state with (strong) $U(1)$ symmetry, in 2d and 3d we can describe a finite-time SW-SSB transition where the system transitions between a regime of quantum dynamics at short times to a long-time dynamical regime which is described by  classical Langevin dynamics corresponding to Model F of Halperin-Hohenberg~\cite{RevModPhys.49.435}.

As in Sec.~\ref{sec:rotor}, we will consider a lattice of quantum rotors, with number operators $\hat{n}_j$ and phases $\hat{\phi}_j$, but now with coherent Hamiltonian dynamics,
\begin{equation}
\hat{H}_0 = \frac{K}{2} \sum_j \hat{n}_j^2  - J \sum_{\langle ij\rangle} \cos(\hat{\phi}_i - \hat{\phi}_j).
\end{equation}

We choose two sets of jump operators to target two different classical steady states,
\begin{equation}
\hat{\rho}_\phi = \int \prod_j d \phi_j e^{-\beta F_\phi(\phi)} | \phi \rangle \langle \phi | ; \hskip0.4cm F_\phi = - J \sum_{\langle ij\rangle} \cos(\phi_i - \phi_j)  
\end{equation}
and 
\begin{equation}
\hat{\rho}_n  = \sum_{\{ n_j \}} e^{-\beta F_n(n)} | \{n\}  \rangle \langle \{n \} |; \hskip0.3cm F_n = \frac{K}{2}  \sum_{j} n_j^2 .
\end{equation}

To target $\hat{\rho}_\phi$ we introduce our first set of (on-site) jump operators,
\begin{equation}
\hat{L}_{\phi_j}^a = e^{- \ii a \hat{n}_j} e^{- \frac{\beta}{4}[ F_\phi ( \hat{\phi}_j + a) - F_\phi (\hat{\phi}_j)]} ,
\end{equation}
which shifts the phase $\phi_j$ by a constant ``a", weighted by the change in classical free energy.
The corresponding term in the Lindbladian is,
\begin{equation}
 {\cal L}_\phi[\hat{\rho}] = \gamma_\phi \sum_j \int_a \partial_a^2 \delta(a) [ \hat{L}^a_{\phi_j}\hat{\rho}  \hat{L}^{a \dagger}_{\phi_j} - \frac{1}{2} \{ \hat{L}^{a \dagger}_{\phi_j} \hat{L}^a_{\phi_j} , \hat{\rho}\} ],
 \end{equation}
 and one can show that ${\cal L}_{\phi} [ \hat{\rho}_\phi] = 0$. We remark that $\partial_a^2 \delta(a)$ has three peaks: positive peaks at infinitesimal positive and negative values of $a$, and a negative peak at $a=0$. Since $\hat{L}^{a = 0}_{\phi_j} = \one$, for which the integrand is identically zero, the distribution selects an infinitesimal positive rotation and an infinitesimal negative rotation from the integral.

To target $\hat{\rho}_n$ we choose jump operators acting on links of the form,
\begin{equation}
\hat{L}^\pm_{n_{ij}}  = e^{\pm \ii (\hat{\phi}_i - \hat{\phi}_j)} e^{-\frac{\beta}{4} [ F_n(\hat{n}_i \pm 1,\hat{n}_j \mp 1) - F_n(\hat{n}_i ,\hat{n}_j )]} ,
\end{equation}
which hops a boson weighted by the change in the classical (charging) free energy.
The corresponding contribution to the Lindbladian is,
\begin{equation}
 {\cal L}_n [\hat{\rho}] = \gamma_n \sum_{\langle i, j \rangle} \sum_{\pm}  [ \hat{L}^\pm_{n_{ij}}\hat{\rho}  \hat{L}^{\pm \dagger}_{n_{ij}} - \frac{1}{2} \{ \hat{L}^{\pm \dagger}_{n_{ij}} \hat{L}^\pm_{n_{ij}} , \hat{\rho}\} ],
\end{equation}
which has the steady state,
${\cal L}_n [ \hat{\rho}_n ] =0$.

The full Lindblad equation of motion is taken as,
\begin{equation}
\partial_t \hat{\rho} = - \ii [ \hat{H}_0, \hat{\rho} ] + {\cal L}_\phi [\hat{\rho}] + {\cal L}_n [\hat{\rho}].
\end{equation}
This dynamics has a strong $U(1)$ symmetry since the unitary operator, $\hat{U}_\theta = e^{\ii \theta \sum_j \hat{n}_j}$
commutes with the Hamiltonian as well as both sets of jump operators,
\begin{equation}
[ \hat{U}_\theta , \hat{H}_0] = 0;  \hskip0.3cm [ \hat{U}_\theta , \hat{L}_{\phi_j}^a] = 0;  \hskip0.3cm [ \hat{U}_\theta ,\hat{L}^\pm_{n_{ij}} ] = 0.
\end{equation}

In the limit with $J=K=0$ we have no coherent Hamiltonian dynamics and our two sets of jump operators simplify to $\hat{L}^{\pm}_{n_{ij}} = e^{\pm \ii (\hat{\phi}_i - \hat{\phi}_j)}$, and $\hat{L}^a_{\phi_j} = e^{- \ii a \hat{n}_j}$. The former is the jump operator that we explored in detail in Sec.~\ref{sec:rotor}.  

\subsection{Keldysh path integral for full Lindbladian dynamics}
We now consider a Keldysh type path integral for the dynamics of the density matrix.
First consider the Hamiltonian dynamics,
$\hat{\rho} (t) = e^{- \ii \hat{H}_0t} \hat{\rho}(0) e^{\ii \hat{H}_0 t}$,
and focus on the propagator for the matrix elements of the density matrix; $\rho(\phi,\phi^\prime;t) \equiv \langle \phi | \hat{\rho}(t) | \phi^\prime \rangle$,
\begin{equation}
\rho(\phi_f,\phi_f^\prime ;t) = \int d \phi_0 d \phi_0^\prime {\cal K}(\phi_f,\phi_f^\prime, t;\phi_0 , \phi_0^\prime, 0) \rho(\phi_0,\phi_0^\prime;0),
\end{equation}
with
\begin{equation}
{\cal K}_0(t,0) = \sum_{W_R,W_L} \int \prod_j  D\phi_{Rj}(t) D\phi_{L j}(t) Dn_{Rj}(t) Dn_{L j}(t) e^{S_0(t)},
\end{equation}
where $W_R,W_L$ are integer winding numbers, and the functional integrals are over real fields,
$\phi_{R/L}, n_{R/L}$ running from $[-\infty,\infty]$.  The boundary conditions on the phase-fields at the initial and final times are,
\begin{equation}
\phi_{Rj}(0)= \phi_{0j}; \hskip0.2cm \phi_{Lj}(0)= \phi_{0j}^\prime;
\end{equation}
\begin{equation}
    \phi_{Rj}(t)= \phi_{fj} + 2\pi W_{Rj}; \hskip0.2cm \phi_{Lj}(t)= \phi_{fj}^\prime + 2\pi W_{Lj} .
\end{equation}

The ``action" is given by
\begin{equation}
S_0(t) = \int_0^t d t^\prime [ \sum_j \ii (n_{Rj} \dot{\phi}_{Rj} -  n_{Lj} \dot{\phi}_{Lj}) + \ii H_0(\phi_{R},n_{R}) - \ii H_0(\phi_L,n_L)].
\end{equation}
Next, we add in the Lindbladian dissipation terms. To simplify matters, we assume that $J$ and $K$ are small such that we may approximate to linear order in these parameters. Then, the full action is $S = S_0 + S_\phi +S_n$, where
\begin{equation}
S_\phi = \gamma_\phi \int_t \sum_j [ - \tilde{n}_j^2  - \ii \beta J \tilde{n}_j  \sum_{i \in j} (\sin(\phi_{Ri} - \phi_{Rj}) -
\sin(\phi_{Li} - \phi_{Lj} ))], 
\end{equation}
with $i \in j$ denoting $i$ being a nearest neighbor of $j$, and
\begin{equation}
S_n = \gamma_n \int_t \sum_{\langle ij \rangle} [ 2\cos(\tilde{\phi}_i - \tilde{\phi}_j) - \ii \beta K (n_i-n_j) \sin(\tilde{\phi}_i - \tilde{\phi}_j)].
\end{equation}
Here, we have defined ``classical" and ``quantum" fields,
\begin{equation}
\phi = (\phi_R+\phi_L)/2; \hskip0.3cm \tilde{\phi} = \phi_R - \phi_L,
\end{equation}
and 
\begin{equation}
n = (n_R+ n_L)/2; \hskip0.3cm \tilde{n} = n_R - n_L.
\end{equation}
We also note that 
\begin{equation}
\ii (n_{Rj} \dot{\phi}_{Rj} - n_{Lj} \dot{\phi}_{Lj}) = \ii \tilde{n}_j \dot{\phi}_j + \ii n_j \dot{\tilde{\phi}}_j ,
\end{equation}
so that the quantum number field, $\tilde{n}_j$,
and the classical phase field, $\phi_j$ are canonically conjugate pairs, while likewise for the classical number field, $n_j$ and the quantum phase field, $\tilde{\phi}_j$.

Under the strong $U(1)$ symmetry, $\phi_R \rightarrow \phi_R + \theta$, the quantum field transforms
appropriately,
\begin{equation}
\tilde{\phi} \rightarrow \tilde{\phi} + \theta ,
\end{equation}
and under the weak symmetry, $\phi_{R/L} \rightarrow \phi_{R/L} + \theta$
the classical field transforms, but not the quantum field;
\begin{equation}
\phi \rightarrow \phi + \theta; \hskip0.3cm \tilde{\phi} \rightarrow \tilde{\phi} .
\end{equation}
Thus $e^{ \ii\phi}$ carries weak $U(1)$ charge, while $e^{ \ii \tilde{\phi}}$ carries strong, but not weak
charge.  Long-ranged order in the R\'enyi-1 correlator of the field $e^{\ii \tilde{\phi}}$ signals a breaking of the strong down to the weak symmetry.
If $\langle e^{ \ii \phi } \rangle  \ne 0$ the weak symmetry breaks.  

\subsection{Breaking the strong to weak symmetry and emergence of Model F classicality}

As in Sec.~\ref{sec:rotor} we take our initial state to be pure with $\prod_j | n_j=0 \rangle$.  
In the simpler rotor model studied therein, for $d \ge 2$ we found a finite time transition, $t_c$, where the strong symmetry was spontaneously broken down to the weak symmetry. We anticipate that this will likewise occur in the more complicated model studied in this Section.   
In that case, for $t>t_c$ we can perform a spin-wave expansion for the quantum field, $\tilde{\phi}$, expanding the  
cosine and sine potentials for small
spatial gradients of $\tilde{\phi}_j$,
i.e.,
\begin{equation}
\cos(\tilde{\phi}_i  - \tilde{\phi}_j) \approx 1 - \frac{1}{2} (\tilde{\phi}_i  - \tilde{\phi}_j)^2 ,
\end{equation}
and 
\begin{equation}
\sin(\tilde{\phi}_i  - \tilde{\phi}_j) \approx (\tilde{\phi}_i  - \tilde{\phi}_j).
\end{equation}
In $d\geq 2$ we are thus precluding vortices in the quantum phase field, which correspond, for example, to a vortex in $\phi_R$ but not $\phi_L$. In 1d at finite-time, the presence of quantum phase slips, say in $\phi_R$ but not $\phi_L$, invalidates the spin-wave expansion.

After making the spin-wave expansion we can introduce Hubbard-Stratanovich noise fields, $\eta(t), \xi(t)$
to decouple the terms $\delta S_\phi  = - \gamma_\phi \int_t \sum_j \tilde{n}_j^2$ and 
$\delta S_n = - \gamma_n \int_t \sum_{\langle i, j \rangle} (\tilde{\phi}_i - \tilde{\phi}_j)^2$.
The full action then becomes,
\begin{equation}\label{eq:model_F_final_action}
S = \int_t \sum_j i \tilde{n}_j [\dot{\phi}_j - \partial F/\partial n_j  + \beta \gamma_\phi \partial F/\partial \phi_j - \eta_j]
+ \int_t \sum_j i \tilde{\phi}_j  [ \dot{n}_j + \partial F/\partial \phi_j - \beta \gamma_n \nabla_j^2 \partial F/\partial n_j - \xi_j] + S_\eta + S_\xi,
\end{equation} 
where $S_\eta$ and $S_\xi$ are quadratic in $\eta$ and $\xi$.
Upon integration over $\tilde{n}_j$ and $\tilde{\phi}_j$ we thereby arrive at 
two coupled classical Langevin equations,
\begin{equation}
\dot{\phi}_j = \partial F/\partial n_j  - \beta \gamma_\phi \partial F/\partial \phi_j + \eta_j,
\end{equation}
\begin{equation}
\dot{n}_j = - \partial F/\partial \phi_j + \beta \gamma_n \nabla_j^2 \partial F/\partial n_j + \xi_j .
\end{equation}
Here, the classical free energy is,
\begin{equation}
F = \frac{1}{2} K \sum_j n_j^2 - J \sum_{\langle i, j \rangle} \cos(\phi_i - \phi_j) ,
\end{equation}
which is of the identical form as the original Hamiltonian, except here both $\phi$ and $n$ are classical fields.  
The noise terms have correlators,
\begin{equation}
\langle \eta_i(t) \eta_j(t^\prime) \rangle = 2 \gamma_\phi \delta_{ij} \delta(t-t^\prime),
\end{equation}
\begin{equation}
\langle \xi_i(t) \xi_j(t^\prime) \rangle = 2 \gamma_n \nabla^2_i \delta_{ij} \delta(t-t^\prime).
\end{equation}
This is a lattice ``hard-spin" (rotor) model corresponding to Halperin-Hohenberg Model F dynamics \cite{RevModPhys.49.435}.
The classical dynamical steady state satisfies,
${\cal P}_{eq} [\phi,n] \propto e^{-\beta F(\phi,n)}$.

It is worth emphasizing that the field $n_j$ in the classical hydrodynamical description takes on all real values - that is, it is not quantized.
This ``unquantization" necessarily occurs when
there is a SW-SSB transition (as there is in 2d and 3d), since under the spin-wave approximation in the quantum phase field, $\tilde{\phi}$, the conjugate classical density field, $n$, becomes continuous. 

In the superfluid phase, with $\langle e^{ \ii\phi} \rangle \ne 0$, we can perform a spin-wave expansion
in the classical phase field.
After taking the spatial continuum limit, we end up with a damped Goldstone hydrodynamic mode, with dispersion,
\begin{equation}
\omega_k = \pm  \sqrt{JK} k + \ii (\gamma_\phi + \gamma_n) k^2 + \cdots  .
\end{equation}
In the normal phase we can drop the boson hopping terms, $J \rightarrow 0$ and $\gamma_\phi \rightarrow 0$,
whence, after taking the spatial continuum limit, we end up with a noisy diffusion equation,
\begin{equation}
\dot{n} = D \nabla^2 n + \xi(x,t); \hskip0.4cm  D = \beta K \gamma_n ,
\end{equation}
as well as a spatially local phase temporal-diffusion equation of motion,
\begin{equation}
\dot{\phi} = K n(x) + \eta(x,t) .
\end{equation}

In contrast to our spin-$1/2$ model where classical diffusion sets in for earlier times before SW-SSB, here we might anticipate non-trivial quantum dynamics for $t< t_c$, with completely classical diffusive dynamics only setting in for $t>t_c$. Moreover, 
if we change the Lindbladian
dynamics to explicitly break the strong $U(1)$ symmetry, the density relaxes quickly
(as does the order parameter)
and we also will not have a propagating hydrodynamic Goldstone mode in the superfluid,
while the phase dynamics will have a diffusive mode.  

\subsection{Wigner function and steady state density matrix}\label{sec:ModelF:Wigner}

As noted above, in the long-time steady state the probability distribution function over the classical fields $\phi$ and $n$ approaches an equilibrium distribution,
${\cal P}_{eq} \propto e^{-\beta F(\phi,n)}$. As we show in Appendix~\ref{app:wigner}, this distribution function is proportional to the Wigner distribution, $W(\phi,n)$, associated with the long-time fully quantum density matrix $\hat{\rho}_\infty$, with the understanding that here $n_j$ lives on the reals (due to our SW-SSB expansion/approximation) while the {\it exact} density matrix will have $n$ being (half-)integer.
If this is the case, we can Fourier transform to obtain expressions for the (quantum) density matrix,
\begin{equation}
 \left\langle \phi + \frac{\tilde{\phi}}{2} \right| \hat{\rho}_\infty \left| \phi - \frac{\tilde{\phi}}{2} \right\rangle = \int \prod_j dn_j  e^{\ii \sum_j n_j \tilde{\phi}_j} W(\phi,n).
\end{equation}
The integration over real $n_j$ then gives, 
\begin{equation}
 \left\langle \phi + \frac{\tilde{\phi}}{2}  \right| \hat{\rho}_\infty \left| \phi - \frac{\tilde{\phi}}{2}  \right\rangle \propto e^{\beta J \sum_{\langle ij \rangle} \cos(\phi_i - \phi_j)}   e^{- \sum_j \tilde{\phi}_j^2/2K} ,
\end{equation}
which shows a suppression of the off-diagonal matrix elements in the $\phi$-basis.  Notice that the right-hand-side is not $2\pi$-periodic in $\tilde{\phi}_j$, which it would be if we had not made the spin-wave expansion/approximation after assuming SW-SSB.

In the density basis, $n$,  the analogous expression is,
\begin{equation}
\left\langle n + \frac{\tilde{n}}{2} \right| \hat{\rho}_\infty \left| n - \frac{\tilde{n}}{2} \right\rangle = \int \prod_j d\phi_j e^{-\ii \sum_j \phi_j \tilde{n}_j} W(\phi,n).
\end{equation}
While the $\phi_j$ integrals cannot be performed in general, if we break the weak symmetry and go into  the superfluid, replacing $\cos(\phi_i - \phi_j)\approx 1 - (\phi_i - \phi_j)^2 /2$, we have
\begin{equation}
    \left\langle n + \frac{\tilde{n}}{2} \right| \hat{\rho} \left| n - \frac{\tilde{n}}{2} \right\rangle \approx e^{-\frac{K}{2} \sum_j n_j^2} e^{-\frac{1}{2J} \int_k | \tilde{n}_k|^2/k^2}.
\end{equation}

To demonstrate (conventional) ``off-diagonal long range order"  ODLRO in the superfluid, we consider the correlator,
\begin{equation}
    G(x) \equiv {\rm Tr}(e^{\ii( \hat{\phi}_{x}-\hat{\phi}_{0})} \hat{\rho}_\infty) \propto \int \prod_j dn_j \left\langle n+ \frac{\tilde{n}(x)}{2} \right| \hat{\rho}_\infty \left| n- \frac{\tilde{n}(x)}{2} \right\rangle,
\end{equation}
with $ \tilde{n}_j(x) = \delta (x_j-x) - \delta (x_j) $, corresponding to ``moving" a boson from site $0$ to site $x$.  Plugging this into above gives,
\begin{equation}
    G(x) \approx e^{-\frac{1}{2J} \int_k \frac{2(1-\cos(kx))}{k^2}},
\end{equation}
as could be obtained directly upon using $\hat{\rho}_\infty$ expressed in the phase basis.  In $d=2$ this gives power-law ODLRO $G(x) \sim |x|^{-1/2\pi J}$.

\section{SW-SSB in classical systems}\label{sec:classical}

\subsection{Generalities}

In the previous sections, we discussed SW-SSB as a sharp transition of a quantum system that can occur during decoherence, beyond which classical physics emerges. A hallmark of this classical regime is Model-F dynamics, which appears only after the SW-SSB transition. Within the SW-SSB phase, the system loses its memory of initial particle positions as well as the discrete nature of individual particles, and classical hydrodynamics becomes a legitimate effective description. In this section, we will adopt classical hydrodynamics as an effective theory for the system deep inside the SW-SSB phase. We then verify that the R\'enyi correlators can be evaluated within this hydrodynamic framework. Our main results are as follows:

\begin{itemize}
\item For spatial dimensions d=2 and d=3, the R\'enyi correlators computed from the hydrodynamic equations are consistent with our earlier analyses, i.e., they exhibit long-range or quasi-long-range order associated with SW-SSB.

\item For d=1, the R\'enyi correlators obtained from hydrodynamics show exponentially decaying correlations. However, fixing the exponential decay constant requires one to pick a ``quantum'' of charge, thus implicitly introducing some notion of charge discreteness (while hydrodynamics is inherently a continuum theory). Moreover, despite the exponential decay of R\'enyi correlators, the CMI computed within hydrodynamics is long-range, inconsistent with what we expect in a strongly symmetric phase. Together, these results suggest that in 1d hydrodynamics is not a faithful description of 
an underlying system with discrete charges, even at late times and large scales: information-theoretic quantities always remain sensitive to discreteness. (We note that at short times, it was proved for discrete classical systems with locally finite configuration spaces that the CMI remains finite, again inconsistent with hydrodynamics~\cite{zhang2025stability}.)

\end{itemize}

We begin by reviewing the definition of SW-SSB for classical distributions; see also Sec.~\ref{sec:diagnostics:observables:SWSSB}. We consider a classical system of particles indexed by density configuration $\bn = n(x)$, whose probability distribution is $P(\bn)$. We assume that we only have access to information of the positions of the particles, without labeling each individual particle. Within the diagnoses of SW-SSB discussed in the previous sections, the Bhattacharyya coefficient $B(P(\mathbf{n}): P(\mathbf{n}')) \equiv \sum\nolimits_{\bn} \sqrt{P(\mathbf{n}) P(\mathbf{n'})}$ is a standard measure of the statistical similarity of two classical distributions.  Here, the distribution $P(\mathbf{n'})$ is defined with $\mathbf{n}'$ given by $\bn$ after ``moving one particle'' from $0$ to $r$, in a sense that will be made precise below.

While the Bhattacharyya coefficient is associated with the indistinguishability of distributions with shifted charge profiles, 
it is itself not explicitly associated with any symmetry features. In order to manifest the connection to the strong and weak symmetry, one can map the classical distribution to the following fictitious quantum state in the doubled space: \beqn |\Psi^{(Q)} \rrangle  \sim \sum_{\bn} \psi(\bn)^{(Q)}
|\bn\rangle_1 \otimes |\bn\rangle_2 , \quad
\psi(\bn)^{(Q)} = P(\bn)^{Q/2}. \eeqn
The strong and weak symmetries are now defined as the separate and diagonal symmetry transformations of the doubled space. We can
also define the R\'{e}nyi-correlators: \beqn C^{(Q)}(r) &\sim& \llangle
\Psi^{(Q)} | e^{- \ii q (\phi_1 + \phi_2)_0} \, e^{+ \ii q (\phi_1 + \phi_2)_r} | \Psi^{(Q)} \rrangle \sim \sum_{\bn} P(\bn)^{Q/2} P(\bn^\prime )^{Q/2}  , 
\eeqn where $\phi$ is the conjugate variable of $n$, and if $n$ is continuous (integer), then the charge $q$ also needs to be continuous (integer). Here, $\bn^\prime(x) = \bn(x) + q \delta (x) - q \delta (x-r)$. The R\'enyi correlators can be defined for any $Q$, but the most accurate diagnosis for SW-SSB is given by $Q =1$: when $C^{(1)}(x)$ has long range (or quasi-long-range) correlation, i.e. $\lim_{r\ra \infty}C^{(1)}(r)$ does not decay exponentially, the classical system exhibits SW-SSB. Note that R\'enyi-1 correlator $C^{(1)}$ is precisely the Bhattacharyya coefficient.

It is also illuminating to write the R\'enyi correlators in the ``first-quantized" form, when the particles are discrete (that is, $\bn$ is an integer, say given by $\bn = 0,1$): 
\beqn C^{(Q)}(r) &\sim& \int d^dx_2 \cdots d^d x_n \, \left( P(0, x_2,
\cdots x_n) \right)^{Q/2} \, \left( P(r, x_2, \cdots x_n) \right)^{Q/2}. \eeqn Here the probability distribution function is assumed to be symmetrical under exchange of any two particles, because we are assuming the particles are not labelled.  As such, it can be viewed as a (boson) wave function, and the R\'{e}nyi-correlators take the same form as the more familiar ODLRO of ordinary quantum systems. 

Mapping the classical distribution to a fictitious quantum systems has the following benefits:

\begin{itemize}

\item The classical-quantum mapping endows a classical probability distribution with well-defined notions of symmetry and symmetry breaking, and allows us to classify classical distributions within the familiar Ginzburg-Landau (GL) paradigm. When the classical distribution undergoes a phase transition, the mapping also provides a natural framework to describe this transition in GL terms. Indeed, in the examples discussed in the previous sections, the transition of the distribution $P(\mm)$ falls into a generalized class of conventional phase transitions.

\item The mapping also highlights a connection between the ``emergent indistinguishability'' of classical particle world lines and the familiar indistinguishability of quantum particles. The indistinguishability of quantum bosons underlies the existence of superfluid or ODLRO in ordinary quantum systems. Correspondingly, SW-SSB is possible only when an emergent indistinguishability arises among classical particle world lines. In our setting, this emergent indistinguishability manifests itself through a transition in the winding-number fluctuations $\langle W^2\rangle$, which we have used as a numerical signature of SW-SSB.

\item As was noted in Ref.~\cite{spinliquid}, the doubled state associated with a classical distribution takes the form of a ``Gutzwiller-projected wave function", a widely studied trial wave function in condensed matter theory. This mapping therefore connects very different areas of physics and provides useful insights for the behavior of distributions. For example, motivated by the connection to Gutzwiller wave functions, one naturally expects that if $P(\bn)$ is a classical density distribution generated by an insulating wave function, then the Bhattacharyya coefficient $B\left(P(\mathbf{n}):P(\mathbf{n}')\right)$ decays exponentially with $x$, whereas if $P(\bn)$ is generated by a metallic wave function, then $B\left(P(\mathbf{n}):P(\mathbf{n}')\right)$ saturates to a constant at large $x$.

\end{itemize}

\subsection{Classical fluctuating hydrodynamics}

We take as our canonical example classical hydrodynamics in $d$-dimensions; other classical Gaussian theories can be treated with similar methods. In the hydrodynamic limit, the dynamics of the continuum density field $n_t(x)$ obeys the fluctuating hydrodynamic equation 
\begin{equation}\label{eq:MFT}
    \partial_t n_t( x) = D \nabla^2 n_t(x) + \nabla\cdot \zeta_t(x), \hspace{1cm} \langle \zeta_t( x)\zeta_s( x')\rangle = \gamma_n \delta(t-s)\delta(x-x').
\end{equation}
Here $D$ is the diffusion constant, and $\zeta_t( x)$ is a Gaussian noise field, with the variance $\gamma_n$ fixed by the fluctuation-dissipation theorem. 
We define the structure factor  
$S_t( k) = \langle n_t( k)n_t(- k)\rangle,$
whose solution from the hydrodynamic equation with initial condition $S_0( k)$ is given by
\begin{equation}\label{eq:kcovariance}
    S_t( k) = e^{-2Dk^2
t}\,S_0( k) + \frac{\gamma_n}{2D}\left(1-e^{-2Dk^2 t}\right).
\end{equation}
Below we will assume that the initial charge configuration is uniform and that $n_t( x)$ is defined relative to this global offset i.e.~$n_0( x) = 0$, and $S_0( x) = 0$. Then in real space, the structure factor takes the form 
\begin{equation}\label{eq:xcovariance}
    S_t(x-x') = \frac{\gamma_n}{2D}\delta(x-x') -\frac{\gamma_n}{2D} \frac{1}{(8\pi D t)^{d/2}} e^{-(x-x')^2/8Dt}.
\end{equation}
Eq.~\eqref{eq:xcovariance} expresses the diffusion of charge correlations in the hydrodynamic limit, as discussed in Sec.~\ref{sec:1d:correlators}. At time $t$, the probability functional is Gaussian: 
\beqn\label{eq:Gaussian}
P_t(n(x)) \propto \exp\left( -\frac12\int d^d x\, d^d x'\ 
n(x)\,V_t(x - x') \,n(x') \right),
\eeqn
\beqn
V_{t}(x - x') =
S_t^{-1}(x - x') = \int \frac{d^dk}{(2\pi)^d} \, S_t^{-1}(k) \, e^{\ii
k \cdot (x - x')}. 
\eeqn
Equivalently, all $n$-point correlation functions can be calculated in terms of $S_t$ via Wick's theorem.

\subsubsection{R\'enyi correlators}

We now turn to the computation of R\'enyi correlators for the Gaussian hydrodynamic theory. For a general Gaussian distribution given by Eq.~\eqref{eq:Gaussian}, the
evaluation is quite straightforward: \beqn C_t^{(Q)}(r) \sim \exp\left(
\frac{q^2 Q^2}{4} \, V_t(r) \right). \label{general} \eeqn If $V(r)$ is a long range
interaction, like the Coulomb, $C^{(Q)}(r)$ will decay
exponentially in 1d (strongly symmetric), and with a power-law in
2d (quasi-long-range SW-SSB). This is the case for the hydrodynamic correlator Eq.~\eqref{eq:kcovariance} (with $S_0(k) = 0)$: 
\beqn C_t^{(Q)}(r) \sim \exp\left( \frac{q^2 Q^2 D}{ 2 \, \gamma_n} \int
\frac{d^dk}{(2\pi)^d} \, \frac{e^{\ii k\cdot r}}{1 - e^{- 2 D k^2
t}} \right). \eeqn 
If we are interested in the long distance behavior of the R\'enyi correlator, we
can look at the small-$k$ behavior of the quantity above, 
\beqn C_t^{(Q)}(r) \sim \exp\left( \frac{q^2 Q^2}{4 \gamma_n t} \int
\frac{d^dk}{(2\pi)^d} \, \frac{e^{\ii k\cdot r}}{k^2} \right).
\eeqn 
Its scaling in each dimension is roughly: \beqn {\rm d} = 1
&:& C_t^{(Q)}(r) \sim \exp\left(- \frac{c}{t} |r| \right), \cr\cr
{\rm d} = 2 &:& C_t^{(Q)}(r) \sim \frac{1}{|r|^{c/t}}, \cr\cr {\rm 
d} = 3 &:& C_t^{(Q)}(r) \sim {\rm const}. \label{eq:Renyiscalings} \eeqn
This scaling matches the long-time behavior of the decohered spin and rotor models discussed in previous sections: in 1d, the R\'enyi correlators decay exponentially with a decay length that grows linearly with time, while in d=2 (resp. d=3), R\'enyi are (quasi)-long-range ordered.

It is interesting to explore the possibility of a finite-time SW-SSB transition, driven by the hydrodynamics equation. For example, in 2d and 3d, if one starts with a state (a density distribution) with strong $U(1)$ symmetry, and evolves the system according to hydrodynamics, is there a SW-SSB transition at finite $t$? The solution Eq.~\eqref{eq:kcovariance} suggests that this is not going to happen, even if we start with a distribution with strong symmetry, e.g. $V_0(x) \sim - |x|$, the long wavelength behavior of structure factor $S_t(k)$ is always dominated by the second term of Eq.~\eqref{eq:kcovariance} in ${\rm d} > 1$, therefore the SW-SSB emerges at infinitesimal $t$. On the other hand, for ${\rm d } = 1$, if we start with an initial distribution with SW-SSB, like $V_0(x) \sim e^{- m |x|}$, the structure factor $S_t(k)$ is dominated by $S_0$ in Eq.~\eqref{eq:kcovariance}, therefore the SW-SSB will persist. 

\subsubsection{R\'enyi-1 correlator in 1d; numerical evaluation}

Below we perform an explicit numerical calculation of the R\'enyi-1 correlator, for the one-dimensional case within the hydrodynamic description. The density field is defined on a discrete lattice of $L$ sites, with the structure factor given by Eq.~\eqref{eq:kcovariance} with the quantized momenta $k_n = 2\pi n/L$, $0\leq n< L$. We take a lattice spacing $a$ with $x_j = j a$. We will compute the Bhattacharyya distance, $\mathcal{B}_r(P^{+-}_q (\bm n  ):P^{-+}_q(\bm n)) = -\log B(P^{+-}_q (\bm n  ):P^{-+}_q(\bm n))$; here $B$ is the Bhattacharyya coefficient (R\`enyi-1 correlator) with charge $q$ ``moved'' across the system by distance $r$. Concretely, we have defined conditional probability distributions;
\begin{equation}
    P^{+-}_q (\bm n  ) = P(\bm n'|n(0)=q,n(r)=-q), \hspace{1cm} P^{-+}_q (\bm n  ) = P(\bm n'|n(0)=-q,n(r)=q),
\end{equation}
with $\bm n'$ labelling the density on all sites except $x =  0,r$. 

The Bhattacharyya distance can be efficiently computed for Gaussian distributions, and in our case gives
\begin{equation}\label{eq:Bhattacharyya}
    \mathcal{B}_r(P^{+-}_q (\bm n  ):P^{-+}_q(\bm n)) = \frac{1}{2} \sum_{ij}' \mu_q (x_i) S_q^{-1}(x_i-x_j)\mu_q (x_j),
\end{equation}
where $\mu_q$ is the mean density field in the conditional state $P^{+-}_q (\bm n  )$, and $S_q^{-1}$ the conditional covariance matrix. The sum runs over all $i,j \neq 0,r/a$. We refer to Appendix \ref{app:Bhattacharyya} for further details. The important thing to note is that after conditioning, the density average $\mu_q$ is displaced into a dipole-like configuration, with negative $\mu$ in the vicinity (set by the correlation length $\xi \approx \sqrt{Dt}$) of $x=0$ and positive $\mu$ in the vicinity of $x=r$. The long-distance behavior $r\gg t$ of $S_q^{-1}$ approximates the 1d Coulombic interaction $S_q^{-1} \approx |x-x'|/t$. The Bhattacharyya distance therefore measures the energy of the $\mu_q$-dipole configuration, which therefore goes as $\mathcal{B}_r(t)\approx r/t$ in 1d. 

We confirm this physical picture numerically, computing the Bhattacharyya distance for several different times as shown in Fig.~\hyperref[fig:gaussianCMI]{\ref*{fig:gaussianCMI}(a)}. We plot the coefficient rescaled by a factor of $t/cL$, where $L=3000$ and $c=a^2q^2/\gamma_n$ follows from dimensional analysis. The finite-size scaling collapse $\mathcal{B}_r(t) = \frac{1}{4t} \frac{r}{L}(1-\frac{r}{L})$ is immediate (orange dashed line). We have checked this result for larger $L$. In the limit of infinite system size, we observe the result
\begin{equation}
   \mathcal{B}_r(t) =\frac{a^2q^2}{\gamma_n}\frac{r}{2t} \sim \frac{r}{\xi^{(1)}}.
\end{equation}
Here, $\xi^{(1)}(t)$ is the decay length for the fidelity and R\'enyi-1 correlator, which grows linearly with time, $t$, as we found in previous sections for our 1d microscopic discrete particle models.
This is perhaps surprising, since 1d continuum hydrodynamics does not emerge from our microscopic models, unless we ignore the  discrete phase-slip events which will be present in the dynamics. 
We can well ask whether information theoretic quantities like CMI
for the 1d continuum hydrodynamics will be different from what we found in our microscopic discrete particle models. As we shall now show, this is indeed the case.

\subsubsection{Conditional mutual information}

Although the classical diagnostics of SW-SSB for the continuum Gaussian hydrodynamics in the previous section yield a lengthscale which grows linearly in time, the CMI behaves in a qualitatively distinct way. 

We consider a 1d geometry where $A\cup B\cup C$ is a proper subset of the system, and $B$ separates $A$ from $C$, as shown in Fig.~\ref{fig:1d-CMI}. We denote the size of $B$ by $R_B=|B|$. To compute the CMI for the hydrodynamic theory, we again make use of the Gaussian property of the correlations. We treat the structure factor as a matrix $\bm S_{ij} = S(x_i-x_j)$ and define the corresponding covariance matrix restricted to a region $X$ as $\bm S^X$ --- for example the matrix of region $A\cup B$ is $\bm S^{AB}$ (our notation is slightly different from the previous section due to the geometry we consider here). When $\bm {n}$ is normally distributed with covariance $\bm S$, the differential entropy of a subsystem $A$ is given by \cite{ahmed1989entropy} $H^A = \frac{1}{2}\log \|2\pi e\bm S^A\|$, where $\|\cdot\|$ denotes the matrix determinant. Using the CMI definition $I(A:C|B) = H^{AB}+H^{BC}-H^B-H^{ABC}$, we obtain the useful formula
\begin{equation}\label{eq:gaussianCMI}
    I(A:C|B) = \frac{1}{2}\log \frac{\|\bm S^{AB}\|\|\bm S^{BC}\|}{\|\bm S^{ABC}\|\|\bm S^B\|}.
\end{equation}

As a consequence of the charge conservation law $\int dx S(x,t) = 0$, the spectrum of the full system covariance matrix is gapless with a quadratic zero, $S(k)\approx a k^2$ at small $k$.  This feature leads to surprising behavior in the CMI -- namely algebraic decay in space, as we will show below. From a mathematical point of view, this result can be understood from the theory of Toeplitz determinants, and in particular corrections to subsystem entropies which arise due to the ``Fisher-Hartwig singularities'' in certain critical systems \cite{fisher1969toeplitz, basor2005toeplitz, vidal2003entanglement, jin2004quantum, peschel2009reduced}. Here we will give a simple argument in order to show the essential mechanism. 

To see the effect of the gapless covariance on the CMI, we replace $S(k)$ with the approximation 
\begin{equation}\label{eq:Fisher}
    S_0(k) = c(2-2\cos k),
\end{equation}
where $c = 2\gamma_n Dt$. This dispersion matches the true one for $k\ll (2Dt)^{-1/2} \ll 1$. Intuitively, we expect that correlations contributing to the CMI over a range $R_B$ will be long-wavelength modes with $k \lesssim R_B^{-1}$. Therefore we expect our approximation to hold for $R_B \gg (2Dt)^{1/2} \gg 1$. The covariance $S_0(k)$ has the advantage that the resulting $\bm S$ is a tri-diagonal matrix $S_{i,i} = 2c$, $S_{i,i+1} = S_{i,i-1} = -c$, and its determinant can be computed from the recurrence relation $D(R_B) = 2D(R_B-1)-D(R_B-2)$, giving $D(R_B) = R_B+1$. The resulting entropy for region $B$ of size $R_B$ is
\begin{equation}
    H^B_0 = \frac{1}{2}R_B\log (2\pi e a)  + \frac{1}{2}\log (R_B+1).
\end{equation}
Aside from a volume law term $H^B_0 \propto R_B$ (which is sensitive to UV modes but does not contribute to the CMI), we see that the entropy obtains a universal logarithmic correction due to the low energy modes. This directly results in the CMI decaying algebraically in $R_B$: taking $A$ and $C$ each as a single site for simplicity, we find for $R_B\gg 1$
\begin{gather}
    I(A:C|B) = \log(R_B+2) -\frac{1}{2}\log(R_B+1) - \frac{1}{2}\log(R_B+3) = \frac{1}{2}R_B^{-2} +O(R_B^{-3}).
\end{gather}
While the CMI is expected to show a non-universal correction in the region $R_B \lesssim (2Dt)^{1/2}$, the long-distance decay will be governed by the power law $I(A:C|B) \approx \frac{1}{2}R_B^{-2}$ for $R_B \gg (2Dt)^{1/2}$. In Appendix \ref{app:renyiCMI}, we perform a calculation of the R\'enyi-2 CMI for the 1d decohered rotor model, finding agreement with the Gaussian power law decay.

We demonstrate the predicted scaling numerically, considering a finite ring of $L$ sites and evaluating entropies via matrix determinants. We fix $A$ and $C$ each to be regions of a single site, and compute the CMI for increasing separations $R_B$ using formula \eqref{eq:gaussianCMI}. Results are shown in Fig.~\hyperref[fig:gaussianCMI]{\ref*{fig:gaussianCMI}(b)}, for several different times, and rescaled according to $I(A:C|B) = t^{-1} f(R_B/\sqrt{t})$. We find a short-distance regime $R_B \lesssim (2Dt)^{1/2}$ where the CMI is $O(t^{-1})$, and a crossover to the predicted algebraic scaling, including the prefactor $1/2$.

\begin{figure}
    \centering
    \includegraphics[width=0.9\linewidth]{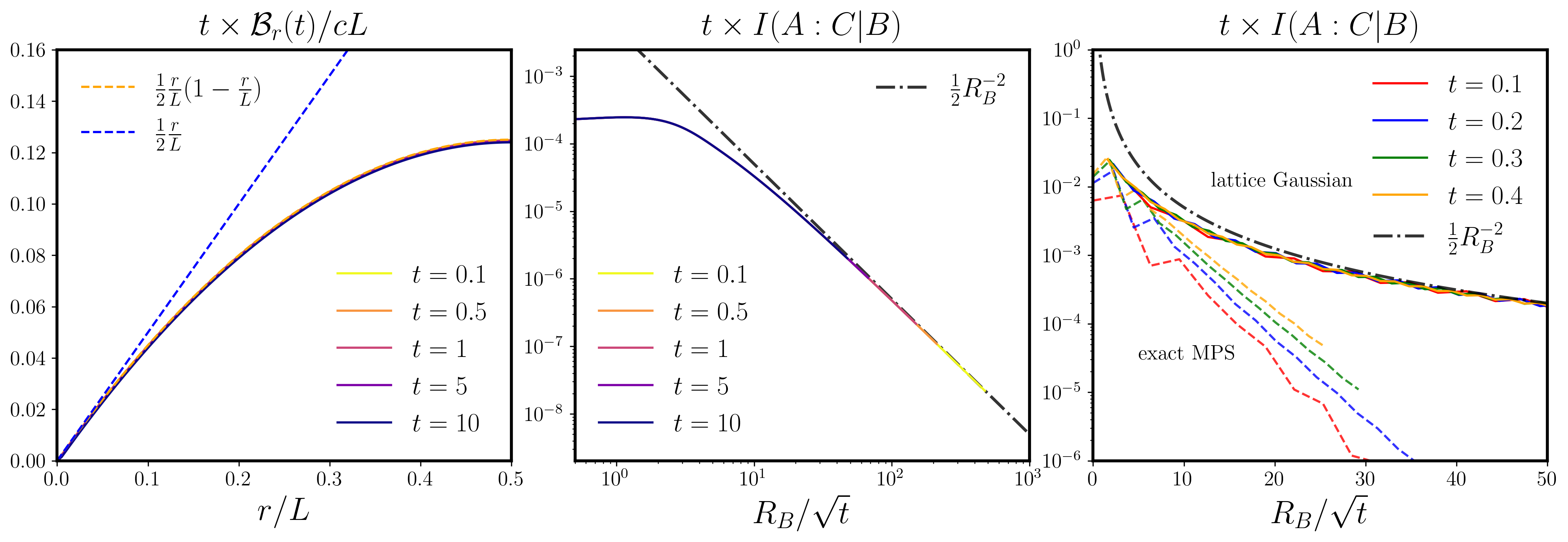}
    \caption{(a) Bhattacharyya distance computed according to Eq.~\eqref{eq:Bhattacharyya}. We show collapsed data $t \times \mathcal{B}^r(t)/cL$, with $c = a^2q^2/\gamma_n$, as a function of $r/L$ for different times, and a system size $L=3000$. The blue dashed line is the curve $y(x) = x/2  $ and gives the infinite size scaling limit, from which we obtain $\mathcal{B}_r(t) \sim  r/2t$. 
    (b) CMI calculated in the Gaussian hydrodynamic theory, for several different times and a system size $L=3000$. The CMI exhibits a universal collapse onto the power law decay $I(A:C|B) \approx \frac{1}{2}R_B^{-2}$ for $R_B \gg (2Dt)^{1/2}$. (c) The CMI computed using the Gaussian hydrodynamic formula Eq.~\eqref{eq:gaussianCMI} applied to the exact lattice covariance matrix (solid lines), and the CMI computed from MPS for the spin-$1/2$ (SSEP) simulations (dashed lines). The Gaussian formula gives the power-law decay (dot-dashed line), while the spin-$1/2$ CMI decays exponentially at large distances.}
    \label{fig:gaussianCMI}
\end{figure}

Our result should be compared to the numerically obtained CMI for the spin-1/2 model obtained in Section~\ref{sec:spin-1/2:1d}. There, we found an intermediate power-law regime for $ (2Dt)^{1/2} \lesssim R_B \ll t$, followed by exponential decay on a scale set by the Markov length, $\xi^M \propto t$. The numerical exponent was found to have a maximum value $\approx 1.4$ at the latest accessible times, while the Gaussian prediction gives an exponent of $2$; nevertheless we expect the algebraic decay can be traced to the emergence of a continuum hydrodynamic regime, with the discrepancy attributed to finite time deviations. In the hydrodynamic Gaussian theory, the CMI decays algebraically even for $R_B \gg t$.  
Note, however, that when starting from the microscopic spin-1/2 model, the diffusive hydrodynamic limit is only justified for $a_{\rm UV} \ll R_B \sim (Dt)^{1/2} \ll Dt$ with $a_{\rm UV}$ the UV spatial cutoff. To highlight this point, we perform exact MPS simulations for short times, starting from the N\'eel state, and exactly evaluate the CMI for small $R_B$.
We also compute the CMI using the Gaussian formula Eq.~\eqref{eq:gaussianCMI} applied to the exact lattice covariance matrix, which can be obtained either from the MPS or from an exact calculation. The results are displayed in Fig.~\hyperref[fig:gaussianCMI]{\ref*{fig:gaussianCMI}(c)}: the CMI computed for the spin-$1/2$ model exhibits a clear exponential decay with increasing $R_B$, whereas the CMI within the Gaussian hydrodynamics approximation matches the expected algebraic decay. 

We note that one fundamental difference between the spin-1/2 model and the hydrodynamic description is that the particles are discrete in the former, whereas the hydrodynamic density field, $n(x)$, takes on a continuum of values. In 
Section~\ref{sec:ModelF}
it was shown that the ``spin-wave" approximation for the quantum phase field, $\tilde{\phi}$, is tantamount to ignoring the discreteness of the particles.  It is plausible that on spatial scales with $r \ll t$ the spin-wave approximation might be valid, but on the longest scales phase slips cannot be ignored, and the spin-wave approximation which is necessary for obtaining the continuous density field hydrodynamic description is not valid - so does not give a faithful description of the spin-1/2 model dynamics.
In any event, it is striking that the exponential decay of the CMI in the spin-1/2 model is not obtained in hydrodynamics. 

\section{Summary, discussion and outlook}\label{sec:discussion}

\subsection{Summary}

In this paper we considered three complementary models for the Lindbladian dynamics of systems with $U(1)$ symmetry, focusing on SW-SSB and emergent classicality.  The spin-$1/2$ model with nearest-neighbor $S^+_i S^-_j$ jump operators is perhaps the simplest model for such physics, and we numerically computed information-theoretic diagnostics of SW-SSB, including the R\'enyi-1, the R\'enyi-2 correlators in 1d and 2d, as well as the conditional mutual information and accuracy of charge decoding in 1d. 

We also studied a quantum rotor model under Lindblad evolution with the strong $U(1)$ symmetry in both 1d and 2d. In both dimensions, we uncovered a universal relation between the R\'{e}nyi-1 and R\'{e}nyi-2 correlator. This relation implies that in 1d, though the system has no $U(1)$ SW-SSB, the correlation lengths of the R\'{e}nyi-1 and R\'{e}nyi-2 correlators satisfy $\xi^{(1)} = 2 \, \xi^{(2)} \sim t$ at large $t$, consistent with our numerical simulation for the spin-$1/2$ model. In 2d, there is a finite-time SW-SSB transition into a phase with quasi-long-range algebraic R\'{e}nyi-1 and R\'{e}nyi-2 correlators.   In this phase the scaling dimension of the R\'{e}nyi-1 correlator is half of that of the R\'{e}nyi-2 correlator. Our RG calculation based on the rotor model shows that the $U(1)$ SW-SSB driven by the Lindblad evolution belongs to the same BKT-like universality class that controls the ``charge-sharpening'' transition in monitored 1+1d systems with $U(1)$ symmetry~\cite{SharpeningEFT}.  

Thirdly, we studied Lindbladian dynamics for rotors that included coherent Hamiltonian evolution and carefully chosen jump operators to target two different long-time steady states.  For this model, we found that in 2d and 3d for times larger than the finite-time SW-SSB transition the dynamics became classical, described by Langevin dynamics corresponding to Model F of Halperin and Hohenberg. In this instance, the finite-time SW-SSB transition in 2d and 3d is actually a transition where classicality emerges! 

In all three of our microscopic models we have demonstrated  that in 2d (and 3d) there is a finite-time SW-SSB transition,
and at times thereafter classical continuum hydrodynamics emerges, wherein the discreteness of the particles is unimportant and the particle density could be described by a continuous field.  In 1d, due to the presence of phase slips 
continuum hydrodynamics is not expected to be emergent at any finite time, at least for information-theoretic quantities that are nonlinear in the state/probability distribution, such as CMI.

In Sec.~\ref{sec:classical} we formulated SW-SSB directly within the classical continuum hydrodynamic theory, and computed SW-SSB correlators and CMI.  In 2d and 3d we found that SW-SSB occurs at infinitesimal time (no finite time transition); this is not surprising since the very existence of hydrodynamics required that SW-SSB had already occurred.
In d=1, despite our belief that continuum hydrodynamics does not emerge, we used the theory to compute R\'enyi correlators and CMI.
We found that although the R\'{e}nyi correlators decay exponentially, with a length scale that grows linearly in time, the CMI within continuum hydrodynamics decays subexponentially at all times: thus, the Markov length in continuum hydrodynamics is always infinite in 1d, despite the absence of SW-SSB.
These results are inconsistent with the behavior we found for 1d microscopic models where both R\'{e}nyi correlators and CMI fall off exponentially, and indicate that 1d continuum hydrodynamics does not faithfully capture the long-distance and late-time  physics of microscopic models with discrete charges in 1d. This is consistent with our expectations, however, since phase slips in 1d microscopic models always preclude the emergence of continuum hydrodynamics.

\subsection{Experimental protocol} 

To measure R\'enyi correlators in quantum systems, we can perform shadow tomography of the density matrix~\cite{aaronson2018,Huang_2020}, or we can use the 
``quantum-classical" estimator~\cite{PhysRevX.13.021026}.
We also propose an alternative variational approach, described below. This variational approach gives an upper bound to the R\'enyi-1 correlator. It applies directly to classical distributions. For a quantum state, one first replaces the quantum state by a classical bit-string distribution in the computational basis (which is fixed by $U(1)$ symmetry), by applying a fully dephasing channel. This dephased R\'enyi-1 correlator gives an upper bound to the quantum R\'enyi-1 correlator as well, by the following argument. The R\'enyi-1 correlator measures the Holevo fidelity $\mathcal{F}(\rho, \mathcal{Q}_x(\rho))$ between two states $\rho$ and $\mathcal{Q}_x(\rho)$, where $\mathcal{Q}_x(\cdot) = \sigma^+_x \sigma^-_0 \cdot \sigma^+_0 \sigma^-_x$~\cite{weinstein2025efficient}. Importantly, the linear operation $\mathcal{Q}$ commutes with the dephasing channel $\mathcal{D}$ (since it maps diagonal entries to diagonal entries), so $\mathcal{Q}_x(\mathcal{D}(\rho)) = \mathcal{D}(\mathcal{Q}_x(\rho))$. Thus the fully dephased R\'enyi-1 correlator can be written as $\mathcal{F}(\mathcal{D}(\rho), \mathcal{D}(\mathcal{Q}_x(\rho))) \geq \mathcal{F}(\rho, \mathcal{Q}_x(\rho))$ because fidelity is non-decreasing under quantum channels.

Beyond being simpler to estimate, the classical R\'enyi correlator is easier to probe in near-term experiments, since it only requires computational-basis snapshots. For example, in a quantum gas microscope experiment~\cite{QGmicroscope,QGmicroscope2, doi:10.1126/science.abk2397,FCScoldatoms}, one takes a simultaneous snapshot of all the particles in the system, leading to a classical probability distribution over classical configurations. Alternatively, one could consider SW-SSB in purely classical systems such as active matter (treating the classical particles as indistinguishable in practice, see discussion below).

Our approach exploits the information-theoretic interpretation of the R\'{e}nyi-1 correlator as a measure of distinguishability. Specifically, the R\'{e}nyi-1 correlator is equivalent to the Bhattacharyya coefficient (BC) between the original distribution $P(\mathbf{n})$ and a displaced distribution $P'(\mathbf{n}) = P(\mathbf{n}')$, where $\mathbf{n}'$ is obtained by displacing a single particle within the configuration by certain distance. The BC quantifies the similarity between these two distributions: as BC $\to 0$, the distributions become perfectly distinguishable, allowing an ideal classifier to achieve a zero failure rate. Conversely, a finite BC implies an overlap, ensuring that any classifier will suffer from a non-zero failure rate.

Based on this observation, we outline the following protocol:

\begin{itemize}
    \item Collect snapshots of the density configurations $\mathbf{n}$, or other relevant local degrees of freedom, from the experimental system.

    \item For each observed configuration $\mathbf{n}$, generate a new configuration $\mathbf{n}'_x$ by shifting a single particle from a  random origin by a distance $x$. This procedure generates a displaced distribution $P'_x(\mathbf{n}) = P(\mathbf{n}'_x)$.

    \item Train a classifier to distinguish between the ensembles $P$ and $P'_x$ and record the resulting failure rate $R(x)$. Because any realized classifier may be sub-optimal, $R(x)$ provides an experimental upper bound on the R\'{e}nyi-1 correlator (i.e., an upper bound on the ideal failure rate). Specifically, the Bayes error bound relates the failure rate to the BC (the R\'{e}nyi-1 correlator) via \cite{devroye2013probabilistic}:
    \begin{equation}
        R(x) \geq \frac{1}{2} \left( 1 - \sqrt{ 1 - |C^{(1)}(x)|^2} \right).
    \end{equation}
    A direct consequence of this relation is that if the failure rate $R(x)$ decays exponentially with $x$, the R\'{e}nyi-1 correlator $C^{(1)}(x)$ must also be short-ranged, signaling strongly symmetric, or no SW-SSB.
    
\end{itemize}

The classical conditional mutual information can, analogously, be measured by a decoding experiment. One takes a snapshot of all the particles in the system, and feeds the configuration of region $B$ into a decoder (either one of the human-constructed decoders we discussed in Sec.~\ref{sec:1d:decoding}, or a suitable machine-learning algorithm). The decoder's task is to use this information to predict the number of particles in $A$. As we showed in Sec.~\ref{sec:1d:decoding}, its success rate is close to unity when the size of region $B$ exceeds the Markov length. Since a general decoder will be suboptimal, this protocol yields an upper bound on the Markov length.

\subsection{Discussion and outlook}

Our results suggest a natural conjecture as to what properties a classical world might need to have for its dynamics to exhibit SW-SSB transitions. 
The objects in this classical world might be distinguishable particles, continuum waves, or discrete but indistinguishable objects, like the particles in a lattice gas. 
If the classical particles are distinguishable, then SW-SSB cannot occur, which is apparent from our world-line/trajectory discussion for SSEP. Because worldlines can be tagged by the particle ``labels,'' any positional uncertainty can only spread as far as an individual (labelled) particle, and the state is strongly symmetric at the longest distances at all times. In the opposite limit of a continuous classical density field evolving according to (linear) diffusive hydrodynamics, as we showed SW-SSB is present at all times or never present, as  discussed in Sec.~\ref{sec:classical}. 
We thus conjecture that ``discrete and indistinguishable'' is a necessary (but not sufficient) condition required to have ``intrinsic'' SW-SSB {\it transitions}, be it a classical or a quantum world. Of course, even in a world of distinguishable particles, SW-SSB might \emph{appear} to take place, for an imperfect observer without access to the charge labels, but it would not correspond to any fundamental property of the world.

In the quantum world, where particles are truly indistinguishable, our results have implications for the status of continuum hydrodynamics as a theory. Conventionally, hydrodynamics is regarded as an approximate description, which involves coarse-graining and losing information. However, the existence of SW-SSB transitions suggests that continuum hydrodynamics might be valid in a precise information-theoretic sense after the SW-SSB timescale: continuum hydrodynamics contains all the information about charge that can possibly be locally accessed past the SW-SSB timescale. (Note that at finite time after the SW-SSB time scale, this continuum hydrodynamic description will include infinitely many couplings describing nonlinearities, higher-derivatives {\it etc.}; conventional hydrodynamics like Eq.~\eqref{eq:MFT} involving only a few parameters emerges only at long times.) Conventionally, the validity of the continuum description does not depend on dimensionality: our results suggest that there are natural questions about state reconstruction that lie beyond hydrodynamics, but only in one dimension. Importantly, the fact that even the continuum diffusion equation does not exhibit SW-SSB in one dimension shows that one cannot fully identify the diffusion mode with the Goldstone mode of $U(1)$ SW-SSB, as has been suggested~\cite{huang2025hydrodynamics} (outside the steady-state setting explored in that work); in one dimension, the diffusion mode is well defined even though SW-SSB is absent. It would be interesting to revisit recent proposals on the onset of hydrodynamics~\cite{PhysRevLett.115.072501, delacretaz2025bound, le2023observation} in light of these results. 

Our focus was entirely on SW-SSB of the simplest continuous symmetry, $U(1)$. In this case, the SW-SSB phase need not have long-range entanglement. For nonabelian symmetries like $SU(2)$, even the maximally mixed state in a given charge sector is known to be long-range entangled~\cite{NonAbelianChannels,NonAbelianChannels2}. How this long-range entanglement emerges under $SU(2)$-symmetric dissipative dynamics, whether it exhibits singularities at the SW-SSB transition, and whether SW-SSB in $SU(2)$ symmetric systems is related in some way to monitored $SU(2)$ dynamics~\cite{monitoredSU2} remain open questions. In the case we explored, the failure of continuum Model-F hydrodynamics to emerge in 1d could be directly attributed to charge discreteness being a relevant perturbation, but it is not clear how far this feature generalizes. Understanding the {\it dynamics} of SW-SSB transitions for higher-form symmetries (relevant, e.g., to the error correction threshold for topological codes) remains another interesting open question. Already for $U(1)$ symmetry the SW-SSB and ordinary SSB transitions are in different universality classes, but the nature of SW-SSB phases and phase transitions with different symmetry groups could be even richer.  Finally, while most work on SW-SSB so far has focused on lattice models with internal symmetries, it would be interesting to explore the dynamics of SW-SSB for continuous spatial symmetries like translation or rotation. For example, the Navier–Stokes equations—hydrodynamics based on momentum conservation—might emerge as a consequence of SW-SSB of translational symmetry. A more thorough discussion of this connection could shed light on the sharp onset of hydrodynamics in nature. We leave a detailed exploration of this possibility for future work. \\

{\bf --- Acknowledgment}\\

We acknowledge numerous natural intelligence discussions with, among others, Ehud Altman, Yimu Bao, Tarun Grover, Timothy Hsieh, David Huse, Ali Lavasani, Jong Yeon Lee, Shang Liu, Ruochen Ma, Shengqi Sang, Zack Weinstein, Carolyn Zhang, Frank Zhang, and Yizhi You. We also benefit from OpenAI for helpful discussions and assistance with clarifying physical intuition and checking steps in several derivations. 
Use was made of computational facilities purchased with funds from the National Science Foundation (CNS-1725797) and administered by the Center for Scientific Computing (CSC). The CSC is supported by the California NanoSystems Institute and the Materials Research Science and Engineering Center (MRSEC; NSF DMR 2308708) at UC Santa Barbara. This research was also partly done using services provided by the OSG Consortium~\cite{osg1,osg2,osg3,osg4}, which is supported by the National Science Foundation awards \#2030508 and \#2323298. This work was supported by the Simons
Collaboration on Ultra-Quantum Matter, which is
a grant from the Simons Foundation (651457, M.P.A.F.
and J.H.). M.P.A.F. was also supported by a
Quantum Interactive Dynamics grant from the William
M. Keck Foundation. H.H. was supported in part by Grant No. NSF PHY-1748958 to the Kavli Institute for Theoretical Physics (KITP), the Heising-Simons Foundation, and the Simons Foundation (216179, LB). C.X. is supported by the Simons Foundation through the Simons Investigator program. R.V. acknowledges partial support from the U.S. Department of Energy, Office of Science, Basic Energy Sciences, under Award
No. DE-SC0023999, and acknowledges the hospitality of  KITP during the program ``Learning the Fine Structure of Quantum Dynamics'', supported by NSF PHY-2309135. S.G. was supported by NSF Award No. OMA-2326767. T.G.K. acknowledges support from the Gordon and Betty Moore Foundation under GBMF7392 and from the National Science Foundation under grant PHY-2309135 to the Kavli Institute for Theoretical Physics (KITP).

\bibliography{refs}

\appendix

\section{Derivation of diagonal-sector evolution for spin-1/2 model}\label{app:diagonal_lindbladian}
In this Appendix, we provide the steps leading to the calculation of Eq.~\eqref{eq:imaginaryevolution}. Here, it useful to work with a vectorized form of the density matrix $\h{\rho}$. We will call this ``doubled state'' $\kett{\rho}$, defined such that $\braakett{s,s'}{\rho} = \mel{s}{\rho}{s'}$. Accordingly, we may view our Lindbladians as operators on this doubled Hilbert space. Defining $\h{S}_{L,i}^{\pm}$ and $\h{S}_{R,i}^{\pm}$ to act on the left (ket) and right (bra) Hilbert spaces respective, we may write
\begin{align}
\h{\calL} = \gamma \sum_{\expval{ij}} [\h{L}_{L,ij} \h{L}_{R,ij}^*
- \frac{1}{2}(\h{L}_{L,ij}^\dag \h{L}_{L,ij} + \h{L}_{R,ij}^T \h{L}_{R,ij}^*) + (i \leftrightarrow j)].
\end{align}
We also remark that in $\ket{{\pm}s}$ basis, $\h{L}_{L,ij}^* = \h{L}_{L,ij}$ and $\h{L}_{L,ij}^T = \h{L}_{L,ij}^\dagger$ (and the same for $L_{R,ij}$). A short calculation then yields
\begin{align}
\mathcal{L} = \gamma \sum_{\langle ij\rangle}
\Big(
S_{L,i}^+S_{L,j}^- S_{R,i}^+S_{R,j}^-
+ S_{L,i}^-S_{L,j}^+ S_{R,i}^-S_{R,j}^+
+ S_{L,i}^z S_{L,j}^z+ S_{R,i}^z S_{R,j}^z-\tfrac12
\Big).
\end{align}
We remark that this operator has a weak symmetry $S^z_{L,i} - {S}^z_{R,i}$
at each site $i$. Therefore, if we restrict to the diagonal sector where $S^z_{L,i} = {S}^z_{R,i} =: S^z_i$
, we obtain the effective Lindbladian
\[
\mathcal{L}^{\textsf{diag}} = 2\gamma \sum_{\langle ij\rangle}
\Big(\mathbf{S}_i\cdot\mathbf{S}_j - \tfrac{1}{4}\Big),
\]
which is, of course, equivalent to imaginary time evolution under the Heisenberg model. We remark that the dynamics has an enhanced $SU(2)$ symmetry in this \mbox{spin-$1/2$} case, though this is not true when analogous procedures are applied to higher spin dynamics.

When we consider states in the diagonal sector, we will use the notation $\ket{\rho}$ to emphasize that we have reduced to a single copy of Hilbert space (albeit with a nonstandard normalization condition, namely $2^{L/2}\braket{+\cdots+}{\rho} = 1$). We remark that even if we did not select an initial state in the diagonal sector, this would still be the most important sector at late times. In other sectors, where $S^z_{L,i} - S^z_{R,i} = \pm 1$ at one or more sites, the hopping terms connected to these sites always vanish leading to a finite gap relative to the diagonal sector. Consequently, the diagonal sector describes the slowest dynamics with \emph{any} initial state.

\section{Comparison of TCI data and sampled data for spin-1/2 model}\label{app:TCI_vs_sampling}
R\'enyi-$1$ correlations in 1d can be obtained from MPS numerics using either of two methods, both described in Sec.~\ref{sec:spin-1/2:numerics}. First, an MPS form for $\ket{\sqrt{\rho}}$ can be inferred from $\ket{\rho}$ via tensor cross-interpolation (TCI). Second, R\'enyi-$1$ correlations may be sampled directly from $\ket{\rho}$. In Fig.~\ref{fig:TCI_vs_sampled}, we compare the two methods, verifying that they give consistent results.

\begin{figure}[h]
    \centering
    \includegraphics[width=0.5\linewidth]{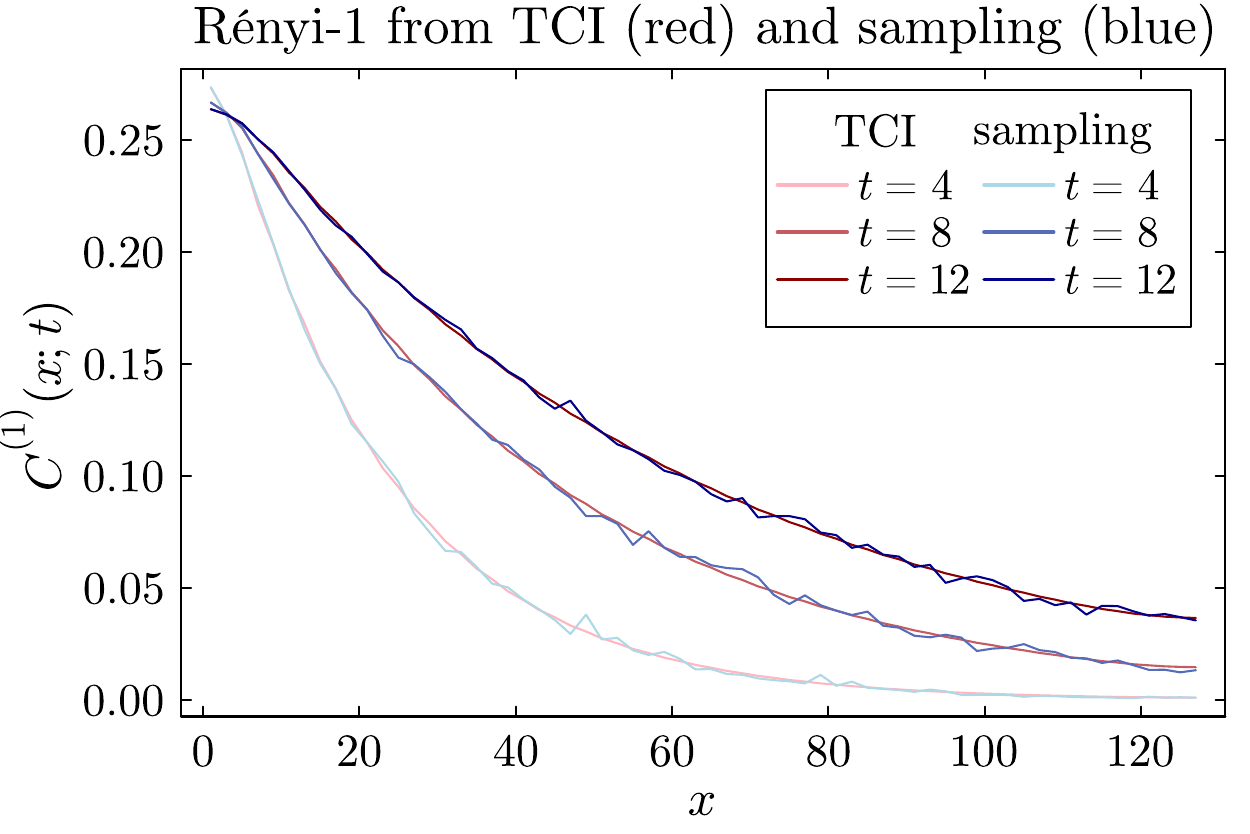}
    \caption{R\'enyi-$1$ correlations obtained in 1d from TCI and from direct sampling. We find that the two methods agree.}
\label{fig:TCI_vs_sampled}
\end{figure}

\section{QMC for the decohered spin-1 model in 2d}

\begin{figure}[!t]
    \centering
    \includegraphics[width=0.45\linewidth]{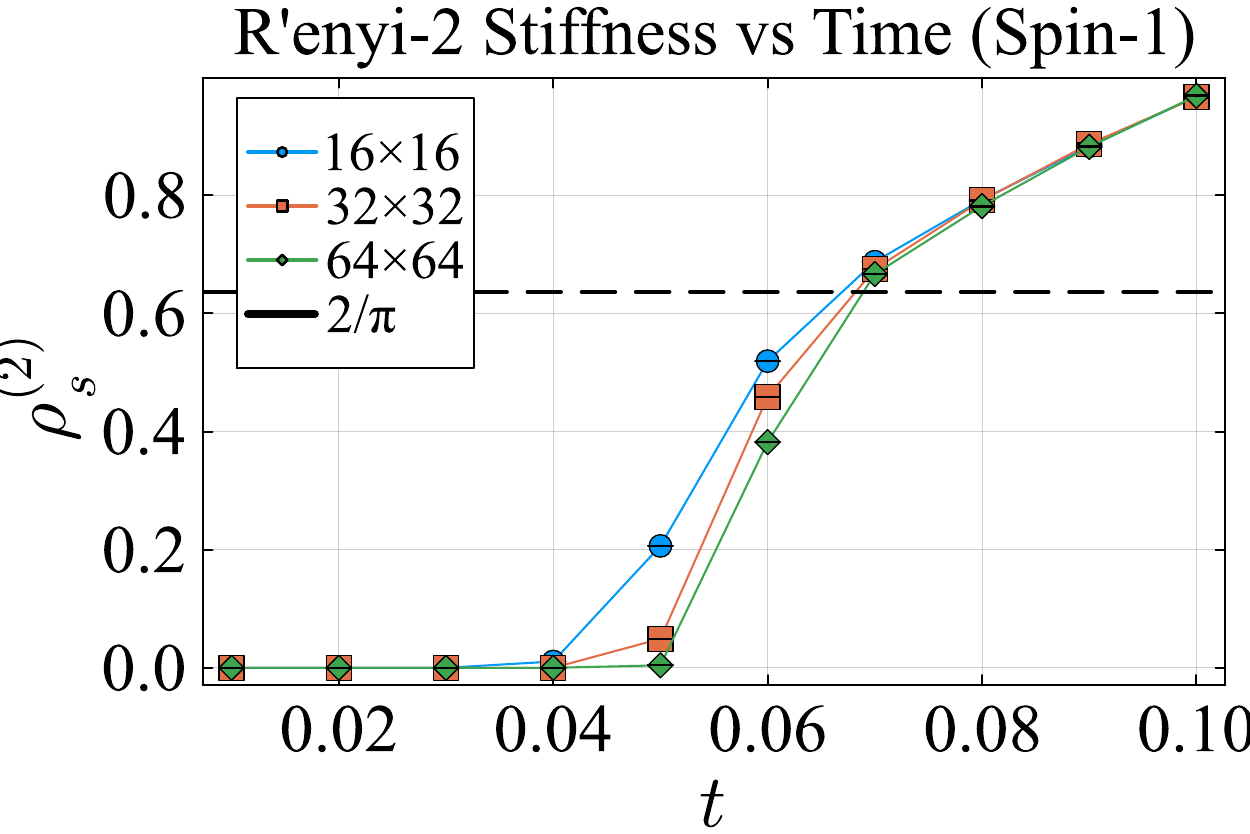}
\includegraphics[width=0.45\linewidth]{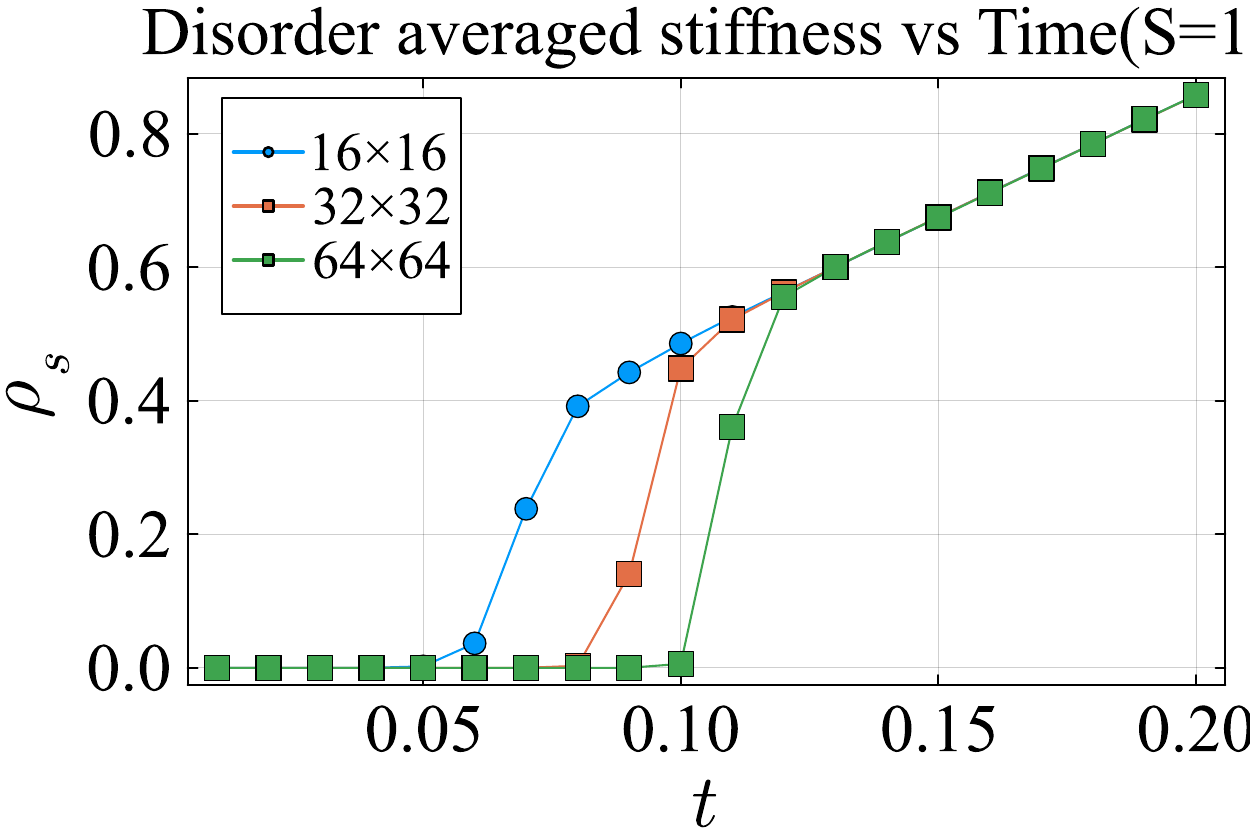}
    \caption{The R\'{e}nyi-2 and disorder average (R\'{e}nyi-1) stiffness of the decohered spin-1 model. The behavior is qualitatively similar to the decohered spin-1/2 model plotted in Fig.~\ref{fig:stiffness}: the R\'{e}nyi-2 stiffness has an evident $2/\pi$ jump, while the disorder averaged stiffness exhibits stronger finite size effects. }    \label{stiffnessspin1}
\end{figure}

In section \ref{sec:spin-1/2:2d} we presented the QMC study of a spin-1/2 system under Lindbladian evolution, with jump operator $\h{L}_{ij}=\h{S}_i^+\h{S}_j^-$. We found that the R\'{e}nyi-2 stiffness exhibits a jump with size $2/\pi$, consistent with the ordinary BKT transition, while the disorder-averaged (R\'{e}nyi-1) stiffness has a more complicated behavior with a stronger finite size effect, as the critical point drifts significantly with increasing system size. In section \ref{sec:rotor} we analyzed a quantum rotor model under decoherence, and the RG calculation indicates that the SW-SSB transition for R\'{e}nyi-2 quantities should indeed belong to the ordinary BKT transition universality class, while the R\'{e}nyi-1 quantities interestingly will be influenced by an extra term in the RG flow of the vortex-dipole fugacity (Eq.~\eqref{RGxy}). This extra term leads to a much longer correlation length near the transition, and hence stronger finite size effects.

To further verify this picture, we also performed QMC for a spin-1 system.
We choose the jump operator $\hat{L}_{ij} = \hat{S}_i^+\hat{S}_j^-$, same as the spin-1/2 model, and our initial state is $\prod_i |S^z_i = 0\rangle$. The spin-1 model has a significant difference from its spin-1/2 counterpart: there is no longer a $SU(2)$ symmetry in the diagonal subspace. However, its stiffness behaves very similarly (Fig.~\ref{stiffnessspin1}): the R\'{e}nyi-2 stiffness has a jump at $2/\pi$, while the disorder averaged stiffness still sees a pronounced finite size effect. 

\section{More details about decohered rotor model in 1d}\label{app:rotor}

In this section we provide more details of the rotor model in Section~\ref{sec:rotor}. We would first like to evaluate factor $A(x)_t$ of Eq.~\eqref{AB1d}. The factor $A(x)_t$ is the partition function of 1d Coulomb gas with charge configurations $\{m\}^A$; which includes two test charges $\pm 1/2$ at position $i$ and $j$ with distance $x$. We first note that the energy of a Coulomb gas can be computed as the total energy of the electric field in the space, and therefore we introduce the electric field $E_{\bar{i}}$ variables defined as \beqn
    E_{\bar{i}} = \sum_{k = 1}^i m^A_i.
\eeqn or reversely \beqn
    m_i^A = E_{\bar{i}} - E_{\bar{i} - 1}.
\eeqn The Coulomb gas energy in terms of electric field $E_{\bar{i}}$ can be expressed as \beqn
  \frac{1}{2 \td}\sum_{j<k} m^A_j \, m^A_k \, |x_j-x_k|   = \sum_{\bar{n} = 1}^{L - 1} \frac{1}{2 \, \tilde{t}} E_{\bar{n}}^2.
\eeqn Note that $E_{\bar{n}}$ takes half integer value when $\bar{n}$ is between $i$ and $j$, and take integer values otherwise.

The factor $A(x)_t$ now becomes \beqn
    A(x)_t \sim \sum_{\{E_{\bar{n}}\}} \prod_{\bar{n} }^{L-1} \exp\left( -
    \frac{E_{\bar{n}}^2}{2 \, \tilde{t}} \right),
\eeqn The reason $A(x)_t$ depends on $x$ is because $E_{\bar{n}}$ takes half-integer (integer) values inside (outside) of the segment $\bar{n} \in (i, j)$. To evaluate the $x$ dependence of $A(x)_t$, we will need two identities: \beqn \sum_{k = -\infty}^{\infty}e^{ - k^2/(2 \, \tilde{t})} = \theta_3\left(0,\frac{\ii}{2\pi \tilde{t} }     \right), \quad \sum_{k=-\infty}^{\infty}e^{ - (k+\frac{1}{2})^2 /(2 \, \tilde{t}) } = \theta_2\left( 0, \frac{ \ii }{2\pi \tilde{t}} \right). \eeqn Therefore $A(x)_t$ depends on $x$ in the following way:
\beqn
    A(x)_t \sim \prod_{n=i}^{j-1} \left( \frac{\theta_2\left(0,\frac{\ii }{2\pi \td} \right)}{\theta_3 \left(0,\frac{ \ii }{2\pi \td} \right)} \right), \eeqn and $B(x)$ is given by $ B(x) \sim \exp(- \frac{|i-j|}{8 \, \td})$. Combining $A(x)$ and $B(x)$ gives a correlation length of the R\'enyi-1 correlator: \beqn C^{(1)}(x)_t = A(x)_t B(x)_t \sim e^{- x/\xi}, \cr\cr \xi^{-1} \sim -\log \left(\frac{\theta_2(0,\frac{ \ii }{2\pi \td})}{\theta_3(0,\frac{ \ii }{2\pi \td})} \right) + \frac{1}{8 \, \td}, \eeqn and we can plot the correlation length as a function of $ \td$ (Fig.~\ref{fig:1dexact}).    
\begin{figure}[!h]
\centering
\includegraphics[width=0.6\linewidth]{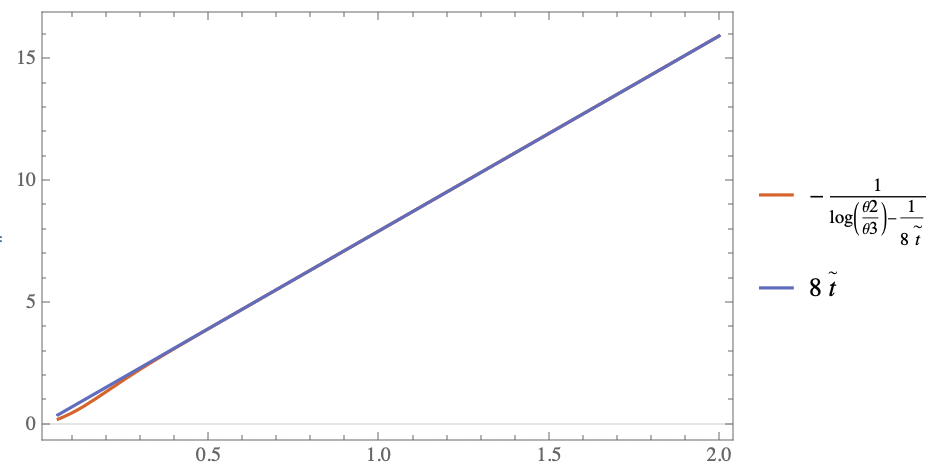}
\caption{The correlation length $\xi$ of the R\'enyi-1 correlator of the 1d decohered rotor model v.s. $\tilde{t}$. }
\label{fig:1dexact}
\end{figure}

Fig.~\ref{fig:1dexact} shows that the correlation length $\xi$ of the R\'enyi-1 correlator converges to $8 \, \td$ rapidly. In fact the asymptotic behavior of $\xi$ in the limit $t\rightarrow \infty$ can be extracted analytically: \beqn \frac{\theta_2(0,\frac{ \ii }{2\pi \td})}{\theta_3(0,\frac{ \ii }{2\pi \td})} \sim 1 - 4 e^{- 2 \pi^2 \, \td} + O(e^{-4 \pi^2 \, \td}), \eeqn therefore with large $\td$, $A(x)_t$ does not decay with $x$, as was expected.

Therefore $\xi/ \td$ saturates rapidly to a constant with increasing $t$, though at short $t$ it is slightly smaller. We stress that this is because we are considering a rotor model whose rotor number is quantized as integers. If the rotor numbers take half-integer values, which is a better description for the spin-1/2 systems simulated numerically in this work, $\xi/\td$ would still saturate to the same constant at long $t$, but it takes a larger value at short $t$. This is consistent with Fig.~\hyperref[fig:1d-renyi]{\ref*{fig:1d-renyi}(a)}, which shows that $\xi/t$ slightly decreases with increasing $t$. 

\section{R\'enyi-$Q$ CMI for the decohered rotor model in 1d}\label{app:renyiCMI}

Below we present a calculation for the R\'enyi-$Q$ conditional mutual information(CMI) for the 1d decohered rotor model. The R\'enyi-$Q$ CMI is defined as
\begin{equation}
    I^{(Q)}(A:C|B) = S^{(Q)}(AB)+S^{(Q)}(BC)-S^{(Q)}(B) - S^{(Q)}(ABC),
\end{equation}
or more explicitly
\begin{equation}
    I^{(Q)}(A:C|B) = -\ln\left(\frac{\Tr[\hat\rho_{AB}^Q]\Tr[\hat\rho_{BC}^Q]}{\Tr[\hat\rho_{ABC}^Q]\Tr[\hat\rho_B^Q]}\right).
\end{equation}
We consider the entire system to be a long and finite chain with $L = R_A+R_B+R_C$ sites with open boundary conditions at the two ends. We partition the system into three consecutive regions $A,B,C$, where region $B$ consists of $R_B$ consecutive sites. 

To calculate the R\'enyi-$Q$ CMI we need to evaluate the reduced density matrices of the subregions of the chain. We do this within the doubled-state formalism. We can expand an arbitrary density matrix in the number occupation basis in the doubled Hilbert space as
\begin{equation}
    \hat{\rho} = \sum_{n_L,n_R}\rho(\{n_L\},\{n_R\})|\{n_L\}\rangle|\{n_R\}\rangle.
\end{equation}
In our setup, the density matrix $\hat\rho_{ABC}$ is always diagonal in the $\{n\}$ basis, i.e.
\begin{equation}
    \{n_L\} = \{n_R\}.
\end{equation} We therefore simplify the notation to
\begin{equation}
    |\{n\}\rangle \equiv |\{n_L=n\}\rangle|\{n_R=n\}\rangle.
\end{equation}
Staying within the diagonal subspace of the doubled Hilbert space means we are dealing with classical probability distributions, and obtaining the reduced density matrices is equivalent to calculating the marginal probability in these subregions. To illustrate the procedure we start with the reduced density matrix $\hat\rho_{AB}$, which is obtained from tracing out the degrees of freedom in region $C$. 
\begin{equation}
\begin{aligned}
    P(\{n_{AB}\}) = &\sum_{\{n_C\}}\langle \{n_{AB}\},\{n_C\}|e^{t\sum_{ij} \cos(\hat\phi_i-\hat\phi_j)}|\{0\}\rangle\\
    =&\langle \{n_{AB}\},\{\phi_C=0\}|e^{t\sum_{ij} \cos(\hat\phi_i-\hat\phi_j)}|\{0\}\rangle,
\end{aligned}
\end{equation}
Here we used $|\{0\}\rangle$ to label the initial state with zero number occupation everywhere, and we use the notation
\begin{equation}
    \sum_{n_i}\langle n_i| = \langle\phi_i = 0|,
\end{equation}
where $\hat\phi_i$ is the angular operator canonically conjugate to the number operator $\hat n_i$. Performing the summation over occupation numbers in region $C$ fixes $\phi_i=0$ for all sites $i\in C$. And $\Tr[\hat\rho_{AB}^Q]$ becomes the partition function
\begin{equation}
    \Tr[\hat\rho_{AB}^Q] = \sum_{\{n_{AB}\}}P(\{n_{AB}\})^Q\equiv Z^{(Q)}_{AB},
\end{equation}
and $Z^{(Q)}_{AB}$ is given by
\begin{equation}
    Z^{(Q)}_{AB} =\sum_{\{n_{AB}\}} \int \prod_{k=1}^{Q}\prod_{i\in AB}d\phi^k_i \ \exp(t\sum_{k = 1}^Q\sum_{ij\in AB}\cos(\phi^k_i-\phi^k_j) + \ii \sum_{ i \in AB} n_i \phi^k_i).
\end{equation}
Note that the link connecting $B$ and $C$, with $i \in B$ and $i+1 \in C$ now becomes
\begin{equation}
    \cos(\phi_i-\phi_{i+1}) \rightarrow \cos(\phi_i),
\end{equation}
since $\phi_{i+1} = 0$. Including site $i+1$ therefore imposes a fixed boundary condition for $\phi$
at the right boundary of region B.
\\

We can further perform the summation over $\{n_{AB}\}$, which imposes a constraint $\sum_{k = 1}^Q \phi^k_{i} = 0$. We eliminate $\phi_i^{Q}$ using the constraint
$\sum_{k=1}^Q \phi_i^k = 0$,
writing
$\phi_i^{Q} = -\sum_{k=1}^{Q-1} \phi_i^k$. 

\begin{equation}
    Z^{(Q)}_{AB} = \int \prod_k^{Q-1}\prod_{i\in AB}d\phi^k_i \ \exp[t\sum_{k = 1}^{Q-1}\sum_{ij\in AB}\cos(\phi_i^k - \phi_j^k) + t\sum_{ij \in AB}\cos(\sum_{k = 1}^{Q-1} \phi_i^k - \sum_{k = 1}^{Q-1}\phi_j^k)].
\end{equation}

We can similarly evaluate the partition function for the other subregions. The partition function for region $BC$ is
\begin{equation}
    Z^{(Q)}_{BC} = \Tr[\hat\rho_{BC}^Q] ,
\end{equation}
with a fixed boundary condition for $\phi$ at the left boundary of region B,
\begin{equation}
    \cos(\phi_{i-1}-\phi_i)\rightarrow \cos(\phi_{i}),
\end{equation}
where $i-1\in A$ and $i\in B$. For the reduced density matrix $\rho_B$, we have
\begin{equation}
    Z^{(Q)}_{B} = \Tr[\hat\rho_{B}^Q],
\end{equation}
with fixed boundary condition for $\phi$ at both the left boundary of $B$ and the right boundary of $B$. 
And finally
\begin{equation}
    Z^{(Q)}_{ABC} = \Tr[\hat\rho_{ABC}^Q], 
\end{equation}
where $\phi$ is not fixed anywhere.

\subsubsection{Transfer matrix formalism}

It is most convenient to use the transfer matrix formalism to evaluate the partition function for the 1d system we are considering.
To get a better analytical handle, hereafter we adopt the Villain formalism. 
In terms of transfer matrix, $Z^{(Q)}_{AB}$ reads
\begin{equation}
    Z^{(Q)}_{AB} = \int \prod_{k=1}^{Q-1} \prod_{i\in AB}d\phi^k_i \ \prod_{i\in AB}\left(T(\sum_{k=1}^{Q-1}\phi_i^k, \sum_{k=1}^{Q-1}\phi^k_{i+1}) \prod _{k=1}^{Q-1}T(\phi^k_i, \phi^k_{i+1})\right),
\end{equation}
and for the Villain model the transfer matrix is
\begin{equation}
    T(\phi^k_i,\phi^k_{i+1}) = \sum_n \exp(-\frac{t}{2}(\phi^k_i-\phi^k_{i+1}+2\pi n)^2).
\end{equation}
Within the transfer matrix formalism, the boundary conditions become initial and final states on which the transfer matrices act on. Since we are summing over all $\phi$, the open boundary condition becomes
\begin{equation}
    \langle n=0| =\frac{1}{2\pi} \int d\phi_0 \langle\phi_0|, 
\end{equation}
and a fixed boundary condition for $\phi=0$ can be written as
\begin{equation}
    \langle \phi=0| = \sum_n \langle n|.
\end{equation}
The four partition functions now read
\begin{equation}
\begin{aligned}
    Z^{(Q)}_{ABC} =& \langle n=0|\bar{T}^{R_A + R_B + R_C-1}|n=0\rangle, \\
    Z^{(Q)}_{B} =& \langle \phi=0| \bar{T}^{R_B+1}|\phi=0\rangle, \\
    Z^{(Q)}_{AB} = & \langle n=0|\bar{T}^{R_{A} + R_B}|\phi=0\rangle, \\
    Z^{(Q)}_{BC} = & \langle \phi=0|\bar{T}^{R_B + R_C}|n=0\rangle, 
\end{aligned}
\end{equation}
where $\bar{T}$ stands for the total transfer matrix for the $Q-1$ species of $\phi$'s. 

The partition function is a product of transfer matrices across the chain, one can perform a basis transformation and go to the eigenstate basis of the transfer matrices.
The eigenvalue for a single transfer matrix is 
\begin{equation}
\begin{aligned}
    \lambda_n = \int d\phi \  e^{\ii n\phi}T(\phi) = \sqrt{\frac{2\pi}{t}}e^{-\frac{n^2}{2t}}.
\end{aligned}
\end{equation}
And then the eigenvalue of  $\bar{T}$ can be expressed as
\begin{equation}
\begin{aligned}
    \lambda^{(Q)}_{\vect{n}} =& \int e^{\ii \vect{n}\cdot \vect{\phi}}\ T(\vect{\phi})\\
    =& \sum_r \lambda_r \prod_{k=1}^{Q-1} \lambda_{n^k-r}.
\end{aligned}
\end{equation}
where $\vect{n} = (n^1,...,n^{Q-1})$. For convenience we will also use $\vect{\phi} = (\phi^1,...,\phi^{Q-1})$.

Now the partition functions can be expressed in terms of these eigenvalues of the transfer matrix as 
\beqn
    Z^{(Q)}_{ABC} &=& \langle \vect{n}=\vect{0}|\bar{T}^{R_A + R_B + R_C}|\vect{n}=\vect{0}\rangle = \lambda_{\vect{0}}^{R_A + R_B + R_C-1}, \cr\cr
    Z^{(Q)}_{B} &=& \langle \vect{\phi} = \vect
    0| \bar{T}^{R_B+1}|\vect{\phi}=\vect{0}\rangle = \sum_{\vect{m}}\lambda_{\vect{m}}^{R_B + 1}, \cr\cr
    Z^{(Q)}_{AB} &= & \langle\vect{n}=\vect{0}|\bar{T}^{R_{A} + R_B}|\vect{\phi}=\vect{0}\rangle = \lambda_{\vect{0}}^{R_A + R_B}, \cr\cr
    Z^{(Q)}_{BC} &= & \langle \vect{\phi}=\vect{0}|\bar{T}^{R_B + R_C}|\vect{n}=\vect{0}\rangle  = \lambda_{\vect{0}}^{R_B + R_C}. \label{ZQ}
\eeqn

\subsubsection{R\'enyi-2 CMI}

We now consider the R\'enyi-2 CMI. The eigenvalue of the transfer matrix admits closed form expressions as
\begin{equation}
\begin{aligned}
    \lambda^{(2)}_m =& \sum_r \exp(-\frac{r^2}{2t} - \frac{(r-m)^2}{2t})\\
    =&e^{-\frac{m^2}{4t}}\sum_r \exp(-\frac{(r-\frac{m}{2})^2}{t})\\
    =&e^{-\frac{m^2}{4t}}\theta_3(0,\frac{\ii}{\pi t}), \quad \mathrm{m \  even}\\
    &e^{-\frac{m^2}{4t}}\theta_2(0,\frac{\ii}{\pi t}),\quad  \mathrm{m \  odd}
\end{aligned}
\end{equation}
Now we have
\begin{equation}
    \begin{aligned}
        \frac{Z^{(2)}_{ABC}Z^{(2)}_{B}}{Z^{(2)}_{AB}Z^{(2)}_{BC}}  =&\sum_{m'} e^{-\frac{(2m')^2(R_B + 1)}{4t}} + \sum_{m'}e^{-\frac{(2m'+1)^2(R_B+1)}{4t}}\left(\frac{\theta_2(0,\frac{ \ii}{\pi t})}{\theta_3(0,\frac{\ii}{\pi t})} \right)^{R_B+1}\\
        =& \theta_3 \left(0, \frac{\ii(R_B+1)}{\pi t} \right) + \left(\frac{\theta_2(0,\frac{\ii}{\pi t})}{\theta_3(0,\frac{\ii}{\pi t}} \right)^{R_B + 1}\theta_2\left(0, \frac{ \ii (R_B + 1)}{\pi t} \right).
\end{aligned}\label{eq:r2partition}
\end{equation}
We are most interested in the asymptotics. We look at the limit where $(R_B+1)/t \gg 1$, then we are allowed to keep just the leading terms which lead to
\begin{equation}
\begin{aligned}
    I^{(2)}(A:C|B) &\sim \log(1+2e^{-\frac{R_B + 1}{t}}+2 \left(\frac{\theta_2}{\theta_3} \right)^{R_B + 1}e^{-\frac{R_B+1}{4t}} + ...)\\
    &\sim 2e^{-\frac{R_B + 1}{4t}}.
\end{aligned}   
\end{equation}
Here we have used the fact that $\theta_2(0,\frac{ \ii}{\pi t})/\theta_3(0,\frac{\ii}{\pi t})$ rapidly approaches $1$ for large $t$. \\

To bridge the previous numerical calculations in Sec:~\ref{sec:classical},we consider the geometry where $A$, $B$, $C$ are all parts of a larger system. In this case the four partition functions all take a similar form:
\begin{equation}
    \begin{aligned}
        Z^{(2)}_{ABC} = &\sum_{m} (\lambda_m^{(2)})^{R_A + R_B + R_C+1} \\
        Z^{(2)}_{AB} =&\sum_{m}(\lambda_m^{(2)})^{R_A + R_B + 1}\\
        Z^{(2)}_{BC} =& \sum_m (\lambda_m^{(2)})^{R_B + R_C+1}
    \end{aligned}
\end{equation}
For the limit $(R_B+1)/t \gg 1$ the CMI still decays exponentially. On the other hand, if we consider the limit $(R_B+1)/t \ll 1$ we may replace the sum in Eq.\eqref{eq:r2partition} with an integral, which leads to
\begin{equation}
    Z_B^{(2)} \sim \sqrt{\frac{\pi t}{R_B+1}} + \sqrt{\frac{\pi t}{R_B+1}}\left(\frac{\theta_2(0,\frac{\ii}{\pi t})}{\theta_3(0,\frac{\ii}{\pi t}} \right)^{R_B + 1}
\end{equation}
The ratio between partition functions in Eq.~\eqref{eq:r2partition} now becomes 
\begin{equation}
\frac{Z_{ABC}^{(2)}Z_{B}^{(2)}}{Z_{AB}^{(2)}Z_{BC}^{(2)}} \sim \sqrt{\frac{(R_A + R_B + R_C)(R_B)}{(R_A + R_B) (R_B + R_C)}}.
\end{equation}
Let's assume that $R_A , R_C \ll R_B$, as the case in the numerical computation in Sec:~\ref{sec:classical},. In this geometry the leading order expression of $I^{(2)}(A:C|B)$ decays as a power-law of $R_B$, precisely consistent with the numerics:
\begin{equation}
    I^{(2)}(A:C|B) \sim \frac{1}{R_B^2}.
\end{equation}

\section{Stiffness and winding number fluctuations}\label{sec:winding}

The stiffness of a superfluid phase is usually defined as \beqn
\rho_s = \lim_{\delta \ra 0} \partial^2_\delta F(\delta), \eeqn
where $F(\delta) = - \ln Z (\delta)$ is the free energy of the
rotor model with twisted boundary condition $\phi(x + L) = \phi(x)
+ \delta$. When evaluating the R\'enyi-2 stiffness, we need to use
the R\'enyi-2 partition function \beqn  Z^{(2)} (\delta)_t &=& \frac{1}{\left( {\cal N}_t \right)^2} \sum_{ \mm } \int \prod_{a =
1,2} D\phi_a \ e^{- \int d^2x \sum_{a=1,2} - \td \cos \left(
\nabla_x \phi_a + \frac{\delta}{L} \right) -
\td\cos(\nabla_y \phi_a)^2} e^{\ii (\phi_{1,i} + \phi_{2,i})
m_i } \cr\cr &\sim& \frac{1}{\left( {\cal N}_t \right)^2} \sum_{\mm} \int \prod_{a =
1,2} D\phi_a \ e^{- \int d^2x \sum_{a=1,2} \frac{\td}{2} \left(
\nabla_x \phi_a + \frac{\delta}{L} \right)^2 +
\frac{\td}{2}(\nabla_y \phi_a)^2} e^{\ii (\phi_{1,i} + \phi_{2,i})
m_i }. \eeqn Summing over $m_i$ leads to constraint $\phi_{1,i} +
\phi_{2,i} = 2\pi \mathbb{Z}$. In the Gaussian phase we ignore the
periodicity of $\phi_a$, and replace the constraint by $\phi_{1,i} +
\phi_{2,i} = 0$. Then the partition function is \beqn Z^{(2)}
(\delta)_t \sim \frac{1}{\left({\cal N}_t \right)^2} \int D\phi \ \exp\left( - \td \, \int d^2 x \frac{\delta^2}{L^2} - \int d^2x \ \td (\nabla_x
\phi)^2 + \td(\nabla_y \phi)^2
\right). \eeqn Here ${\cal N}_t$ is a normalization factor - the path-integral of a single component rotor model with zero twisting and summation over $\mm$. Note that in this path-integral we need to fix the
periodic boundary condition for $\phi(x)$. The R\'enyi-2 stiffness
then follows as $2 \, \td$.

\medskip

The ``disorder-averaged stiffness" is defined through the disorder
averaged winding fluctuation \beqn \bar{\rho}_s = \sum_{\mm}
\frac{1}{L^2}P(\mm) W_{\mm}^2 = \sum_{\mm} P(\mm) \
\lim_{\delta \ra 0}
\partial^2_{\delta} F(\mm, \delta). \eeqn Here $F(\mm, \delta)
= - \ln Z(\mm, \delta)$ is the free energy of the rotor model
at disorder $\mm$ and twisting $\phi \ra \phi + x \delta  /L$.
In the replica field theory, $\bar{\rho}_s$ reads \beqn
\bar{\rho}_s = - \lim_{Q \ra 1} \frac{1}{Q - 1} \sum_{\mm}
Z (\mm)_t \lim_{\delta \ra 0}
\partial^2_\delta \left( Z (\mm, \delta)_t \right)^{Q-1}. \eeqn
We first ignore the limit and derivative, and evaluate the
quantity \beqn && \sum_{\mm} Z (\mm)_t \left( Z (\mm, \delta)_t
\right)^{Q-1} \cr\cr &\sim& \frac{1}{\left({\cal N}_t \right)^Q} \int \prod_{a = 1}^{Q-1} D \phi_a \
\exp\left( - \int d^2x \ \sum_{a = 1}^{Q - 1} \frac{\td}{2} \left(
\nabla_x \phi_a + \frac{\delta}{L} \right)^2 + \frac{\tilde{t}}{2}
(\nabla_y \phi_a)^2 + \frac{\td}{2} (\nabla_\mu \sum_{a = 1}^{Q-1}
\phi_a )^2 \right) \cr\cr &\sim& Z^{(Q)}_t \exp\left( - \frac{\td}{2}
(Q-1) \delta^2 \right). \eeqn Note that we have imposed the
constraint arising from summing over $\mm$, and used the fact that
$\phi$ has a periodic boundary condition. Here $Z^{(Q)}_t$ is the R\'enyi-$Q$
partition function with zero twisting: \beqn Z^{(Q)}_t &=& \frac{1}{\left({\cal N}_t \right)^Q} \int
\prod_{a = 1}^{Q} D \phi_a \ \delta(\sum_a \phi_a) \exp\left( - \sum_{ij} \, \sum_{a = 1}^Q - t \cos(\nabla_{ij} \phi_a) \right) \cr\cr &\sim& \frac{1}{\left({\cal N}_t \right)^Q} \int
\prod_{a = 1}^{Q} D \phi_a \delta(\sum_a \phi_a) \exp\left( - \int
d^2x \, \sum_{a = 1}^Q \frac{\td}{2} (\nabla_\mu \phi_a)^2 \right) \cr\cr &=& \frac{1}{\left({\cal N}_t \right)^Q}
\int \prod_{a = 1}^{Q-1} D \phi_a \exp\left( - \int d^2x \,
\sum_{a = 1}^{Q-1} \frac{\td}{2} (\nabla_\mu \phi_a)^2 +
\frac{\td}{2} (\nabla_\mu \sum_{a = 1}^{Q-1} \phi_a ) ^2 \right).
\eeqn The disorder-averaged stiffness eventually reduces to \beqn
\bar{\rho}_s = - \lim_{Q \ra 1} \frac{1}{Q-1} \ \lim_{\delta \ra 0}
\partial^2_\delta \Big[ Z^{(Q)}_t \exp\left( - \frac{\td}{2}
(Q-1) \delta^2 \right)\Big] = \td. \eeqn We have used the fact that $\lim_{Q \ra 1} Z^{(Q)}_t = 1$, as it is just the total probability of all $\mm$.

\section{Derivation of Model F Wigner function}\label{app:wigner}
In Sec.~\ref{sec:ModelF}, we derived the action for the propagator at $t > t_c$ (after SW-SSB has occurred) and with $J$ and $K$ small. In particular, we obtained the action given in Eq.~\eqref{eq:model_F_final_action} that reduces to classical, stochastic dynamics. By treating the temporal boundary conditions more carefully, we can also use this expression to derive the Wigner function at late times, which we do here.

To begin, we note that in the course of the derivation we integrated by parts to transform $\ii n_j \dot{\tilde{\phi}}_j$ to $-\ii \tilde{\phi}_j \dot{n}_j$. In the process, we also acquire boundary terms such that Eq.~\eqref{eq:model_F_final_action} becomes
\begin{multline}
    S = \int_t \sum_j i \tilde{n}_j [\dot{\phi}_j - \partial F/\partial n_j  + \beta \gamma_\phi \partial F/\partial \phi_j - \eta_j]
+ \int_t \sum_j i \tilde{\phi}_j  [ \dot{n}_j + \partial F/\partial \phi_j - \beta \gamma_n \nabla_j^2 \partial F/\partial n_j - \xi_j] \\ - \int_t \sum_j \left(\frac{\eta_j^2}{4\gamma_\phi} + \frac{\xi_j^2}{4\gamma_n \nabla^2}\right)  + i\sum_j [\tilde{\phi}_j(T)n_j(T) - \tilde{\phi}_j(0)n_j(0)].
\end{multline}
In terms of this action, the state at time $T$ is given by 
\[
&\left\langle\phi(T)-\frac{\tilde{\phi}(T)}{2}\right|\rho(T)\left|\phi(T)+\frac{\tilde{\phi}(T)}{2}\right\rangle\\
&\quad\quad = \sum_{W_R,W_L} \int \prod_{j} D\phi_j(t)D\tilde{\phi}_j(t)Dn_j(t)D\tilde{n}_j(t)D\eta_j(t)D\xi_j(t)e^{S(T)}\left\langle\phi(0)-\frac{\tilde{\phi}(0)}{2}\right|\rho(0)\left|\phi(0)+\frac{\tilde{\phi}(0)}{2}\right\rangle\label{eq:app:modelF_1}\\
&\quad\quad = \sum_{W_R,W_L} \int \prod_{j} dn_j(T) D\eta_j(t)D\xi_j(t)\exp[- \int_t \sum_j \left
(\frac{\eta_j^2}{4\gamma_\phi} + \frac{\xi_j^2}{4\gamma_n \nabla^2}\right)]\exp[i\sum_j \tilde{\phi}_j(T)n_j(T)]\label{eq:app:modelF_2}\\
&\hspace{2cm}\times\prod_j\delta\left[
\dot{\phi}_j - \left(\partial F/\partial n_j  - \beta \gamma_\phi \partial F/\partial \phi_j + \eta_j\right)\right]\delta\left[
\dot{n}_j - \left(- \partial F/\partial \phi_j + \beta \gamma_n \nabla_j^2 \partial F/\partial n_j + \xi_j\right)\right]\label{eq:app:modelF_3}\\
&\quad\quad\propto \int \prod_j dn_j(T) e^{\ii \sum_j \tilde{\phi}_j(T)n_j(T)} e^{-\beta F[\phi(T),n(T)]},\label{eq:app:modelF_4}
\]
where some clarifying remarks are in order. From Eq.~\eqref{eq:app:modelF_1} to Eq.~\eqref{eq:app:modelF_2}-\eqref{eq:app:modelF_3}, we have integrated over all $\tilde{\phi}_j(t)$ except for $\tilde{\phi}_j(T)$ and all $\tilde{n}_j(t)$, with the result that Model F stochastic differential equations are enforced. Then, we have integrated over all $\phi_j(t)$, except for $\phi_j(T)$ which is set externally, and all $n_j(t)$, except for $n_j(T)$ which remains nontrivially coupled to $\tilde{\phi}_j(T)$. For a fixed realization of noise and fixed boundary conditions, only one path for $\phi_j(t)$ and $n_j(t)$ is possible, making these integrations trivial. We emphasize that the measure $\prod_j dn_j(T)$ in Eq.~\eqref{eq:app:modelF_2} and Eq.~\eqref{eq:app:modelF_4} corresponds to $n$ degrees of freedom only at time $T$, not over all time. Finally, to arrive at Eq.~\eqref{eq:app:modelF_4}, we organized the ensemble of realizations of noise into the steady-state probability distribution $e^{-\beta F[\phi(T),n(T)]}$.

The Wigner function for the rotor density matrix at time $t$ is
\[
W[\phi(t), n(t)] = \int \prod_j d\tilde{\phi}_j (t)\left\langle\phi(t)-\frac{\tilde{\phi}(t)}{2}\right|\rho(t)\left|\phi(t)+\frac{\tilde{\phi}(t)}{2}\right\rangle e^{-i\sum_j\tilde{\phi}_j(t)n_j(t)}.
\]
Therefore, it follows that
\[
W[\phi(t), n(t)] &\propto \int \prod_j d\tilde{\phi}_j(t) \left( \int \prod_j dn'_j(t) e^{\ii\sum_j \tilde{\phi}_j(t)n'_j(t)} e^{-\beta F[\phi(t),n'(t)]} \right)e^{-i\sum_j\tilde{\phi}_j(t)n_j(t)}\\
&\propto \int \prod_j dn'_j(t) e^{-\beta F[\phi(t),n'(t)]} \prod_j \delta[n'_j(t) - n_j(t)]\\
&\propto e^{-\beta F[\phi(t),n(t)]}
\]
as we use in Sec.~\ref{sec:ModelF:Wigner}.

\section{Bhattacharyya distance in continuum hydrodynamics}\label{app:Bhattacharyya}

In this Appendix we derive the formula for the Bhattacharyya distance within the Gaussian hydrodynamics, Eq.~\ref{eq:Bhattacharyya}, which is used to generate the numerical data in Fig.~\ref{fig:gaussianCMI}a. 

We first introduce some standard notation which will allow us to discuss conditional distributions of Gaussian-distributed variables. The density is denoted $ n_i \equiv n(x_i)$, with mean $\mu_i = \langle {n}_i\rangle $ and covariance $S_{ij} =  \langle ({n}_i-\mu_i)({n}_j-\mu_j) \rangle $. When the fluctuations are distributed according to the normal distribution, we will write $\bm {n} \sim \mathcal{N}(\bm \mu, \bm S)$. Below we will be interested in the case where a subset of the densities $\bm \bn _2 $ are conditioned to take fixed values. We partition the density as $\bm \bn = [\bm \bn_1, \bm \bn_2]$, and similarly with $\bm\mu = [\bm\mu_1, \bm\mu_2]$, 
\begin{equation}
    \bm S= \begin{pmatrix}
        \bm S_{11} & \bm S_{12} \\ \bm S_{21} & \bm S_{22}
    \end{pmatrix}.
\end{equation}
The distribution of $\bm \bn _1$ conditioned on the values $\bm \bn_2 = \bm q_2 $ is given by a new normal distribution $\bm \bn _{1|\bm q_2} \sim \mathcal{N}(\bm\mu_{|\bm{q}_2}, \bm S_{|\bm{q}_2})$, where 
\begin{equation}
    \bm \mu_{|\bm{q}_2} = \bm\mu_1 +\bm S_{12}\bm S^{-1}_{22}(\bm q_2 - \bm \mu_2), \hspace{1cm}
    \bm S_{|\bm{q}_2} = \bm S_{11} -\bm S_{12}\bm S^{-1}_{22}\bm S_{21}.
\end{equation}
Intuitively, observing fluctuations of $\bm  \bn_2$ further from the mean leads to a stronger revised belief for correlated values of $\bm  \bn_1$. Note that the conditioned covariance $\bm S_{|\bm{q}_2}$ does not depend on the actual values of $\bm q_2$.

Next, we introduce a symmetrised version of the fidelity correlator to probe SW-SSB in the Gaussian framework. We consider an initial density distribution given by the hydrodynamic solution, with $\Sigma_{ij}(t) = \Sigma_n(x_i-x_j,t) $ given by Eq.~\ref{eq:xcovariance} and zero mean $\bm \mu =\bm 0$. We will suppress the time label below for clarity. We will condition the density at two points $x=0$ and $x=r$ i.e.~$\bm \bn_2 = [ n_0,  n_r]$.  The fidelity is taken between two oppositely conditioned states: the first with $\bm \bn_2 = \bm{q}_2 = [q,-q]$, and the second with $\bm \bn_2 = -\bm{q}_2 = [-q,q]$. Writing $P^{+-}_q \equiv \mathcal{N}(\bm\mu_{|\bm{q}_2}, \bm S_{|\bm{q}_2})$ and $P^{-+}_q \equiv \mathcal{N}(\bm\mu_{|-\bm{q}_2}, \bm S_{|-\bm{q}_2})$ the Bhattacharyya coefficient is then given by
\begin{equation}
    \text{BC}_r(P^{+-}_q (\bm n  ):P^{-+}_q(\bm n)
) \equiv \exp({-\mathcal{B}_r(P^{+-}_q (\bm n  ):P^{-+}_q(\bm n))}) = \int d\bm \bn \sqrt{P^{+-}_q (\bm n)P^{-+}_q(\bm n))},
\end{equation}
where we introduced the Bhattacharyya ``distance'' $\mathcal{B}_r$. The Bhattacharyya distance determines the decay of the fidelity as the charge separation $r$ is varied. For a Gaussian state,  $\mathcal{B}_r$ is given by \cite{dodge2003oxford}
\begin{equation}
    \mathcal{B}_r(P^{+-}_q (\bm n  ):P^{-+}_q(\bm n)) = \frac{1}{8}(\bm\mu_{|\bm q_1} - \bm \mu_{|-\bm q_1})^T  \bar{\bm S}^{-1} (   \bm\mu_{|\bm q_1} - \bm \mu_{|-\bm q_1}) + \frac{1}{2} \log \Bigg(\frac{\|\bar{\bm S}\|}{\sqrt{\|{\bm S}_{|\bm q_1}\| \|{\bm S}_{|-\bm q_1}\|}}\Bigg) , 
\end{equation}
where $\bar{\bm S} = \frac{1}{2} (\bm S_{|\bm q_1}+\bm S_{-|\bm q_1})$ and $\|\bm A\| = \text{det}(\bm A)$. Note that in the symmetrised setup, we have $\bm S_{|\bm{q}_2} = \bm S_{|-{\bm{q}}_2}$ as the covariance doesn't depend on the conditioned values, and furthermore $\bm\mu_{|\bm q_1} = - \bm\mu_{|\bm -q_1}$ whenever $\bm \mu = \bm 0$. In this case the  Bhattacharyya distance is equivalent to the Kullback-Leibler divergence of the two conditional distributions and simplifies to 
\begin{equation}
    \mathcal{B}_r(P^{+-}_q (\bm n  ):P^{-+}_q(\bm n)) =  \frac{1}{2} \bm \mu_{|\bm q_1}^T  \bm S_{|\bm q_1}^{-1}\bm \mu_{|\bm q_1},
\end{equation}
which is Eq.~\ref{eq:Bhattacharyya}. The above equation is essentially the energy of a charge-dipole with charges at $x=0$ and $x=r$.

\end{document}